\documentclass[twocolumn]{aastex631} 

\usepackage{csquotes} 
\usepackage{xspace}
\usepackage{xcolor}
\usepackage{amsmath} 

\newcommand{\AL}{$A_L$\xspace}
\newcommand{\ALw}{$A_{L, w}$\xspace}
\newcommand{\ALuw}{$A_{L, uw}$\xspace}

\graphicspath{{./}{figures/}}

\shorttitle{Can Clump Properties Predict Core Distribution in Star Formation? A Statistical Analysis of MHD Simulations}
\shortauthors{Wei-An Chen et al.}

\begin{document}

\title{Can Clump Properties Determine Core Distribution in Star Formation? A Statistical Analysis of MHD Simulations}

\author[0000-0002-8336-2837]{Wei-An Chen}
\affiliation{Graduate Institute of Astrophysics, National Taiwan University, No.\ 1, Sec.\ 4, Roosevelt Rd., Taipei 10617, Taiwan, R.O.C}
\affiliation{Institute of Astronomy and Astrophysics, Academia Sinica, 11F of Astronomy-Mathematics Building, No.\ 1, Sec.\ 4, Roosevelt Rd, Taipei 106216, Taiwan, R.O.C.}

\author[0000-0001-9751-4603]{S. D. Clarke}
\affiliation{Institute of Astronomy and Astrophysics, Academia Sinica, 11F of Astronomy-Mathematics Building, No.\ 1, Sec.\ 4, Roosevelt Rd, Taipei 106216, Taiwan, R.O.C.}

\author[0000-0002-0675-276X]{Ya-Wen Tang}
\affiliation{Institute of Astronomy and Astrophysics, Academia Sinica, 11F of Astronomy-Mathematics Building, No.\ 1, Sec.\ 4, Roosevelt Rd, Taipei 106216, Taiwan, R.O.C.}

\begin{abstract}
	Dense cores, the progenitors of stars, are in sub-pc scale and fragmented from pc-scale clumps.
	However, it is still unclear that how strongly the fragmentation process is affected by the properties of the host clumps, and how these properties influence the core distribution observed in recent millimeter (mm) and sub-mm observations.
	To systematically investigate this relation, we employed MHD simulations of convergent flows to generate a large sample of clumps and analyzed their properties using various techniques. 
	Alignment parameters were used to quantify core distribution, while energy terms were calculated to assess the influence of gravity, magnetic fields, and turbulence.
	We found the core distribution only exhibiting weak correlations between alignment parameters and clump properties.
	For an individual clump, turbulence is believed to significantly contribute to these features by inducing non-homologous collapse and ongoing fragmentation. 
	Nevertheless, for the entire population, more compact core distributions are observed due to the dominance of gravity.
	Overall, these factors suggest that clump properties are not sufficient to accurately determine core distribution. 
\end{abstract}

\section{Introduction} \label{sec:intro}
Observations in millimeter (mm) and sub-mm wavelength have revealed the existence of dense, gravitationally bound cores at sub-parsec scales, which are the precursors of stars and are embedded within structures such as filaments and clumps at $\sim  1$ pc scale. 
The formation of these cores from their host clumps is a process known as fragmentation. 
While gravity plays a crucial role, turbulence, magnetic fields, and stellar feedback from nearby high-mass stars are also believed to influence this process (see review by \citealp{2023ASPC..534..233P}).

To study clump fragmentation, observations of star-forming regions are conducted to investigate the relationship between cores and their host clumps. 
For example, the ASHES survey observed 39 infrared dark clouds (IRDCs) at early stages of high-mass star formation \citep{2019ApJ...886..102S, 2023ApJ...950..148M}. 
Their analysis revealed that core separation, a characteristic outcome of fragmentation, can be well described by thermal Jeans fragmentation \citep{2024ApJ...966..171M}.
Similar results have been reported in other studies of IRDCs \citep{2015A&A...581A.119B, 2018A&A...617A.100B, 2021A&A...649A.113B, 2018ApJ...855...24P, 2019ApJ...871..185L, 2020ApJ...894L..14L, 2024arXiv240706845I}.
However, turbulent fragmentation is thought to be dominant in low-mass star formation regions \citep{2025A&A...695L..25I}.
When considering the role of magnetic fields, \cite{2021ApJ...912..159P} investigated 18 massive dense cores and found a positive correlation between the number of cores at sub-pc scales and the mass-to-flux ratio. 
In contrast, the 20 high-mass star-forming regions studied by the CORE program did not exhibit these correlations but found comparable importance between turbulent and magnetic energy, suggesting the influence of clump evolution \citep{2024A&A...682A..81B}.
These diverse findings highlight the complexity of the fragmentation process.

The effects of the competition between different energy components can also be manifested in core distribution.
Observations by \cite{2019ApJ...878...10T} proposed a classification scheme for core distributions, identifying three distinct patterns: \textit{no fragmentation}, \textit{aligned fragmentation}, and \textit{clustered fragmentation}. 
Aligned fragmentation occurs when cores are distributed along a predominant direction under magnetic field dominance, while clustered fragmentation indicates a random distribution of cores without a preferred direction, which can occur when turbulence is significant. 
No fragmentation suggests that only a single core is present, possibly due to gravity dominating other forces.
This classification of core distribution has been adopted by \cite{2022AJ....164..175C, 2023ApJ...951...68C}, who reported consistent results regarding energy dominance.
However, as these studies are limited to small samples, the universality of this trend remains an open question.

In this work, we extend the discussion of the relationship between clump properties and core distributions by utilizing magnetohydrodynamic (MHD) simulations. 
This approach enables us to obtain a large, unbiased sample of clumps and conduct a systematic analysis.
Recent studies of star formation have shown that high-mass star-forming regions often form within compressed regions generated by shocks from expanding bubbles (e.g., HI shells, HII regions, ionized region, etc.). 
The supersonic nature of these shocks can lead to density enhancements, which can trigger gravitational instability. 
Various factors, including gravity, turbulence, and magnetic fields, shape the subsequent evolution of cloud collapse, core formation, and star formation (see review by \citealp{2023ASPC..534..233P}).
Therefore, to model this process, we employ convergent flow simulation, which involve the collision of two supersonic gas flows. 
As clumps and cores form from the resulting sheet-like structure, we can trace their evolution and properties.
We then use various techniques to derive clump properties and quantify core distribution. 
Additionally, we estimate energy-related parameters to assess their influence. 
These analyses allow us to explore the correlations between core distribution and clump properties.

This paper is structured as follows. 
Section~\ref{sec:Numerical} introduces the numerical setup and describes the initial conditions of the convergent flow. 
Section~\ref{sec:SheetEvo} presents the simulation results, providing an overview of the sheet-like structure. 
In Section~\ref{sec:Clump}, we describe the process used to extract star-forming clumps and analyze their properties in our simulations. 
The evolution of these parameters is also discussed. 
Subsequently, we emulate ALMA observations towards these clumps and quantify the core distributions in Section~\ref{sec:Core}. 
Sections~\ref{sec:ALClump} and \ref{sec:Caveats} discuss the correlations between clump properties and core distributions, as well as the caveats of this study. 
Finally, Section~\ref{sec:Conclusion} summarizes our findings.

\section{Numerical Methods and Simulation} \label{sec:Numerical}
All numerical simulations presented in this work were conducted using the GAMER code\footnote{\url{https://github.com/gamer-project/gamer}} (GPU-Accelerated Magnetohydrodynamics Adaptive Mesh Refinement; \citealp{2018ApJS..236...50Z, 2018MNRAS.481.4815S}). 
GAMER is a well-established framework that employs adaptive mesh refinement (AMR) for high-resolution simulations.
For solving the ideal MHD equations, we utilizes the MUSCL-Hancock scheme (see review by \citealp{toro2009riemann}). 
This scheme is used with the piecewise parabolic method \citep{WOODWARD1984115} for data reconstruction and the Harten-Lax-van Leer discontinuities solver  \citep{MIYOSHI2005315} for resolving Riemann problems at cell interfaces.
Self-gravity calculations are implemented within GAMER. 
The gravitational potential on the coarsest level (root level) is obtained using the fast Fourier transform method from the FFTW library \citep{FFTW3}. 
In refined regions with higher resolution, the gravitational potential is solved independently on each level by employing the successive over-relaxation method \citep{SOR}.
Finally, to maintain a divergence-free magnetic field throughout the simulation, GAMER utilizes the constraint transport (CT) method \citep{1988ApJ...332..659E, 2008JCoPh.227.4123G}.
This method operates on the magnetic fluxes at cell interfaces and is not affected by our sink particle implementation (see below), as only gas properties are removed from the grid during accretion, while the magnetic field itself is conserved and continues to be evolved by the CT scheme.

To overcome the limitations of AMR in capturing the dynamics of collapsing structures, we incorporated sink particles into GAMER. 
These numerical proxies for collapsed objects allow simulations to progress beyond the resolution limit suggested by Truelove's criterion \citep{1997ApJ...489L.179T}: $\lambda_J \geq 4 \Delta x$, where $\lambda_J$ is the local Jeans length and $\Delta x$ is the cell size.
The implementation of sink particle formation follows the criteria established by \cite{2010ApJ...713..269F}. 
When the density of a cell at the highest AMR level surpasses a critical threshold ($\rho_{sink}$), determined using Truelove's criterion, a series of tests are conducted to assess the suitability for sink particle creation. 
These tests ensure the cell resides in a local minimum of gravitational potential and the collapsing structure is gravitationally bound with converging inflowing gas (see \citealp{2010ApJ...713..269F} for more details). 
An additional criterion, based on Equation 5 of \cite{2017MNRAS.468.2489C}, is also included to prevent excessive sink particle formation.
If all criteria are met, a sink particle is created at the cell center and begins accreting gas within the accretion radius ($r_{acc} = 4 \Delta x_{min}$) as described in \cite{2010ApJ...713..269F}. 
For particle evolution, we employed the Triangular-Shaped-Cloud (TSC; \citealp{1988csup.book.....H}) scheme for data interpolation, while the Kick-Drift-Kick \citep{1988csup.book.....H} scheme was used to track particle motion (implementation details are provided in \citealp{2018MNRAS.481.4815S}).
In the context of this study, sink particles are treated as numerical proxies that represent the location and onset of star formation. 
While the criteria for their creation are physically motivated to identify irreversibly collapsing gas, they should not be considered fully realistic protostars, as our simulations do not include crucial physics such as stellar feedback, which we discuss further in Section~\ref{sec:Caveats}.

To solve the full set of MHD equations, an equation of state (EOS) is needed to determine the gas pressure. 
In this work, we employ a barotropic EOS to model the gas evolution across diverse density regimes. 
This approach provides a computationally efficient approximation of the complex thermodynamics in star-forming regions. 
At the lower densities typical of clumps ($\rho < \rho_{ad}$), the gas is optically thin and can efficiently radiate away heat from compression, remaining nearly isothermal. 
However, as a region collapses to form a protostar ($\rho > \rho_{ad}$), it eventually becomes optically thick to its own cooling radiation, which traps thermal energy and causes the temperature to rise adiabatically.

Our EOS is formulated to capture this physical transition from an isothermal to an adiabatic state.
Mathematically, the gas temperature $T$ is related to density $\rho$ as follows:
\begin{equation}\label{eqn:EOS}
	T(\rho) = T_0 (1 + (\frac{\rho}{\rho_{ad}})^{\gamma - 1}),
\end{equation}
where $T_0$ is the initial temperature, $\rho_{ad} = 10^{-14}$ g cm$^{-3}$ is the transition density, which is a canonical choice that represents the approximate density at which the gas becomes optically thick \citep{1998ApJ...495..346M, 2000ApJ...531..350M}, and $\gamma = 5/3$ is the adiabatic index.
The gas pressure $P$ is then given by:
\begin{equation}\label{eqn:Pressure}
	P = \rho c_s^2,
\end{equation}
where $c_s$ is the sound speed at temperature $T(\rho)$.


\subsection{Initial Condition} \label{sec:InitialCondition}
\begin{figure*}[htb!]
	\includegraphics[width=1.0\textwidth]{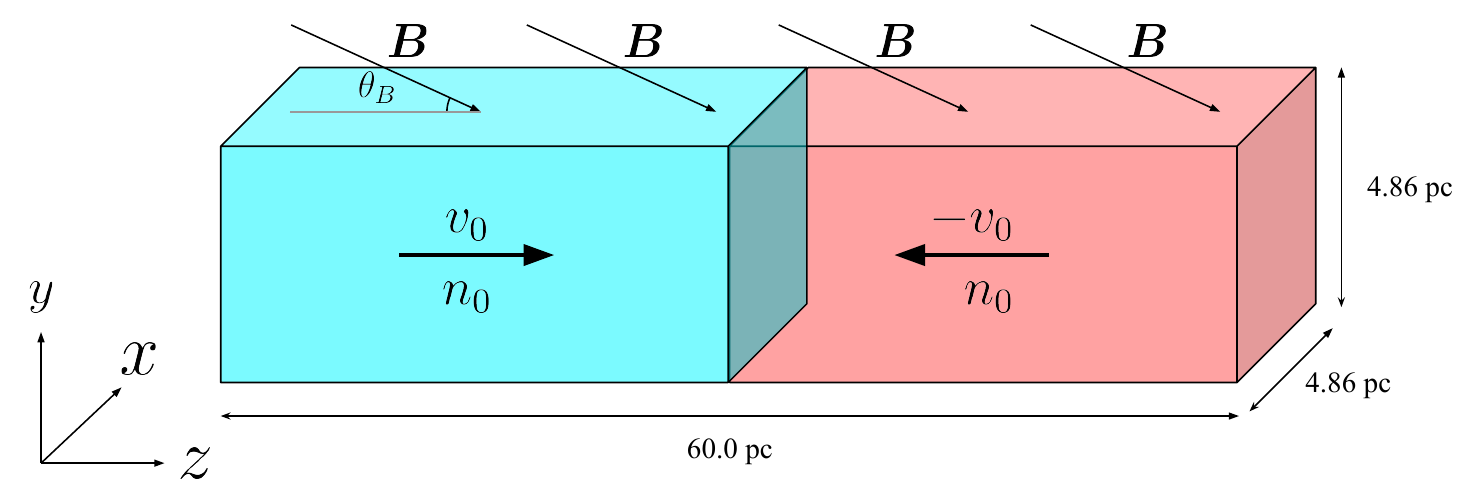}
	\caption{Overview of the initial setup for the convergent flow simulation. The values of the simulation parameters are listed in Table~\ref{tab:SimulationPara}.}
	\label{fig:CFInit}
\end{figure*}

In this work, we employ a convergent flow configuration to form clumps and cores in our simulations.
The simulation domain was established with dimensions of 4.86 pc $\times$ 4.86 pc $\times$ 60 pc, where the long axis aligned with the z-axis represents the direction of gas flow. 
Periodic boundary conditions were applied in the $x$- and $y$-directions, while a diode boundary condition was used in the $z$-direction to simulate a large-scale inflow of turbulent gas. 
For gravity, periodic boundary conditions were implemented in all directions.
An overview of the simulation domain is shown in Figure~\ref{fig:CFInit}.
This simulation effectively replicates the formation of cold and dense regions within molecular clouds. 
These regions can arise from the shock compression of gas driven by expanding bubbles or supersonic turbulence (see review by \citealp{2023ASPC..534..233P}). 
The inhomogeneity of the material and the presence of the shock can trigger instabilities that convert the flow's kinetic energy into turbulent energy, leading to the development of self-sustained and self-consistent turbulence within the sheet-like structure. 
Additionally, the continuous inflow of gas from the large-scale flow supplies mass for the structure's formation and evolution.

For the initial parameters, the gas number density was set to $n_0$ = 135 cm$^{-3}$, the initial temperature was set to $T_0$ = 10 K, and the initial flow velocity was set to $v_0$ = 1.5 km s$^{-1}$, converging towards z = 30 pc (see Figure~\ref{fig:CFInit}).
With these parameters and molecular weight $\mu = 2.3$, the simulations show that a mass of approximately $2\times10^3\ M_{\odot}$ accumulates within a central cubic volume of (4.86 pc)$^3$ after 2 Myr. 

To seed the initial inhomogeneity into the gas density, Burgers turbulence was imposed on $v_0$ with an energy power spectrum $\propto k^{-4}$, and was generated by following the procedure described in \cite{2015MNRAS.449..662L}. 
This turbulence was further scaled to a Mach number of $\mathcal{M} = 2$, and to mitigate any bias from a single turbulence realization, five different random seeds were employed.
The introduced velocity dispersion ($\sim 0.4$ km s$^{-1}$) is smaller than $v_0$, so turbulence in the interface, which most affects the evolution and formation of the clumps and cores, is driven by the convergent flow self-consistently.

The magnetic field was initialized as $\boldsymbol{B} = (0, B_0\sin \theta_B, B_0\cos \theta_B)$, where $B_0$ represents the initial field strength with a value of 5 $\mu$G, resulting an Alfv\'en velocity = 0.62 km s$^{-1}$ and mass-to-flux ratio = 3.08 for the initial density of the gas within a cubic volume of (4.86 pc)$^3$. 
The angle of the magnetic field with respect to the $z$-axis, inclined toward the $y$-axis ($\theta_B$) was varied across three values: $0^{\circ}, 15^{\circ}$, and $30^{\circ}$ (see Figure~\ref{fig:CFInit}). 
As shown in previous studies, different inclination angles can significantly alter structure evolution.
For example, a highly inclined magnetic field can introduce strong magnetic pressure, suppressing the formation of dense structures \citep{2019ApJ...873....6I, 2022ApJ...934..174I}.
In total, we have 15 models, the summary of these parameters and nomenclature for each model can be found in Table \ref{tab:SimulationPara}.

The initial AMR grid consisted of a $48 \times 48 \times 592$ cells, resulting in a base cell size of (0.101 pc)$^3$. 
The simulation was capable of refining regions as needed to meet Truelove's criterion, which we used a minimum resolution of 8 cells per Jeans length to prevent artificial fragmentation. 
The maximum refinement level was set to 7, yielding a minimum cell size of $\Delta x_{min}$ = 163 AU. 
This resolution is significantly smaller than the typical size of 0.1 pc for dense cores, enabling the simulation to resolve fragmentation down to this scale.
To summarize the key resolution and density parameters, our methodology ensures a self-consistent physical framework. 
Our maximum resolution ($\Delta x_{min}$) corresponds to a maximum resolvable density of = $1.7 \times 10^{-17}$ g cm$^{-3}$. 
Our sink formation threshold is set to $\rho_{sink}$ =  $2.7 \times 10^{-16}$ g cm$^{-3}$. 
This value ensures that sink particles are created only in regions that have become numerically unresolved. 
Both of these densities are significantly lower than the adiabatic transition density of $\rho_{ad} = 10^{-14}$ g cm$^{-3}$, confirming that the fragmentation processes we analyze occur entirely within the physically appropriate isothermal regime.
The choice of barotropic EOS (Equation~\ref{eqn:EOS}) is then made in anticipation of higher resolution and zoom-in simulations in the future, where gas is expected to collapse to densities where the adiabatic portion of the EOS becomes important.

Finally, to assess the stability of the simulated structures, we estimated the Jeans length using the initial temperature $T_0$. 
At the initial density $n_0$, the Jeans length was approximately 1.13 pc. 

\begin{deluxetable}{lcc}
	\tablecaption{The initial values for the simulation of convergent flow. \label{tab:SimulationPara}} 
	\tablehead{
		\colhead{Parameter} & \colhead{Value} & \colhead{Nomenclature}
	}
	\startdata
	$n_0$                                 & 135 cm$^{-3}$               & -- \\
	$T_0$                                 & 10 K                                & -- \\
	$\mu$                                & 2.3                                  & -- \\
	$v_0$                                 & 1.5 km s$^{-1}$              & -- \\
	$\mathcal{M}$                    & 2                                     & -- \\
	Num. of turbulence seed     & 5                                    & S1$\sim$5 \\
	$B_0$                                 & 5 $\mu$G                      & -- \\
	$\theta_B$                         & $0^{\circ}, 15^{\circ}, 30^{\circ}$ & T0, T15, T30 \\
	AMR refinement levels        & 7                                    & -- \\
	$\Delta x_{min}$                & 163 AU                            & -- \\
	$\rho_{sink}$                     & $2.7 \times 10^{-16}$ g cm$^{-3}$ & -- \\
	$r_{acc}$                            & 4 $\Delta x_{min}$           & -- \\
	$\rho_{ad}$                        & $10^{-14}$ g cm$^{-3}$ & -- \\
	$\gamma$                         & 5/3                                  & -- \\
	\enddata
\end{deluxetable}

\section{Evolution of the Sheet-like Structure} \label{sec:SheetEvo}
\begin{figure*}[htb!]
	\includegraphics[width=1.0\textwidth]{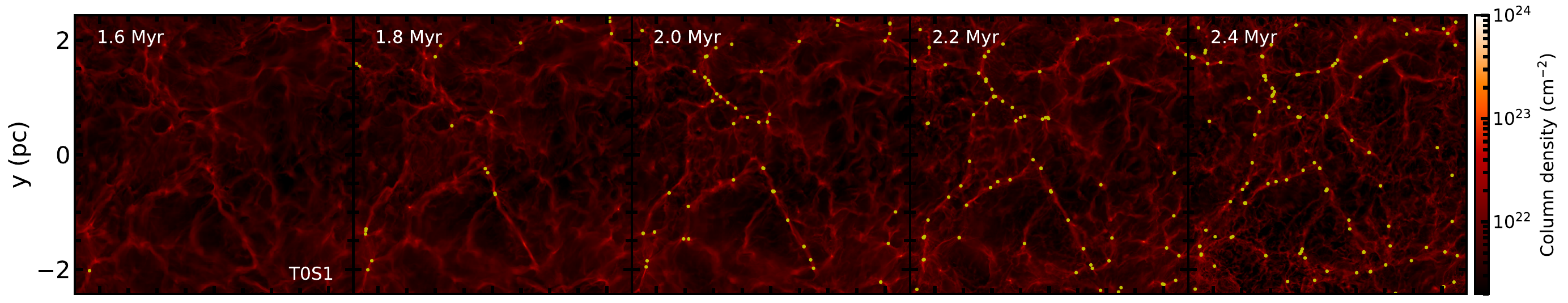}
	\includegraphics[width=1.0\textwidth]{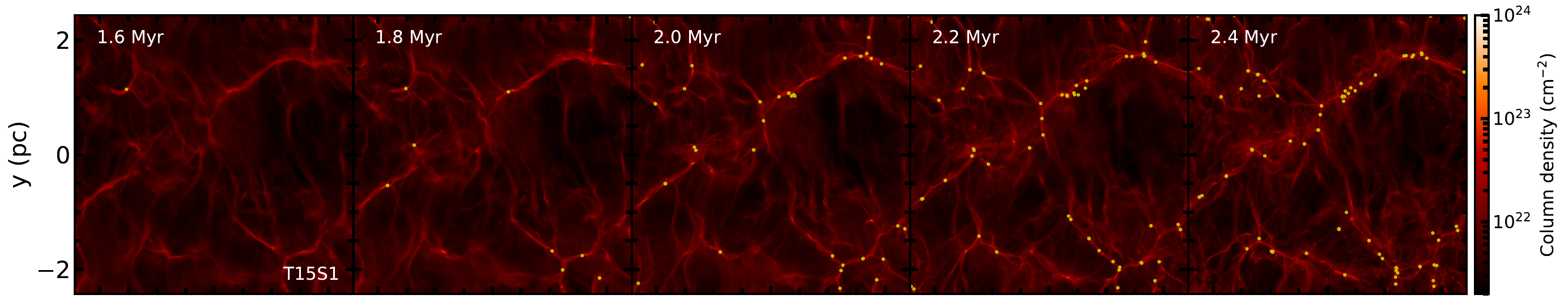}
	\includegraphics[width=1.0\textwidth]{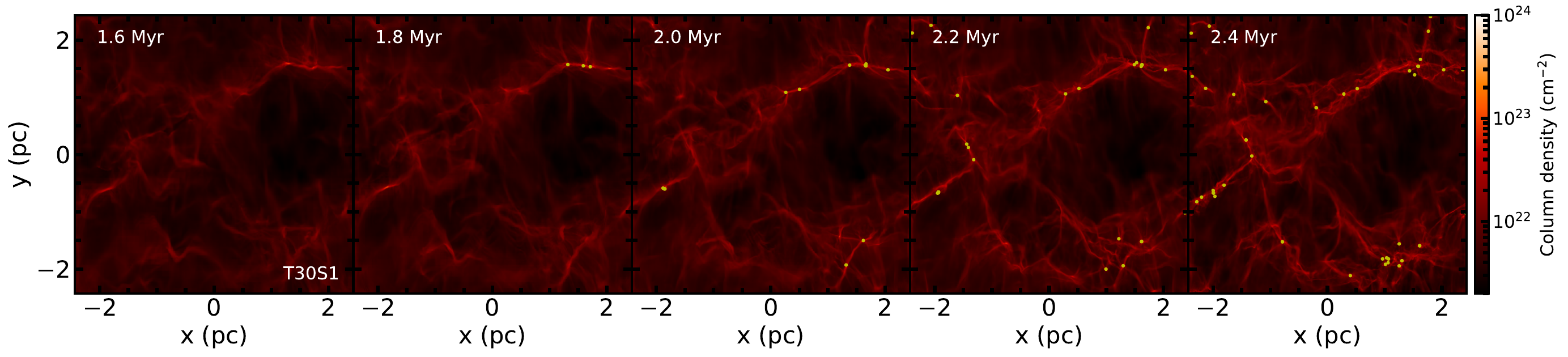}
	\caption{Evolution of column density of the sheet-like structures projected along the $z$-direction for models with the same turbulence seed (i.e., S1). The panels, arranged from top to bottom, correspond to models T0S1, T15S1, and T30S1, representing increasing magnetic field inclinations. The progression from left to right depicts the evolution of column density from 1.6 to 2.4 Myr after the start of simulations. Yellow dots indicate the locations of sink particles.}
	\label{fig:ColEvoS1_xy}
\end{figure*}

The two convergent flows collide at $z = 30$ pc, forming a sheet-like structure confined by the ram pressure. 
Turbulence within the pre-collision gas introduces inhomogeneities into the sheet structure, leading to instabilities at the collision interface \citep{2008ApJ...683..786H}. 
These instabilities subsequently give rise to substructures, which can serve as seeds for their development once gravity becomes dominant (\citealp{1999ApJ...526..279P, 2013ApJ...774L..31I, 2015ApJ...801...77M, 2015MNRAS.453.2471B, 2017MNRAS.465.3483B, 2018PASJ...70S..53I}).
In Figure~\ref{fig:ColEvoS1_xy}, we show the evolution of sheet-like structures projected along the $z$-direction for models sharing the same turbulence seed (i.e., S1) and from 1.6 to 2.4 Myr after the start of simulations \footnote{Unless otherwise stated, all times in this work are relative to the start of the simulation.}. 
Yellow dots represent the locations of sink particles.

From the morphology, the structures at 1.6 Myr in models T15S1 and T30S1 appear similar (the second and third rows of Figure~\ref{fig:ColEvoS1_xy}). 
On the other hand, the dissimilarity with model T0S1 highlight the importance of the inclined magnetic field. 
The inclined field introduces a perpendicular component relative to the gas flow (parallel to the sheet-like structure), facilitating the accretion of gas from the sheet to the gravity-dominated structures initially induced by the instability. 
As the sheets evolve, gas can flow along magnetic field lines more easily, allowing for the formation of more prominent filaments as the sheet evolves (Type-C filament formation in \citealp{2021ApJ...916...83A}). 
This can be seen in models T15S1 and T30S1, in which we observe numerous thin and small filaments (striations) aligned along the $y$-direction, connecting to thicker and larger filaments.

\begin{figure}[htb!]
	\epsscale{1.1}
	\centering
	\includegraphics[width=0.45\textwidth]{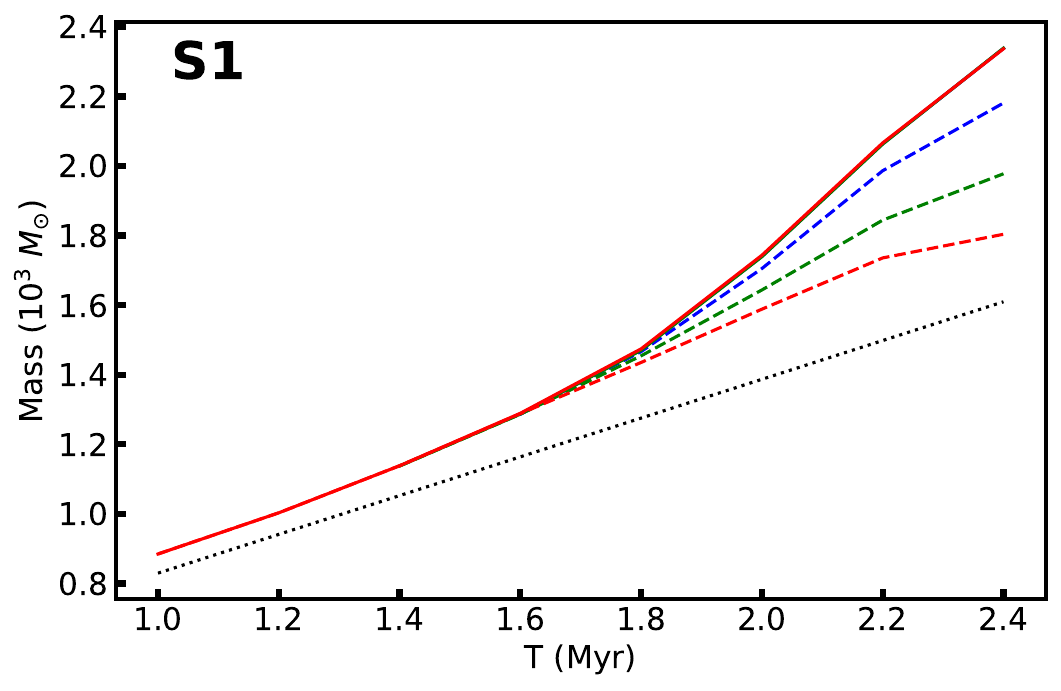}
	\caption{Evolution of sheet mass for models T0S1 (red), T15S1 (green), and T30S1 (blue). The dashed lines represent the growth of gas mass within the region $29.2\ \rm{pc} < z < 30.8\ \rm{pc}$. The solid lines are after including the mass of sink particles, while the black-dotted lines represent the theoretical mass growth for uniform gas inflow without gravitational acceleration.}
	\label{fig:MassGrowth}
\end{figure}

Next, to assess the evolution of the sheet structure and difference introduced by the inclined magnetic fields, we focus on the region $29.2\ \rm{pc} < z < 30.8\ \rm{pc}$, which encompasses the compressed layer.
Figure~\ref{fig:MassGrowth} illustrates the evolution of total mass inside this sheet region over time for models T0S1 (red), T15S1 (green), and T30S1 (blue). 
The dashed lines represent the growth of gas mass, the solid lines are after including the mass of sink particles, and the black-dotted lines represent the case of uniform gas density without gravitational acceleration.
The solid lines in Figure~\ref{fig:MassGrowth} overlap perfectly, demonstrating that the presence of inclined magnetic fields does not significantly influence the convergent flow motion due to its super-Alfv\'enic nature.
Moreover, the growth rates of the solid lines are faster than the black-dotted line, suggesting that the gravity of the sheet has started pulling gas into the structure.
However, the dashed lines representing the gas mass show clear differences among the models. 
For models with higher $\theta_B$, a larger portion of the gas mass remains within the sheet, while only a smaller portion is converted into sink particles. 
This shows that some mechanism is suppressing the formation of sink particles in these models.

This mechanism can origin from the perpendicular component of the magnetic field, which is compressed and amplified by the ram pressure, and it plays a crucial role in suppressing the formation of local dense structures \citep{2008MNRAS.385.1820P, 2011ApJ...742L...9C, 2019FrASS...6....5H}. 
For example, it delays the first sink particle formation for models with higher $\theta_B$ (1.54 Myr, 1.59 Myr, and 1.67 Myr for models T0S1, T15S1, and T30S1, respectively).

\begin{figure}[htb!]
	\epsscale{1.1}
	\centering
	\includegraphics[width=0.45\textwidth]{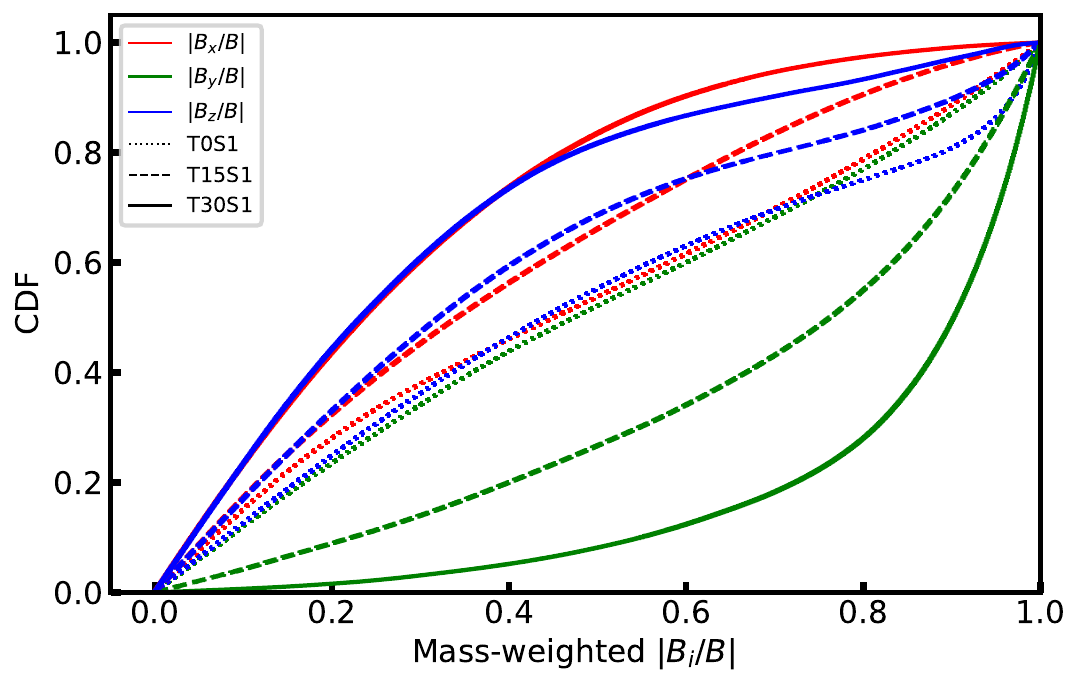}
	\caption{CDFs of the mass-weighted ratio between each component of the magnetic field and its magnitude ($|B_i/\boldsymbol{B}|$) within the sheet region at 1.6 Myr.}
	\label{fig:BiBtot}
\end{figure}

To investigate the compression of the field, we present the cumulative distribution functions (CDFs) of the mass-weighted ratio between each component of the field and its magnitude ($|B_i/\boldsymbol{B}|$) within the sheet region at 1.6 Myr in Figure~\ref{fig:BiBtot}.
For model T0S1 (dotted lines), the $z$-component (blue), which is parallel to the flows, appears slightly dominant. 
In contrast, for models T15S1 (dashed lines) and T30S1 (solid lines), the $y$-component (the perpendicular component, shown in green) is clearly dominant, reflecting the influence of the initial inclined field.
As the sheet continues to be compressed by the ram pressure, the $y$-component for models T15S1 and T30S1 is further amplified, resulting in a shift of the CDF towards higher values (although the evolution is not shown here). 
This indicates that the actual inclination angle of the field within the sheet is increasing, deviating from its initial $\theta_B$.

\begin{figure*}[htb!]
	\epsscale{1.1}
	\centering
	\includegraphics[width=1.0\textwidth]{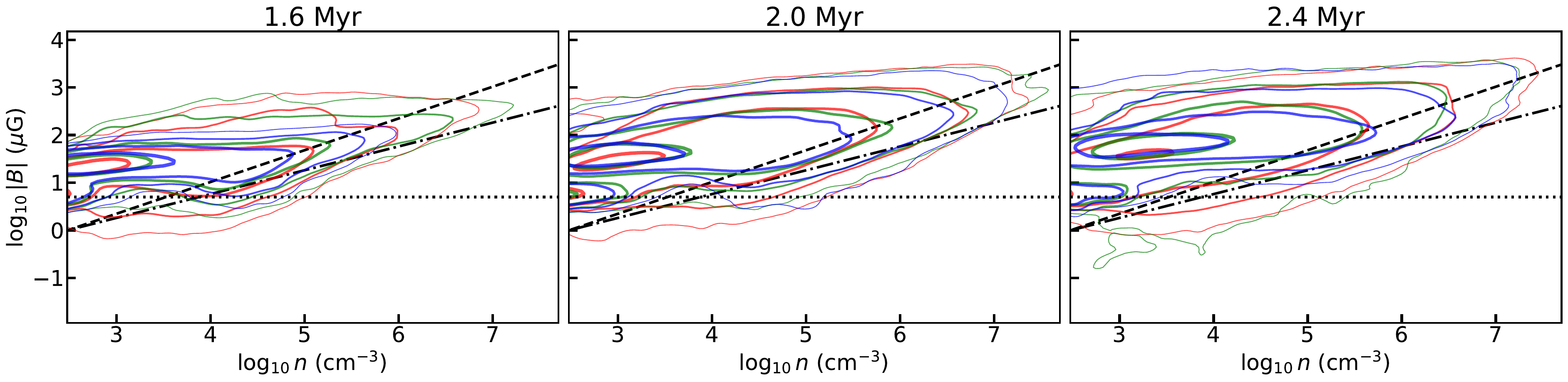}
	\caption{Relationship between magnetic field strength and gas density within the sheet region for models T0S1 (red), T15S1 (green), and T30S1 (blue), shown using 2D kernel density estimation (KDE). The panels are arranged from left to right, representing the relations at 1.6 Myr, 2.0 Myr, and 2.4 Myr, respectively. The black dashed lines indicate the relation $B \propto n^{2/3}$, while the black dot-dashed lines represent $B \propto n^{1/2}$. The black dotted lines mark the initial field strength $B_0 = 5\ \mu$G.}
	\label{fig:DensityBFieldRelation}
\end{figure*}

To provide an overview of the evolution of gas density and magnetic field strength within the sheet region, Figure~\ref{fig:DensityBFieldRelation} illustrates the 2D kernel density estimation (KDE) of the relationship between these two quantities.
The black dashed line represents the relation $B \propto n^{2/3}$, which is predicted for a spherical cloud collapsing under flux-freezing conditions \citep{1966MNRAS.133..265M} and is consistent with observational findings using the Zeeman effect \citep{2010ApJ...725..466C}. The black dot-dashed line represents the relation $B \propto n^{1/2}$, which occurs when the field is inclined to a cylindrical or filamentary cloud \citep{1976ApJ...210..326M}.
From the figure, it is evident that both magnetic field strength and gas density increase over time, and the high-density tails generally follow the predicted power-law relations.
However, at 1.6 Myr, model T30S1 (blue) exhibits a higher proportion of cells with stronger magnetic fields and almost no cells with gas densities exceeding $10^6$ cm$^{-3}$. 
It is only at later times (as shown in the right panel of Figure~\ref{fig:DensityBFieldRelation}) that the high-density tail of model T30S1 becomes comparable to the other two models. 
This clearly demonstrates that the growth of dense structures is delayed in models with higher $\theta_B$.
Furthermore, the distribution of field strengths in model T30S1 remains narrower and peaks at higher values compared to the other two models, indicating that the field is strongly compressed.

These findings collectively suggest that the magnetic field for models with higher $\theta_B$ has a larger impact on the evolution of the sheet and its substructures.
While a full quantitative analysis of the angle distribution between the magnetic field and gas flows and how they influence the structure formation is beyond the scope of this fragmentation-focused study, this mechanism is well discussed by other simulations \citep{2019ApJ...873....6I, 2022ApJ...934..174I}.

%
%

\section{Star-forming Clump Properties} \label{sec:Clump}
In this study, we aim to investigate the relationships between clump properties and core distribution, as well as their evolution over time. 
To achieve this, in this section, we outline the methodologies used to derive clump properties. 
Subsequently, we provide an overview of their evolution.

\subsection{Clump Identification} \label{sec:ClumpIdent}
Inspection of Figure~\ref{fig:ColEvoS1_xy} reveals that sink particles preferentially form within filamentary structures. 
This finding aligns with observation studies, which have shown that the role of filaments as mass reservoirs that promote star formation by feeding material into gravity-dominated substructures (\citealp{2014prpl.conf...27A, 2014ApJ...791..124G, 2024MNRAS.528.1460R, 2018A&A...613A..11W, 2020A&A...642A..87K}, and review by \citealp{2023ASPC..534..233P}).

These substructures, referred to as \textquote{clumps} with typical sizes of approximately 1 pc, represent an intermediate stage in the gas transportation process before being accreted by even smaller and denser structures.
To identify clumps in our simulations, we analyzed snapshots spanning from 1.6 Myr to 2.4 Myr with an interval of 0.2 Myr for model T30. 
For models T0 and T15, the analysis was extend to time range from 1.4 Myr to 2.4 Myr \footnote{For model T0S3, the snapshot at 1.4 Myr was excluded from the analysis due to the inability to identify clumps using our criteria.}. 
This timeframe was chosen considering the formation of the first sink particle, which occurs earlier in models T0 and T15. 
Also, We did not extend the simulations beyond 2.4 Myr, as the absence of stellar feedback in these models would not accurately reflect realistic star-forming environments.

Firstly, we emulated coarse-resolution observations by convolving the column density maps with a 2D Gaussian kernel with a FWHM of $\sim 0.4$ pc. 
This kernel size corresponds to the beam size of the ATLASGAL survey \citep{2009A&A...504..415S} and this process effectively smoothed out fine structures, leaving only the large-scale features.

Next, to identify individual clumps, we employed the dendrogram technique \citep{Rosolowsky2008} implemented within the \texttt{astrodendro} Python package \citep{2019ascl.soft07016R}. 
The dendrogram parameters were set as follows: \textit{min\_value}\footnote{In order to successfully identify clumps for model T0S3, \textit{min\_value} was adjusted to 90\% of the original value for the map at 1.6 Myr. For those maps at 1.4 Myr, 80\% was used} $\sim 5.7 \times 10^{21}$ cm$^{-2}$ (the minimum value to define structures). 
This value is consistent with the high-density filaments identified by \cite{2016A&A...586A.138P}, which have a column density of $\sim 10^{21.7}$ cm$^{-2}$.
\textit{min\_delta} $\sim 5.7 \times 10^{20}$ cm$^{-2}$ (the minimum height of a structure), and \textit{min\_npix} equal to half the beam area (the minimum pixel number for structures). 
The beam area was estimated based on the 2D Gaussian kernel used for convolution. 
After identification, the peak positions, which are insensitive to the exact dendrogram parameters, of the clumps were used to determine their locations.

\begin{figure}[htb!]
	\epsscale{1.1}
	\centering
	\includegraphics[width=0.45\textwidth]{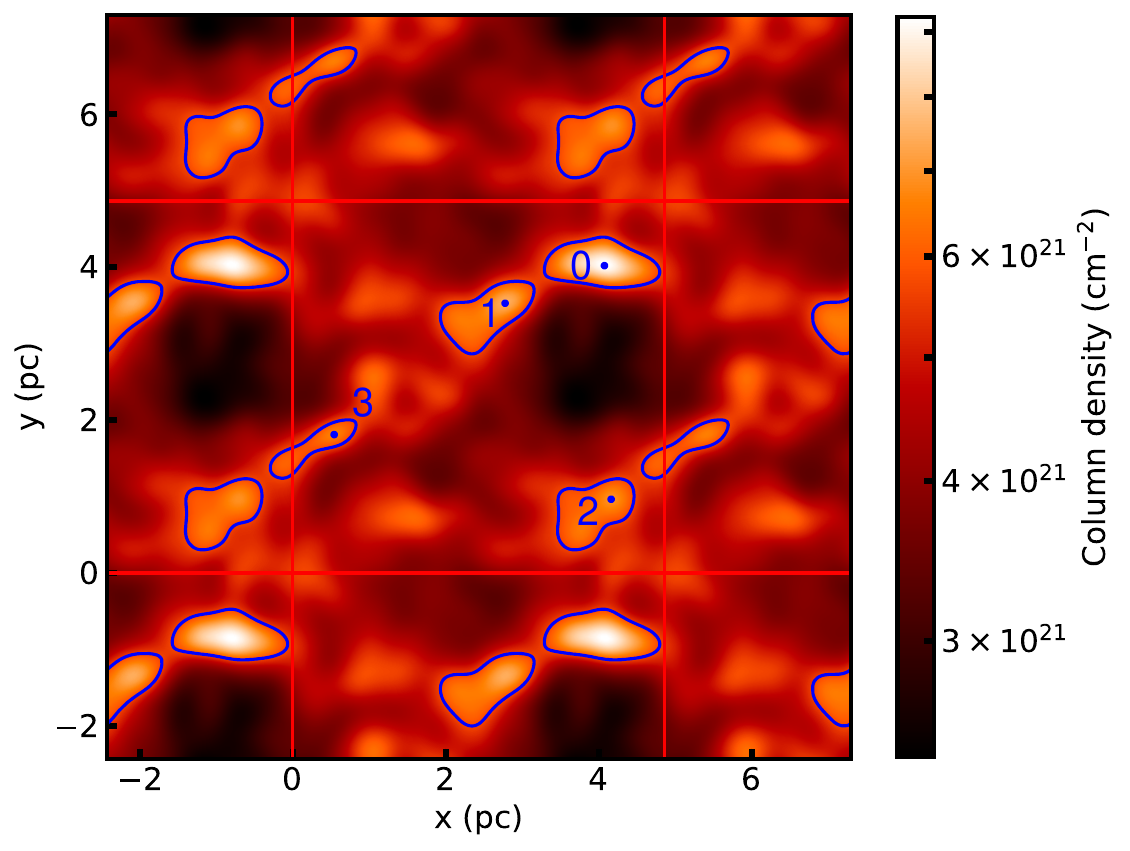}
	\caption{One example of the identified clumps within the convoluted column density map. The blue contours outline the structures identified by the dendrogram, while blue dots mark the peak locations. The red horizontal and vertical lines indicate the boundaries of the simulation domain on the $x$-$y$ plane. Notably, the additional boxes surrounding the central one are copies included to account for the periodic nature of the simulation during the dendrogram calculation.}
	\label{fig:IdentifiedClumps} 
\end{figure}

Within this identification process, we accounted for the periodic boundary conditions in the $x$- and $y$-directions. 
Figure~\ref{fig:IdentifiedClumps} provides an example of the identified clump structures on the column density map for model T30S1 at 1.6 Myr.
In total, there are 484 clumps identified across all models and used snapshots.

\subsection{Evolutionary Sequences} \label{sec:EvoSeq}
To discuss the evolution, we first propose a systematic and reliable methodology for linking clumps identified in Section~\ref{sec:ClumpIdent} across different simulation snapshots.

For clumps observed at two successive simulation snapshots, denoted as $t_1$ (e.g., 1.6 Myr) and $t_2$ (e.g., 1.8 Myr), we first define a radius $R_{\rm{drift}} = 0.5$ pc. 
This radius accounts for the potential movement of the clump's position on the $x$-$y$ plane over time. 
Next, for clumps identified at $t_2$, we search within $R_{\rm{drift}}$ for potential ancestors at $t_1$. 
If only one clump is found within this radius, the two clumps are considered connected in time. 
However, if multiple $t_1$ clumps are identified for a given $t_2$ clump, we calculate the metric $M_{cl}(t_1)/r_{1, 2}$ for each $t_1$ clump. 
Here, $M_{cl}(t_1)$ is the clump mass derived in Section~\ref{sec:MR} for the $t_1$ clump, and $r_{1, 2}$ is the distance between the $t_1$ and $t_2$ clumps on the $x$-$y$ plane. 
The $t_1$ clump with the highest value of this metric is considered connected to the $t_2$ clump. 
The remaining $t_1$ clumps are assumed to have merged with other $t_1$ clumps and evolved into the $t_2$ clump.

Clumps identified at $t_1$ that do not have a corresponding clump at $t_2$ are considered potential candidates for dispersal at later times. 
Conversely, clumps observed at $t_2$ without a counterpart at $t_1$ are likely newly formed clumps that emerged between the two snapshots.
Note that even if a $t_1$ clump does not have a direct counterpart at $t_2$, it may still have merged with another clump. 
This is particularly likely if the host clump is elongated and the $t_1$ clump (the merger) is located near one end of the host clump. 
However, given the large number of simulated clumps, the impact of such mergers on our analysis is expected to be minimal.
In the subsequent analysis, we focus exclusively on clumps that have been successfully linked across different simulation snapshots using this procedure. 

These \textquote{evolutionary sequences} represent the traceable evolution of clumps over time. 
Clumps that were not linked across snapshots are not included in our analysis due to their potentially transient nature.

Furthermore, to facilitate direct comparisons of these sequences, we define the clump age for each clump as follows:
\begin{equation}\label{eqn:ClumpAge}
	t^i_{age} = t^i_{clump} - t_{sink},
\end{equation}
where $t^i_{clump}$ is the snapshot time at which $i^{th}$ clump is identified, and $t_{sink}$ is the time at which the first sink particle is formed in that sequence.
For clumps with $t^i_{age} < 0$, the clump was identified prior to the formation of the first sink particle. 
Conversely, for clumps with $t^i_{age} > 0$, the sink particle has already formed, indicating the potential for local collapse.
In total, we identified 37, 49, and 30 evolutionary sequences for models T0, T15, and T30, respectively.

\subsection{Clump Mass and Size} \label{sec:MR}

To characterize clump properties, we first used mass, size, and density, which are commonly employed parameters. 
Instead of estimating these properties from 2D images, we transferred the AMR grid surrounding the clump region to a uniform grid with a resolution of $\sim 1000$ AU.
A sphere with radius ($R_s$) of 0.5 pc was then defined and centered on the clump's position. 
To determine the 3D clump positions, we first used the 2D ones identified in Section~\ref{sec:ClumpIdent} for the $x$ and $y$ coordinates. 
Subsequently, for each clump, the $z$-coordinate was determined by locating the cell with the highest gas density along the $z$-axis at the corresponding $x$ and $y$ coordinates.

The clump mass ($M_{cl}$) was then defined as the total gas mass within $R_s$. 
Using a 3D grid helps to avoid biases introduced by foreground and background gas in column density maps and mitigates issues arising from fitting arbitrary clump shapes with a single Gaussian.
Next, the clump radius ($R_{cl}$) was defined as the radius of a concentric sphere containing half of the total clump mass $M_{cl}$.

\begin{figure}[htb!]
	\includegraphics[width=0.45\textwidth]{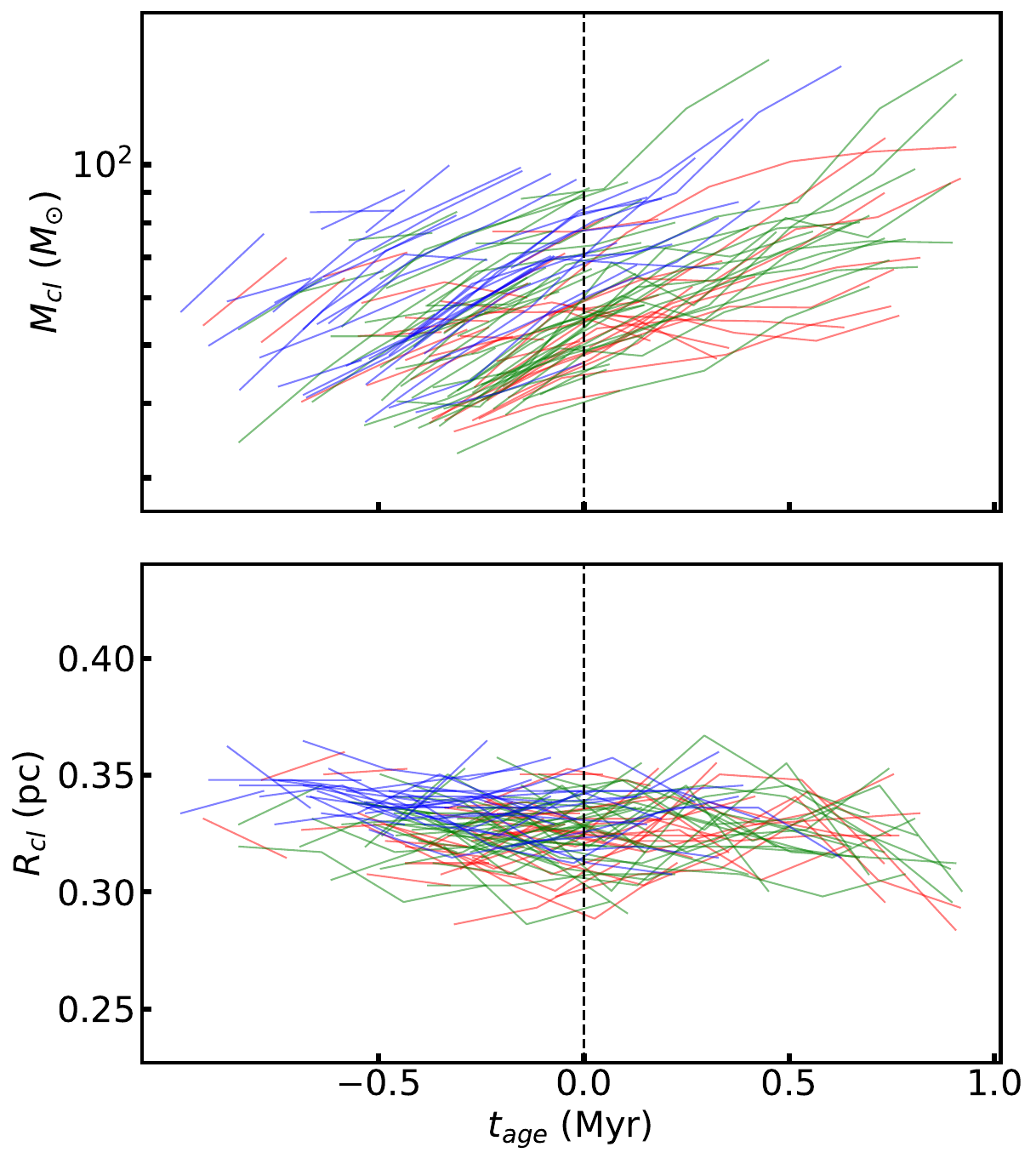}
	\caption{Evolution of clump mass ($M_{cl}$) and radius ($R_{cl}$). Different colors represent the model of origin (red for T0, green for T15, and blue for T30). The vertical black dashed line indicates the time at which the first sink particle forms ($t_{age} = 0$). Note that the sequences without any sink particles are excluded.}
	\label{fig:MR}
\end{figure}

The evolution of $M_{cl}$ and $R_{cl}$ is shown in Figure~\ref{fig:MR}. 
The upper panel demonstrates a clear trend of increasing clump mass over time for all three models. 
Notably, the models with higher $\theta_B$ (T15 and T30) tend to be more massive, which may be attributed to the perpendicular component of the magnetic field within the sheet-like structure, facilitating accretion onto the clumps (Section~\ref{sec:SheetEvo}). 
The mean clump masses are 58, 62 and 66 $M_{\odot}$ for models T0, T15 and T30.
In contrast, the bottom panel shows that clump sizes remain relatively constant, with mean values of $R_{cl}$ = 0.32, 0.33 and 0.34 pc for the three models. 
This intriguing constancy, which occurs despite the ongoing increase in clump mass, can be understood as a fact that the half-mass radius is a metric that is inherently insensitive to the specific shape of the density profile; as we note, a significant steepening of a power-law density profile from a power-law index changing from -1 to -2 results in only a modest ($<$ 30\%) change in the half-mass radius. 
Overall, the combination of increasing mass and stable radius indicates that the mean clump density is increasing over time.


\subsection{Clump Geometry}

The large-scale geometry of a clump is a fundamental observable that may influence how it subsequently fragments. 
It is plausible that a highly elongated clump would fragment into an aligned pattern of cores \citep{2016MNRAS.458..319C, 2017MNRAS.468.2489C, 2020MNRAS.497.4390C, 2016MNRAS.463.4301H, 2018MNRAS.481L...1H, 2015MNRAS.452.2410S, 2017ApJ...848....2H, 2019ApJ...881...97H}, while a more centrally condensed clump might promote a clustered distribution. To test this hypothesis, it is necessary to characterize the clump geometries.
To characterize these properties, we employed the $R_1$ and $R_2$ parameters. 
These parameters quantify the degree of elongation (higher $R_1$) and central condensation (higher $R_2$) of a given structure.
The two values for each clump were obtained based on the unsmoothed column density inside a radius of 0.5 pc centered on clump's position and calculated using the \texttt{RJ-plots} Python library \citep{2022MNRAS.516.2782C}.

While investigating the evolution of these parameters (not shown here), we did not observe a clear evolutionary trend. 
The mean values of $R_1$ are 0.07, 0.09, and 0.08 for models T0, T15, and T30. 
However, individual values can fluctuate significantly, increasing or decreasing by 0.1 or even reaching 0.2 in some cases. 
This dynamic behavior of $R_1$ suggests that clumps can evolve into various shapes, including elongated and quasi-circular configurations.
Notably, the values of $R_2$ were predominantly close to zero, with mean values of 0.07, 0.09, and 0.07 for the three models. 
This indicates that these clumps are not strongly centrally condensed. 
In fact, they typically exhibit multiple cores and clumpy structures within the regions, resulting in moderately positive $R_2$ values \citep{2022MNRAS.516.2782C}.

\subsection{Filament or Hub} \label{sec:SNsub}

In Figure~\ref{fig:ColEvoS1_xy}, we find that stars (sink particles) preferentially form within filamentary structures or hub systems, where multiple filaments converge.
To distinguish between clumps residing in filaments or hub systems and how these structures influence the core distribution, we utilize the filament surface density ($S_N$), a metric introduced by \cite{2020MNRAS.497.4390C}. 
The filament surface density is defined as:
\begin{equation}
	S_{N, j} = \sum^N_i e^{-r_{min, ij}^2/2\sigma_N^2},
	\label{eqn:SNsub}
\end{equation}
where $r_{min, ij}$ represents the shortest distance from the $j^{th}$ pixel to $i^{th}$ filament, and $\sigma_N$ is a bandwidth parameter.
For a pixel located at the junction of filaments, $S_{N, j}$ is typically an integer value, representing the number of filaments connected to the clump. 
However, non-integer values may still occur if nearby filaments do not directly connect to the clump.

To identify filaments within the convoluted column density map, we employed DisPerSE \citep{2011MNRAS.414..350S}. 
DisPerSE was configured with a persistence threshold of $5.7 \times 10^{20}$ cm$^{-2}$.
During the filament identification process, two filament spines were merged if they met at an angle less than $70^\circ$. 
This approach aimed to capture the continuity of filament structures. 
However, for points where more than two filaments converged, an additional criterion was applied: only if one of the filaments was shorter than 100 pixels were the two spines merged. 
This step helped to suppress the excessively identification of short filament segments in complex systems.
Finally, any remaining filament segments shorter than 100 pixels were excluded. 
These processes resulted in the identification of \textquote{filament skeletons}.

In the calculation of $S_N$, we set $\sigma_N$ to 0.17 pc, which is the standard deviation of the Gaussian kernel used in Section~\ref{sec:ClumpIdent}. 
With the previously determined filament skeletons, this approach enabled us to generate $S_N$ maps centered on each identified clump.
To quantify the complexity within the central region of each clump, we determined the maximum $S_N$ value within a radius of 0.5 pc centered on the clump's position. 
The maximum value, denoted as $S_N^{max}$, represents the highest level of filament complexity observed within the central clump region.

The mean and median values of $S_N^{max}$ are 2.5 and 3.0, respectively, for all models, indicating a preference for clump formation in hub-like structures, consistent with observational findings  (\citealp{2009ApJ...700.1609M, 2014prpl.conf...27A,  2014ApJ...791..124G, 2024MNRAS.528.1460R, 2018A&A...613A..11W, 2020A&A...642A..87K, 2024MNRAS.527.4244S}, and review by \citealp{2023ASPC..534..233P}).
When examining the evolution of $S_N^{max}$ (not shown here), it did not reveal a clear trend due to its discrete nature. 
However, when examining the distributions separated by $t_{age} = 0$, we observed a distinct pattern: for models with lower $\theta_B$, as the structures evolve, there is an increasing proportion of clumps found in hubs ($S_N^{max} \geq 3$, i.e., at least three filaments connecting to the clump). 
To determine whether two distributions are statistically similar, we employed the two-sample Kolmogorov-Smirnov (KS) test \citep{1958ArM.....3..469H}. 
For models T0 and T15, the KS test revealed statistically significant difference between the distributions before and after $t_{age} = 0$ (KS statistic = 0.26 and 0.36, $p$-value = 0.02 and $<$ 0.01, respectively).
These results suggest that hub-like configurations are becoming increasingly prevalent.
In contrast, the two distributions for model T30 showed no significant difference (KS statistics = 0.18, $p$-value = 0.56), suggesting that the enhanced magnetic fields may be contributing to the preservation of large-scale structures.

\subsection{Energy Terms} \label{sec:EnergyTerms}
To investigate the interplay of gravity, magnetic fields, and turbulence, we calculated these energies for our simulated clumps.
Similar to the calculation of clump mass and size in Section~\ref{sec:MR}, we directly utilized the 3D grid data to ensure more accurate estimations of these energies.


For the gravitational energy ($E_G$), we focused solely on the self-gravity inside $R_s$, considering the mass distribution within the sphere and neglecting the gravitational potential arising from the surrounding gas. 
However, for a more accurate representation of self-gravity, we included the mass of sink particles within the sphere, as these particles contributed to the gravitational force affecting the gas motion.
Following GAMER's scheme (Section~\ref{sec:Numerical}), we employed the TSC method to distribute the particle mass to neighboring uniform grid cells \citep{1988csup.book.....H}.
The self-gravitational energy of the structure can then be calculated using the following equation:
\begin{equation}
	E_G = \sum_{i = 1}^{N_s} m^d_i \Phi_{s, i},
\end{equation}
where $N_s$ is the number of cells inside $R_s$, $m^d_i$ is the mass of the $i^{th}$ cell after the redistribution of particle mass (note that $m^d_i = m_i$ for cells far from any particle, where $m_i$ is the gas mass), and $\Phi_{s, i}$ is the gravitational potential at the $i^{th}$ cell.
$\Phi_{s, i}$ is given by:
\begin{equation}
	\Phi_{s, i} = -\sum_{j \neq i}^{N_s} \frac{G m^d_j}{|\mathbf{r}_i - \mathbf{r}_j|},
\end{equation}
where $\mathbf{r}_i$ is the position vector of the $i^{th}$ cell, and $G$ is the gravitational constant.
To enhance computational efficiency in calculating $\Phi_{s, i}$ (complexity of $\mathcal{O}(N^2)$), we employed the \texttt{pytreegrav} Python package, which implements the Barnes-Hut algorithm with an opening angle of 0.7, to accelerate the calculation \footnote{For more information on the accuracy compared to a brute-force approach, please refer to the \texttt{pytreegrav} documentation: \url{https://pytreegrav.readthedocs.io/en/latest/index.html}.}.

To calculate the kinetic, magnetic, and thermal energies, we focused on the volume term inside $R_s$. 
These energies are expressed as follows:
\begin{equation}
	E_K = \frac{1}{2} \sum_{i = 1}^{N_s} m_i |\mathbf{v}_i - \mathbf{v_c}|^2,
\end{equation}
\begin{equation}
	E_B = \frac{1}{8 \pi} \sum_{i = 1}^{N_s} |\mathbf{B}_i|^2 (\delta x)^3,
\end{equation}
\begin{equation}
	E_{th} = \frac{3}{2} \sum_{i = 1}^{N_s} P_i (\delta x)^3,
\end{equation}
where $\mathbf{v}_i$ represents the gas velocity at the $i^{th}$ cell, $\mathbf{v_c}$ is the center-of-mass (COM) velocity for the materials inside $R_s$, $|\mathbf{B}_i|$ is the magnitude of the cell-centered magnetic field, $P_i$ is the thermal gas pressure at the $i$-th cell (obtained from Equation~\ref{eqn:Pressure}), and $(\delta x)^3$ is the volume of each uniform cell.

Next, we estimated the turbulent energy. 
We note that the turbulent motion is typically characterized as continuous, non-monotonic, and non-thermal gas motion that contributes to the total gas kinetic energy. 
Therefore, directly calculating the standard deviation of the gas velocity will inevitably include contributions from gravity-induced ordered motion. 
To address this, we employed a \textquote{moving box} technique to measure \textit{local turbulent motion}.
This method involves defining a smaller sphere with a radius ($R_{tur} = 0.05$ pc) centered on the $i^{th}$ cell.
We then calculated the local COM velocity ($\mathbf{v_{c, i}}$) for materials inside $R_{tur}$ to represent the ordered gas motion within the local region.
The turbulent energy can then be determined by:
\begin{equation}
	E_{tur} = \frac{1}{2} \sum_{i = 1}^{N_s} m_i v_{tur, i}^2,
\end{equation}
where $v_{tur, i} = |\mathbf{v}_i - \mathbf{v_{c, i}}|$ is the local turbulent velocity and represents the velocity perturbation of the $i^{th}$ cell relative to $\mathbf{v_{c, i}}$. 
It is important to note that this definition of turbulent velocity is scale-dependent and only used to capture non-ordered gas motions below that scale.
Consequently, increasing $R_{tur}$ will capture stronger non-ordered velocities at larger scales.
Furthermore, the choice of $R_{tur}$ here is motivated by its proximity to the core size, its sufficient resolution within our uniform grid (with a cell size of $\sim 1000$ AU), and its representation of $10\%$ of $R_s$.

Additionally, to characterize the turbulent properties within $R_s$, we defined the turbulence Mach number ($\mathcal{M}_{tur}$), as follows:
\begin{equation}
	\mathcal{M}_{tur} = \frac{\sum_{i = 1}^{N_s} m_i \mathcal{M}_{tur, i}}{\sum_{i = 1}^{N_s} m_i},
\end{equation}
where $\mathcal{M}_{tur, i} = v_{tur, i}/c_{s, i}$ is the turbulent Mach number for the $i^{th}$ cell, calculated using the corresponding temperature to determine the sound speed ($c_{s, i}$).
Furthermore, we also calculated the Alfv\'en velocity ($v_{A, i} = |\mathbf{B}_i|/\sqrt{4 \pi \rho_i}$, $\rho_i$ is the gas mass density) and Alfv\'en Mach number ($\mathcal{M}_{A}$) inside $R_s$ as:
\begin{equation}
	\mathcal{M}_{A} = \frac{\sum_{i = 1}^{N_s} m_i \mathcal{M}_{A, i}}{\sum_{i = 1}^{N_s} m_i},
\end{equation}
where $\mathcal{M}_{A, i} = v_{tur, i}/v_{A, i}$ is the ratio of the turbulent velocity to the Alfv\'en velocity for the $i^{th}$ cell.

\begin{figure*}[htb!]
	\includegraphics[width=1.0\textwidth]{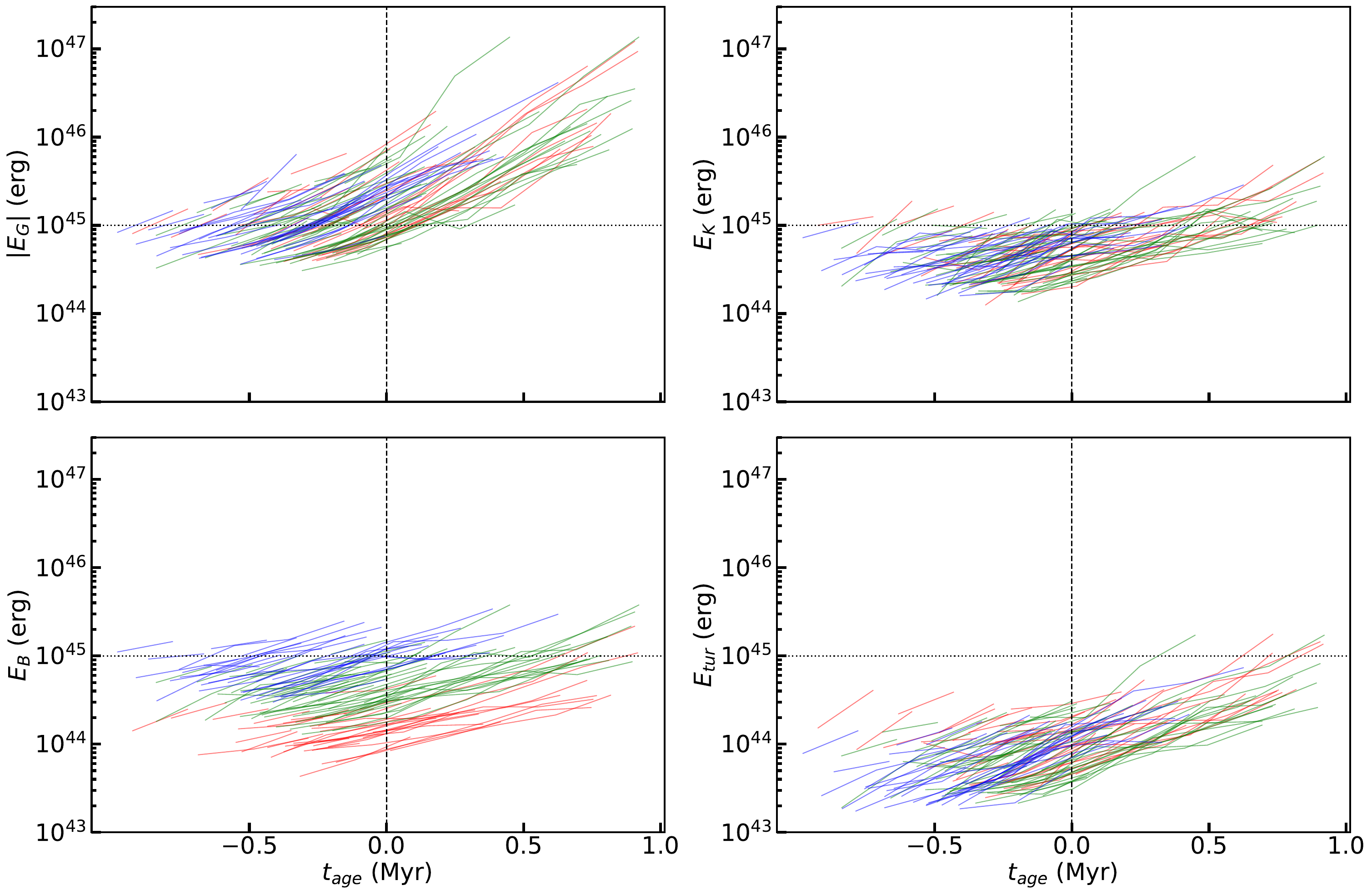}
	\caption{Evolution of gravitational energy ($E_G$) kinetic ($E_K$), magnetic ($E_B$) and turbulent ($E_{tur}$). Different colors represent the model of origin (red for T0, green for T15, and blue for T30). The vertical black dashed line indicates the time at which the first sink particle forms ($t_{age} = 0$). Note that the sequences without any sink particles are excluded.}
	\label{fig:EvoAbsEnergy}
\end{figure*}

\begin{figure*}[htb!]
	\includegraphics[width=1.0\textwidth]{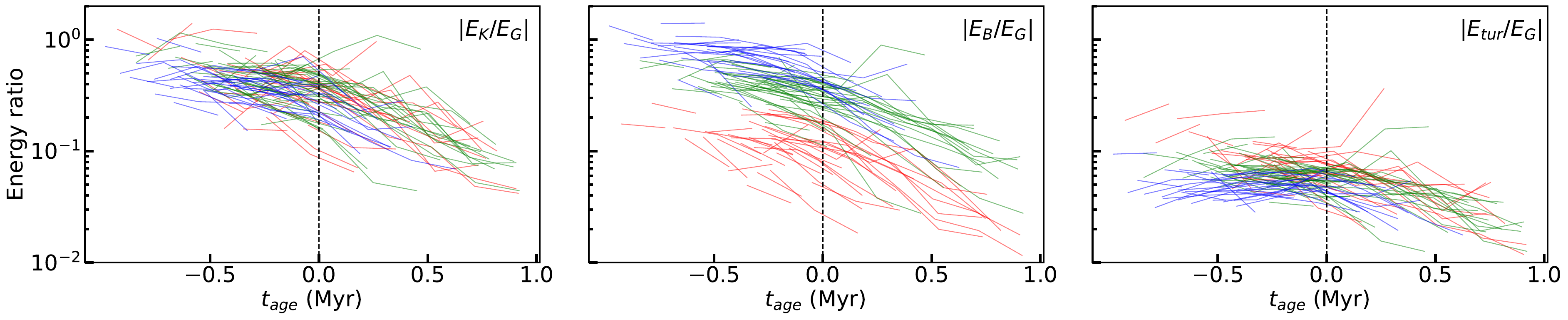}
	\caption{Evolution of kinetic ($E_K$), magnetic ($E_B$) and turbulent ($E_{tur}$) energies relative to gravitational energy ($E_G$). Different colors represent the model of origin (red for T0, green for T15, and blue for T30). The vertical black dashed line indicates the time at which the first sink particle forms ($t_{age} = 0$). Note that the sequences without any sink particles are excluded.}
	\label{fig:EvoEnergy}
\end{figure*}

Figure~\ref{fig:EvoAbsEnergy} shows the evolution of the absolute values of each energy term, while Figure~\ref{fig:EvoEnergy} illustrates the evolution of the ratio of each energy term to $E_G$. 
Note that as $E_{th}$ and $|E_{th}/E_G|$ are significantly smaller, they are not shown for clarity. 
We observe that all energy terms are growing as clumps evolve. 
From the ratios, it is evident that gravity dominates most clumps, and its relative importance increases over time.
A detailed decomposition of the gravitational energy contributions, presented in Appendix~\ref{app:GravDecom}, reveals that this increase in dominance is primarily driven by the gas accretion into the clumps, rather than by gas redistribution or core clustering.

Specifically, the right panel of Figure~\ref{fig:EvoEnergy} shows that $|E_{tur}/E_G|$ generally remains $\sim 0.1$ when $t_{age} < 0$, indicating that turbulence grows alongside gravity prior to the formation of star. 
However, gravity becomes increasingly dominant after $t_{age} = 0$.
Significant difference between models can be found in $|E_B/E_G|$ (middle panel): models with higher $\theta_B$ exhibit larger $|E_B/E_G|$ values throughout their evolution compared to models with lower $\theta_B$. 
This is expected due to the stronger field in these models ( Section~\ref{sec:SheetEvo}).

\begin{figure}[htb!]
	\centering
	\includegraphics[width=0.45\textwidth]{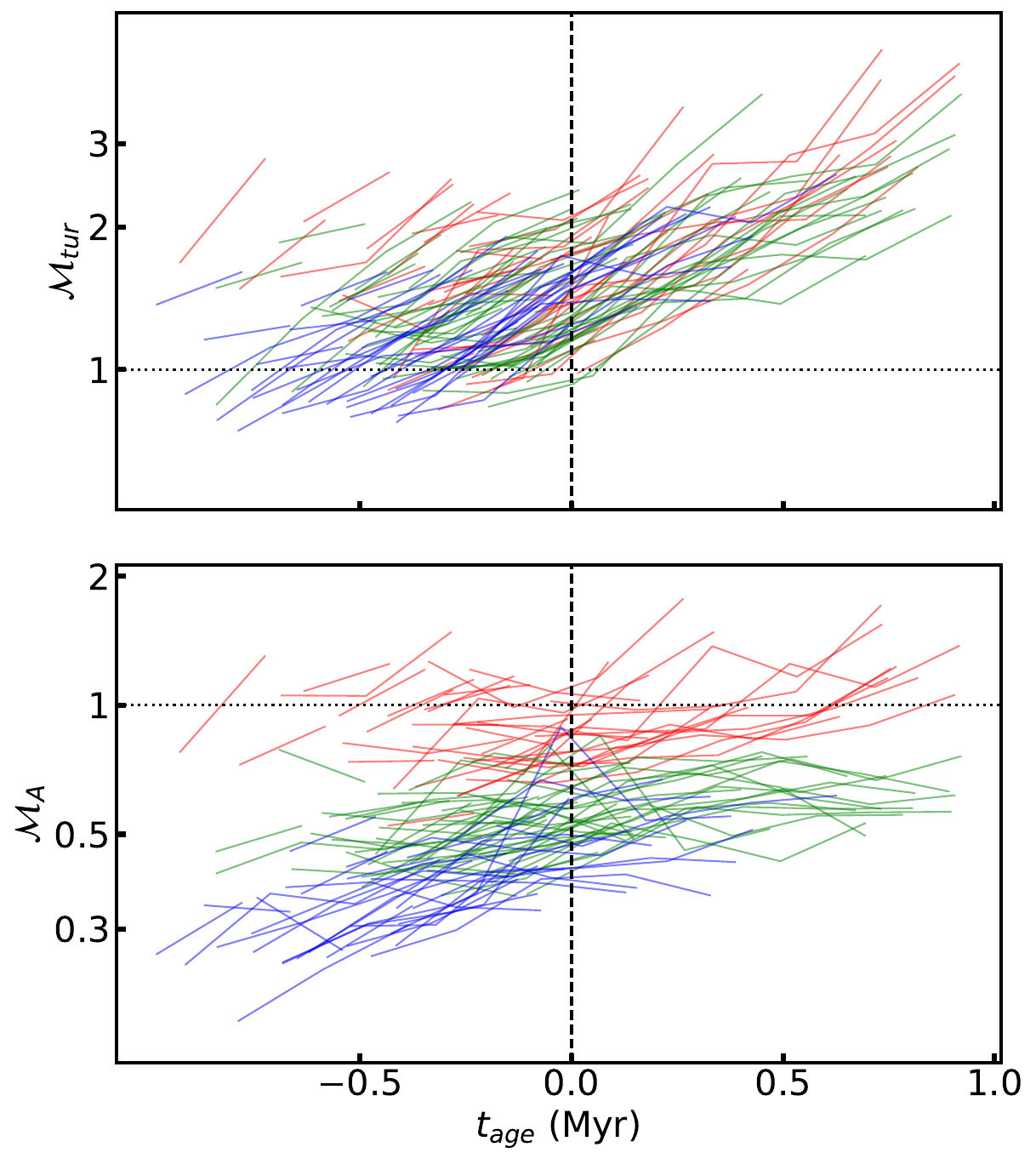}
	\caption{Evolution of turbulence Mach number ($\mathcal{M}_{tur}$, upper panel) and the Alfv\'en Mach number ($\mathcal{M}_{A}$, bottom panel). Different colors represent the model of origin (red for T0, green for T15, and blue for T30). The vertical black dashed line indicates the time at which the first sink particle forms ($t_{age} = 0$), and the horizontal dotted line represents Mach number = 1. Note that the sequences without any sink particles are excluded.}
	\label{fig:EvoLine_Mach}
\end{figure}

The evolution of $\mathcal{M}_{tur}$ and $\mathcal{M}_{A}$ for each model is shown in Figure~\ref{fig:EvoLine_Mach}. 
Both Mach numbers exhibit an increasing trend for all models, indicating that turbulence is growing in all simulations. 
For $\mathcal{M}_{tur}$, all models show a similar trend in values, regardless of their $\theta_B$. 
This similarity may result from the fact that all clumps are dominated by gravity, and turbulence grows alongside gravity.
On the other hand, inspection of $\mathcal{M}_{A}$ in the bottom panel reveals that clumps in model T0 have trans- to super-Alfv\'enic turbulence ($\mathcal{M}_{A} \gtrsim 1$).
This is likely due to the weaker magnetic field dominance in this model. When the magnetic field plays a more significant role (models T15 and T30), $\mathcal{M}_{A}$ is less than one, indicating sub-Alfv\'enic turbulence.

\section{Quantifying Core Distribution} \label{sec:Core}
Observations of clumps have revealed the presence of substructures with sizes of $\sim 0.1$ pc and densities $> 10^5$ cm$^{-3}$. 
These substructures, referred to as \textquote{cores}, are the products of fragmentation within the host clump and serve as the progenitors of stars.
High-resolution observations with ALMA have provided valuable insights into the properties of dense cores, also enabling discussions about the fragmentation mechanisms that produce these cores \citep{2018A&A...617A.100B, 2019ApJ...886..102S, 2021A&A...649A.113B, 2023MNRAS.526.2278A, 2023MNRAS.520.2306T, 2023ApJ...950..148M, 2024ApJS..270....9X, 2024arXiv240808299W}.

Therefore, to enable comparison with ALMA observations, we simulated interferometric observations using synthetic images and subsequently identified and quantified the distribution of cores within these images.

\subsection{Simulating ALMA Observation} \label{sec:SyntheticImg}

We first performed radiative transfer calculations using RADMC-3D \citep{2012ascl.soft02015D} to obtain a dust emission map at a wavelength of 1.3 mm, corresponding to ALMA Band 6 observations. 
The calculations were performed along the $z$-axis, focusing on a domain of approximately 4.86 pc $\times$ 4.86 pc $\times$ 1.62 pc centered at (2.43, 2.43, 30.0) pc.
We adopted the following assumptions for the radiative transfer calculations: a gas-to-dust mass ratio of 100, using gas temperatures for dust, and a dust opacity law from \cite{1994A&A...291..943O} with a value of opacity $= 1.993$ cm$^2$ g$^{-1}$ at 1.3 mm. 
This dust opacity law is suitable for a dust population without ices and an MRN dust size distribution, particularly in environments with gas densities below $10^6$ cm$^{-3}$.

Once we identified clumps in Section~\ref{sec:ClumpIdent}, we extracted sub-regions of size 2.43 pc $\times$ 2.43 pc (corresponding to $1'.86 \times 1'.86$ at a distance of 4.5 kpc) centered on the clump position from the 1.3 mm map to isolate individual clumps. 
This sub-region size was chosen to capture the pc-scale clump structure, consistent with the spatial scales probed by ALMA surveys (e.g., ASHES).
The resulting pixel size of 490 AU (or $0".11$) is smaller than the typical beam size of $\sim 1"$, minimizing potential artifacts in the synthetic images. 

To obtain the synthetic images, these sub-region 1.3 mm maps were used as the \texttt{skymodel} for the \texttt{simalma} task in CASA version 6.6.0.20 \citep{2007ASPC..376..127M}. 
Observation configurations were selected assuming a distance of 4.5 kpc to the clumps. 
This included combining the main 12 m array, the C43-2 configuration, and the ACA. 
A PWV value of 0.75 was set, and mosaic observations with an angular size of $35" \times 35"$ and an integration time of 25 seconds per pointing were simulated. 
These settings resulted in 10 and 3 pointing mosaics for the 12 m array and ACA configurations, respectively, with corresponding observation times of 16.7 and 107 minutes.

In the \texttt{simalma} task, the \texttt{threshold} parameter determines the flux level at which the cleaning process is terminated. 
This threshold varies from map to map.
To select the optimal \texttt{threshold} value, we chose the value that resulted in the distribution of values on the flattened residual map resembling a Gaussian distribution. 
This was verified using the one-sample Kolmogorov-Smirnov test, which assesses the difference between the observed distribution and a theoretical Gaussian distribution.

The resultant images have an angular resolution of $1".52 \times 1".23$ and have $1024 \times 1024$ pixels.
The average noise level is $\sim 0.10$ mJy beam$^{-1}$.
All the synthetic images shown in this work are before the primary beam correction, and a few examples are shown in the first and third columns of Figure~\ref{fig:SYNGallery}.
These images will be used for identifying cores in Section~\ref{sec:GetAL}.


\subsection{Alignment Parameters} \label{sec:GetAL}
Recently, \cite{2025ApJ...979...67C} proposed \textquote{alignment parameters} (generally referred to as \AL) to quantify the degree of core alignment. 
By considering only the positions and fluxes of cores, these parameters provide a robust and objective measure that aligns well with visual inspection.

\begin{figure*}[htb!]
	\includegraphics[width=1.0\textwidth]{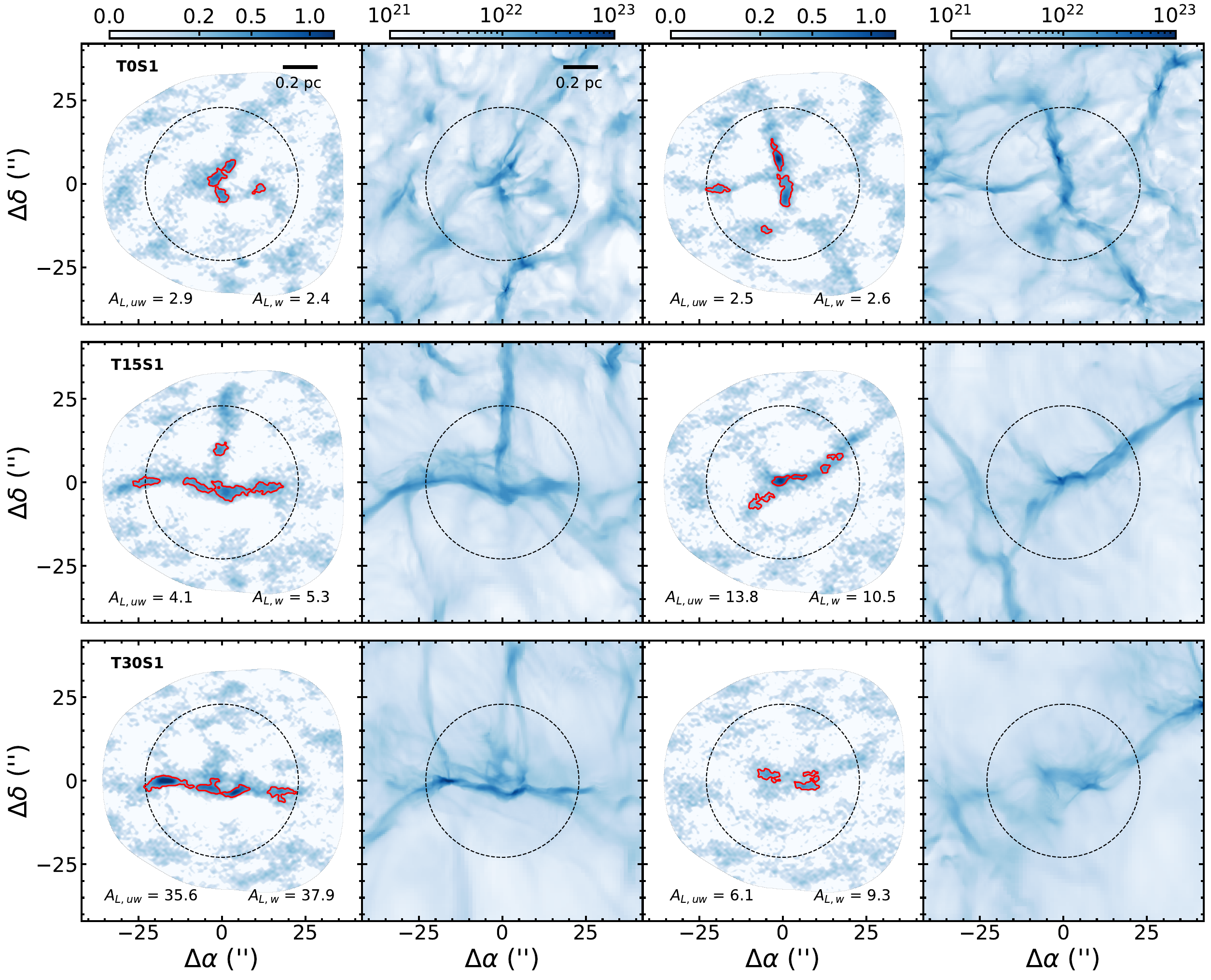}
	\caption{Examples of identified cores (red contours) overlaid on synthetic images (first and third columns) and their corresponding column density maps (second and fourth columns). The synthetic images are in units of mJy beam$^{-1}$ and column density maps are in units of cm$^{-2}$. For better visualization, the color scales are displayed on a square root and logarithmic scale. Each row displays two clumps from models T0S1, T15S1, and T30S1 at 1.6 Myr. The black dashed contours delineate circles with a radius of 0.5 pc centered on the clump centers. The values of \ALuw and \ALw are displayed in the lower side of each clump.}
	\label{fig:SYNGallery}
\end{figure*}

We first employed the dendrogram technique to identify cores within the synthetic images obtained in Section~\ref{sec:SyntheticImg}. 
The dendrogram parameters were set as follows: \textit{min\_value} = 3$\sigma_n$, \textit{min\_delta} = 1$\sigma_n$, and \textit{min\_npix} = one beam area. 
Here, $\sigma_n$ represents the noise level in the corresponding image.
A \textquote{leaf} within the dendrogram represents a structure that lacks any substructure. 
In this study, we define these leaf structures as \textit{cores}.
To ensure consistency with other parameters derived in Section~\ref{sec:Clump}, we considered only cores within a circular region of radius 0.5 pc centered on the clump position.
Figure~\ref{fig:SYNGallery} displays a few examples of the identified cores (red contours) and the corresponding column density maps for clumps at 1.6 Myr.

\begin{figure*}[htb!]
	\includegraphics[width=1.0\textwidth]{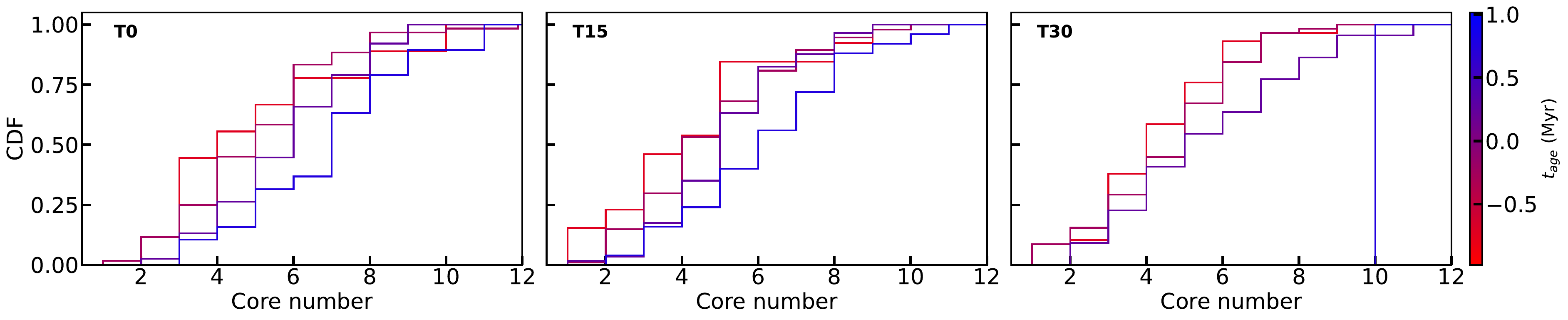}
	\caption{CDFs of identified core number using dendrogram. For each model, the data are binned in equal time interval. The CDFs are color-coded by the time of the bin center.  The left, middle and right panels correspond to models T0, T15, and T30.}
	\label{fig:EvoCDF_CoreNum}
\end{figure*}

Considering all clumps across the evolutionary sequence, the median number of identified cores is 6, 5, and 4 for models T0, T15, and T30. 
As the magnetic field strength increases with higher $\theta_B$, this trend suggests that fragmentation is suppressed in models T15 and T30, consistent with the findings in Section~\ref{sec:SheetEvo}.
Furthermore, we can bin the sample in time. 
Figure~\ref{fig:EvoCDF_CoreNum} shows the evolution of the number of identified cores for each time bin in each model.
The color gradient transitions from red to blue and shifts towards the right, indicating an increase in the number of cores over time.
This supports the conclusion that fragmentation is an ongoing process.

\begin{figure*}[htb!]
	\includegraphics[width=1.0\textwidth]{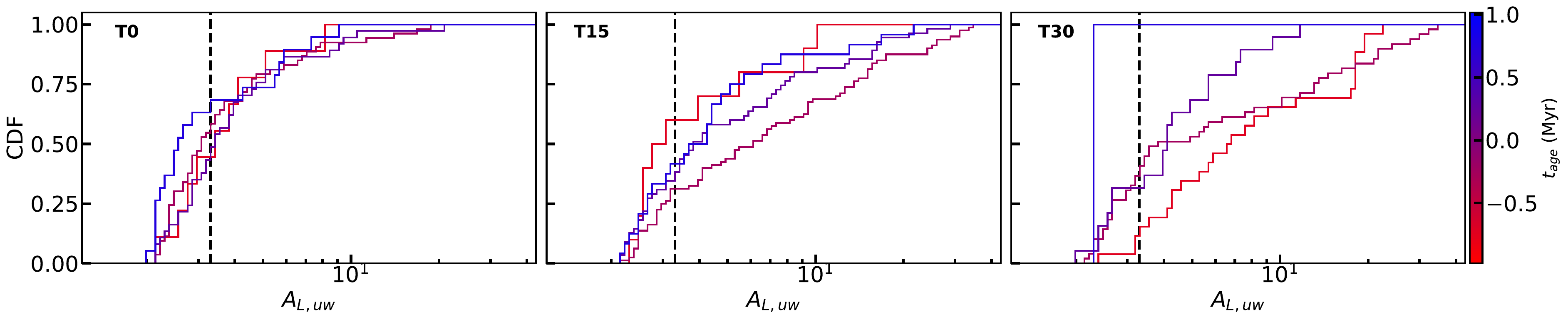}
	\includegraphics[width=1.0\textwidth]{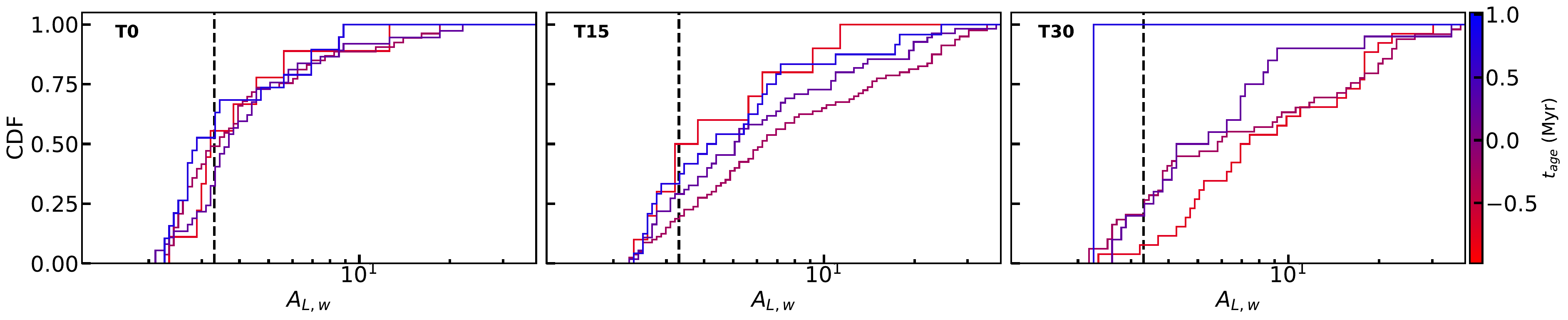}
	\caption{CDFs of \ALuw and \ALw for for each time bin and each model. The figure configuration is the same as Figure~\ref{fig:EvoCDF_CoreNum}. The upper and lower panels display the CDFs for \ALuw and \ALw, respectively. The vertical black-dashed line indicates \AL = 3.3.}
	\label{fig:EvoCDFAL}
\end{figure*}

Next, to quantify the core distribution in each clump, we employed the alignment parameter (\AL) from \cite{2025ApJ...979...67C}. 
The goal of this metric is to intuitively distinguish a thin, \textquote{string-like} arrangement of cores from a more dispersed \textquote{cluster}.
It achieves this by measuring the typical separation between cores in units of the entire distribution's thickness. 
Conceptually, the algorithm uses Principal Component Analysis to determine the shortest axis (minor axis) of the core group. 
For a highly aligned distribution, this \textquote{thickness} is very small, so the core-to-core separations are large in comparison, yielding a high \AL value.
For a clustered distribution, the thickness is large and more comparable to the internal separations, yielding a low \AL value. 
The full mathematical details, including \textit{unweighted} (\ALuw) and flux-\textit{weighted} (\ALw) versions, can be found in \cite{2025ApJ...979...67C} and were implemented using the Python package from \cite{ALPara}.
Additionally, a threshold of 3.3 can be effectively used as a two-label classification tool to separate the sample into two distinct groups.
For each model, the sample, after excluding clumps with only one or two cores, was binned again in time. 
Table~\ref{tab:StatAL} presents the range of $t_{age}$, the number of clumps, the mean and median values of \AL, and the fraction of clumps classified as clustered ($f_{\rm{clustered}}$) for each time bin. 
Figure~\ref{fig:EvoCDFAL} shows the CDFs of \AL for each time bin and each model.

\begin{deluxetable*}{@{\extracolsep{4pt}}lcccccccc@{}}
	\tablecaption{Statistical results of \AL for each time bin and each model. \label{tab:StatAL}}
	\tablehead{
	\multicolumn{3}{c}{} & \multicolumn{3}{c}{\ALuw} & \multicolumn{3}{c}{\ALw} \\
	\cline{4-6} \cline{7-9}
	\colhead{Model} & \colhead{$t_{age}$ range (Myr)} & \colhead{Clump num.}  & \colhead{Mean} & \colhead{Median} & \colhead{$f_{\rm{clustered}}$ (\%)} & \colhead{Mean} & \colhead{Median} & \colhead{$f_{\rm{clustered}}$ (\%)}
	} 
	\startdata
	T0 & $<$ -0.50 & 9 & 3.9 & 3.4 & 44.4 & 4.6 & 3.3 & 44.4 \\
	-	& -0.50 - 0.00 & 53 & 4.5 & 3.1 & 54.7 & 5.1 & 3.5 & 49.1	\\
	-	& 0.00 - 0.50 & 37 & 4.6 & 3.5 & 43.2 & 5.2 & 3.7 & 29.7	\\
	-	& $>$ 0.50 & 19 & 3.7 & 2.6 & 63.2 & 4.1 & 2.9 & 52.6	\\
	\hline
	T15	& $<$ -0.50 & 10 & 4.5 & 3.0 & 60.0	 & 5.1 & 3.6 & 50.0 \\
	-	& -0.50 - 0.00 & 80 & 9.7 & 6.2 & 30.0 & 10.5 & 6.4 & 18.8	\\
	-	& 0.00 - 0.50 & 55 & 6.8 & 3.9 & 34.5 & 8.2 & 5.2 & 29.1	\\
	-	&  $>$ 0.50 & 24 & 5.6 & 4.0 & 41.7	 & 6.5 & 4.3 & 33.3 \\
	\hline
	T30	& $<$ -0.50 & 26 & 9.8 & 6.8 & 11.5 & 10.7 & 7.4 & 7.7	\\
	-	& -0.50 - 0.00 & 49 & 9.3 & 3.8 & 36.7 & 10.3 & 5.9 & 20.4	\\
	-	& 0.00 - 0.50 & 20 & 6.7 & 4.2 & 30.0 & 7.4 & 5.0 & 20.0	\\
	-	& $>$ 0.50 & 1 & 2.3 & 2.3 & 100.0 & 2.3 & 2.3 & 100.0	\\
	\enddata
	\tablecomments{
	Clumps with only one or two clumps are excluded.
	}
\end{deluxetable*}

Unlike the trend observed in the number of identified cores (Figure~\ref{fig:EvoCDF_CoreNum}), there is no clear monotonic evolution in \AL across all models. 
Only models T15 and T30 exhibit a tentative trend towards lower \AL values over time. 
This trend is evident in the decreasing mean, median, and increasing $f_{\rm{clustered}}$ values in Table~\ref{tab:StatAL}. 
More specifically, for \ALuw, when comparing the distributions of clumps in the intervals $0.00$ Myr $< t_{age} < 0.50$ Myr and $t_{age} <$ -0.50 Myr, the two-sample KS test reveals KS statistics of 0.14, 0.23, and 0.33, with p-values of 0.99, 0.40, and 0.07 for models T0, T15, and T30, respectively.
These results indicate that the distributions at different time intervals become more statistically dissimilar for models with higher $\theta_B$. 
A similar trend is also observed for \ALw, suggesting a gradual increase in core clustering, which is consistent with the recent findings by ALMAGAL survey \citep{2025arXiv251012892W}. 
More discussion about core clustering can be found in Section~\ref{sec:Chaotic}.

Additionally, the overall shape of the CDFs for models T15 and T30 suggests a higher prevalence of high \AL values compared to model T0.
This difference in the \AL distributions between the models, especially at early times, is a direct consequence of the initial magnetic field geometry. In the high-$\theta_B$ models, the strong, sheet-parallel magnetic field channels accreting gas into coherent filaments (Section~\ref{sec:SheetEvo}). 
The subsequent gravitational fragmentation of these well-defined filaments naturally produces cores in an aligned pattern, leading to higher \AL values. 
This strong guiding mechanism is absent in the T0 model, where fragmentation is more isotropic from the beginning. 
As we will discuss in Section~\ref{sec:Chaotic}, this initial, magnetically-imprinted alignment is subsequently disrupted by gravitational evolution and the growth of turbulence, causing the distributions to become more clustered over time.

Finally, to assess the dynamic behavior of \ALuw, we measured its variability. 
For any three consecutive clumps within an evolutionary sequence identified in Section~\ref{sec:EvoSeq}, we quantified the variability using the metric:
\begin{equation}\label{eqn:SAL}
	\mathcal{V}_{AL} = \frac{(A_L(t_2) - A_L(t_1))(A_L(t_3) - A_L(t_2))}{(A_L(t_1)A_L(t_2)A_L(t_3))^{2/3}},
\end{equation}
where $t_1$, $t_2$, and $t_3$ represent consecutive timeframes within the evolutionary sequence. 
Only sequences with at least three timeframes were considered, and sequences with more than three timeframes yielded multiple $\mathcal{V}_{AL}$ values.
A positive value of $\mathcal{V}_{AL}$ indicates a monotonic increase or decrease in \AL over the three timeframes.
Conversely, a negative value of $\mathcal{V}_{AL}$ indicates a decrease or increase in \AL from $t_1$ to $t_2$, followed by an inverse change from $t_2$ to $t_3$.
The magnitude represents the degree of variability.

\begin{figure}[htb!]
	\includegraphics[width=0.45\textwidth]{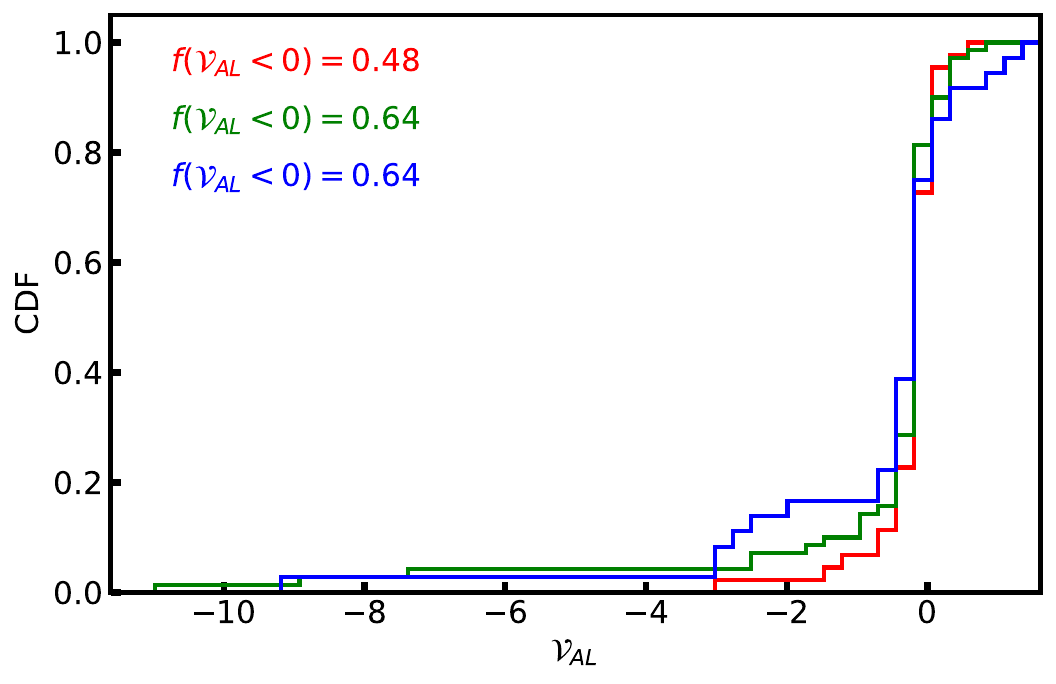}
	\caption{CDFs of $\mathcal{V}_{AL}$ (Equation~\ref{eqn:SAL}) for each model. Different colors represent the model of origin (red for T0, green for T15, and blue for T30). $\mathcal{V}_{AL}$ is a metric used to measure the variability of \ALuw over time. The fraction of negative $\mathcal{V}_{AL}$ values ($f(\mathcal{V}_{AL} < 0)$) is displayed and colored by model.}
	\label{fig:StabilityAL}
\end{figure}

The CDFs of $\mathcal{V}_{AL}$ using \ALuw for each model are shown in Figure~\ref{fig:StabilityAL}. 
The fraction of negative $\mathcal{V}_{AL}$ values ($f(\mathcal{V}_{AL} < 0)$) is also displayed for each model. 
We observed that all models have $f(\mathcal{V}_{AL} < 0) \gtrsim 0.5$, indicating that \ALuw frequently experiences unpredictable changes during evolution. 
Additionally, models with higher $\theta_B$ exhibit a longer tail towards more negative values in the $\mathcal{V}_{AL}$ distribution, suggesting that \ALuw is undergoing more dramatic changes in these cases. 
This is because their initially aligned, high-\AL configurations are inherently fragile and more susceptible to disruption by the chaotic gravitational and turbulent processes as the clump evolves towards a more clustered state, which will be further discussed in Section~\ref{sec:Chaotic}.
Similar behavior is observed for \ALw.

\begin{figure*}[htb!]
	\includegraphics[width=1.0\textwidth]{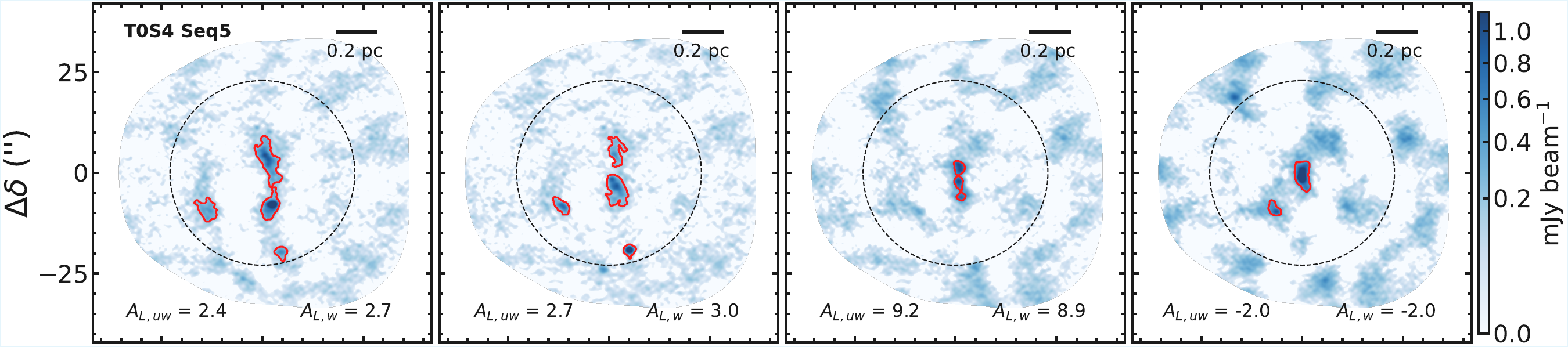}
	\includegraphics[width=1.0\textwidth]{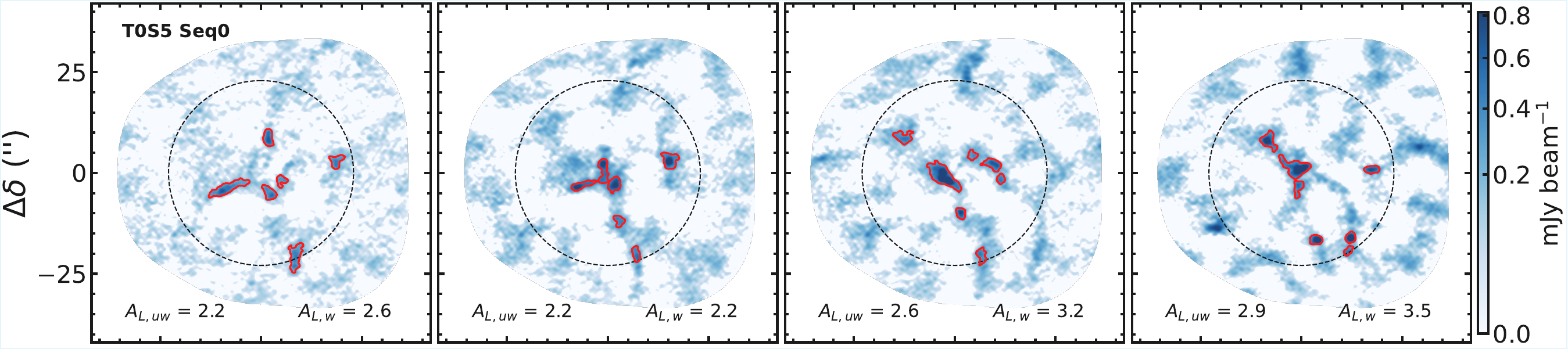}
	\includegraphics[width=1.0\textwidth]{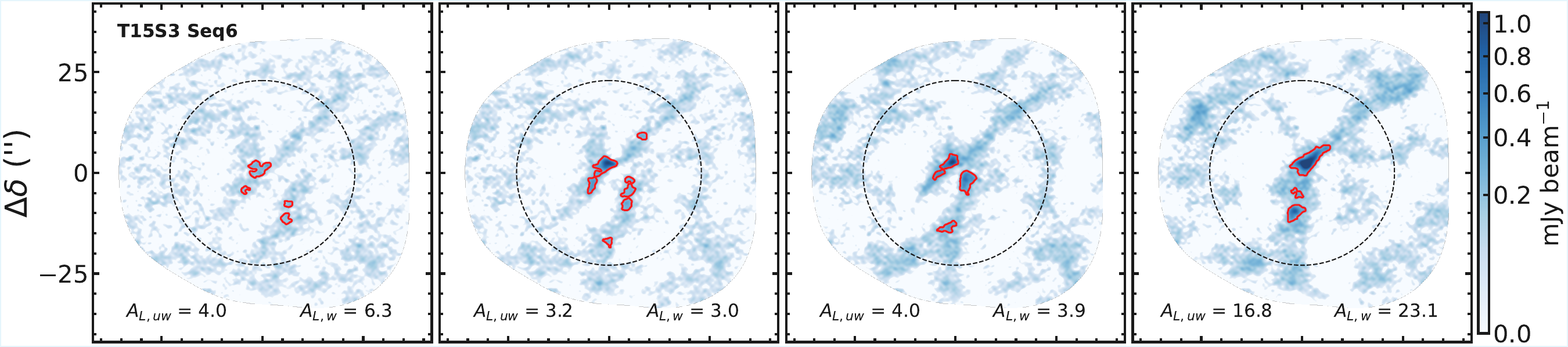}
    \includegraphics[width=1.0\textwidth]{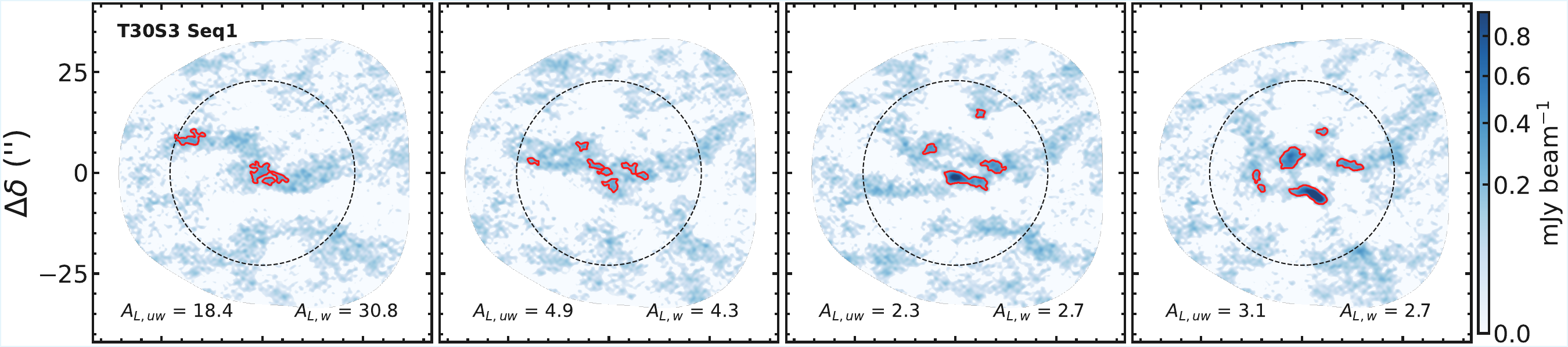}
    \includegraphics[width=1.0\textwidth]{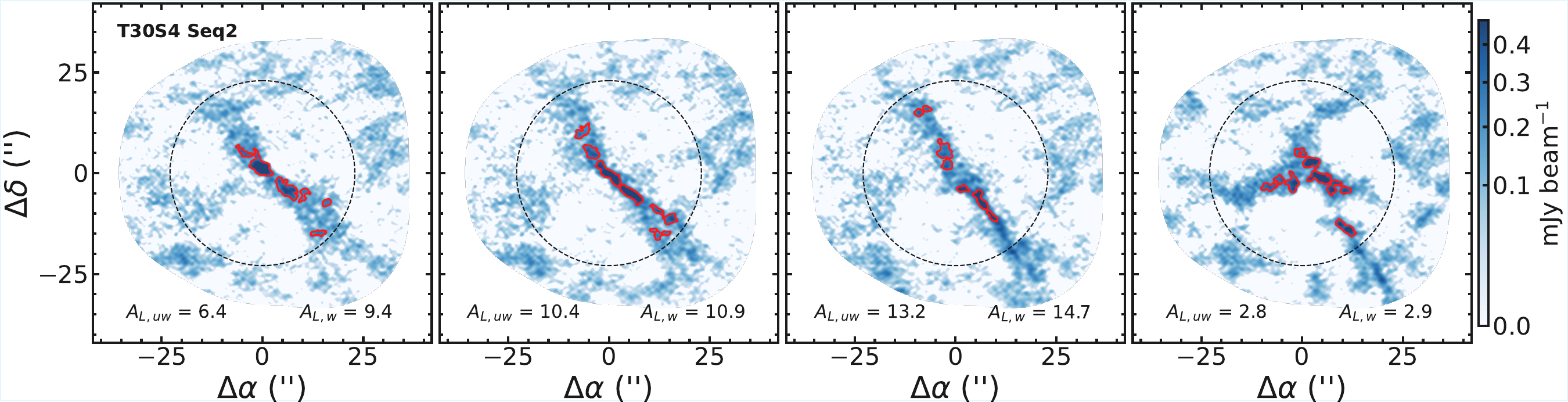}
	\caption{Evolution of core distributions and \AL. Each row tracks the evolution of a single clump (from left to right) with a time interval of 0.2 Myr between panels. Red contours, overlaid on the synthetic images, identify the cores found by the dendrogram. The corresponding \AL value is noted in each panel (see Section~\ref{sec:GetAL}).}
	\label{fig:ALContEvo}
\end{figure*}

Equation~\ref{eqn:SAL} can also be applied to the core number to investigate the evolution variability of the core number ($\mathcal{V}_{\rm{core}}$), which all models reveal $f(\mathcal{V}_{\rm{core}} < 0) \lesssim 0.5$.
This result confirms that the core population does not evolve monotonically.
When correlating $\mathcal{V}_{\rm{core}}$ with $\mathcal{V}_{AL}$, we find no significant correlation (all models show a Kendall's $\tau < 0.15$, see Section~\ref{sec:ALClump} for the test details).
These results demonstrate that the complex, unpredictable evolution of \AL is not driven simply by the net change in core number.
Instead, the evolution of \AL is also influenced by the specific interactions between cores. 
Figure~\ref{fig:ALContEvo} provides a direct visual illustration of these dynamics. 
The figure shows identified cores on the synthetic images of the same clump at three sequential times (each separated by 0.2 Myr).
These snapshots capture the very mechanisms controlling \AL and showing clear examples of core merging, core migration, and ongoing fragmentation occurring simultaneously. 
These processes continuously alter the relative positions and spatial configuration of the cores in a way that is not captured by the total core count alone.

In conclusion, we employed \AL to quantify core distribution. 
Our results indicate that while there is no clear monotonic evolutionary trend for individual clumps, there is an overall tentative tendency towards lower \AL values compared to the distribution at earlier stages when considering the entire population.
Additionally, models with higher $\theta_B$ tend to exhibit higher \AL at an earlier stage but are also more unstable. 

Overall, these findings suggest that core distribution exhibits noticeable time-dependent and \textquote{dynamic} behavior. 
This raises the question of whether there is a relationship between the instantaneous measurement of core distribution and large-scale clump properties.
Section~\ref{sec:Chaotic} will discuss potential mechanisms driving this dynamic behavior.

\section{Core Distribution and Clump Properties} \label{sec:ALClump}
The primary objective of this study is to investigate the potential correlations between core distribution using alignment parameters (\ALuw and \ALw) and the clump properties derived in Sections~\ref{sec:Clump} and \ref{sec:Core}.

Since observation data can't distinguish the initial conditions under which clumps form, we analyzed all identified clumps for these correlations.
In total, 450 clumps were included in the analysis. 
After excluding clumps with only one or two cores, we retained 404 clumps for statistical testing.
To measure the strength of correlations, we employed Kendall's rank correlation, a non-parametric statistical measure of association between two datasets \citep{KendallTest}. 
Given the sufficient number of data points, we considered a Kendall's $\tau$ value with an absolute value greater than 0.6 as \textquote{strong}, between 0.3 and 0.6 as \textquote{moderate}, and less than 0.3 as \textquote{weak}.

As this section focuses on the overall correlations between \AL and clump properties, an analysis of the differences between clumps with distinct fragmentation patterns is presented in Appendix~\ref{app:CorrALThers}.

\subsection{Correlations with Observables} \label{sec:CorrelationALObs}
In the work of \cite{2025ApJ...979...67C}, the authors applied \AL to ASHES sample and reported weak correlations with the identified core number and clump density, and no correlations with $M_{cl}$, $R_{cl}$, $R_1$, and $R_2$. 
However, due to the limited sample size and lack of information about the magnetic field, conclusive statements regarding these correlations could not be drawn. 
In this work, the larger sample size should allow us to derive more robust statistical relations and explore the underlying mechanisms.

\begin{figure*}[htb!]
	\includegraphics[width=1.0\textwidth]{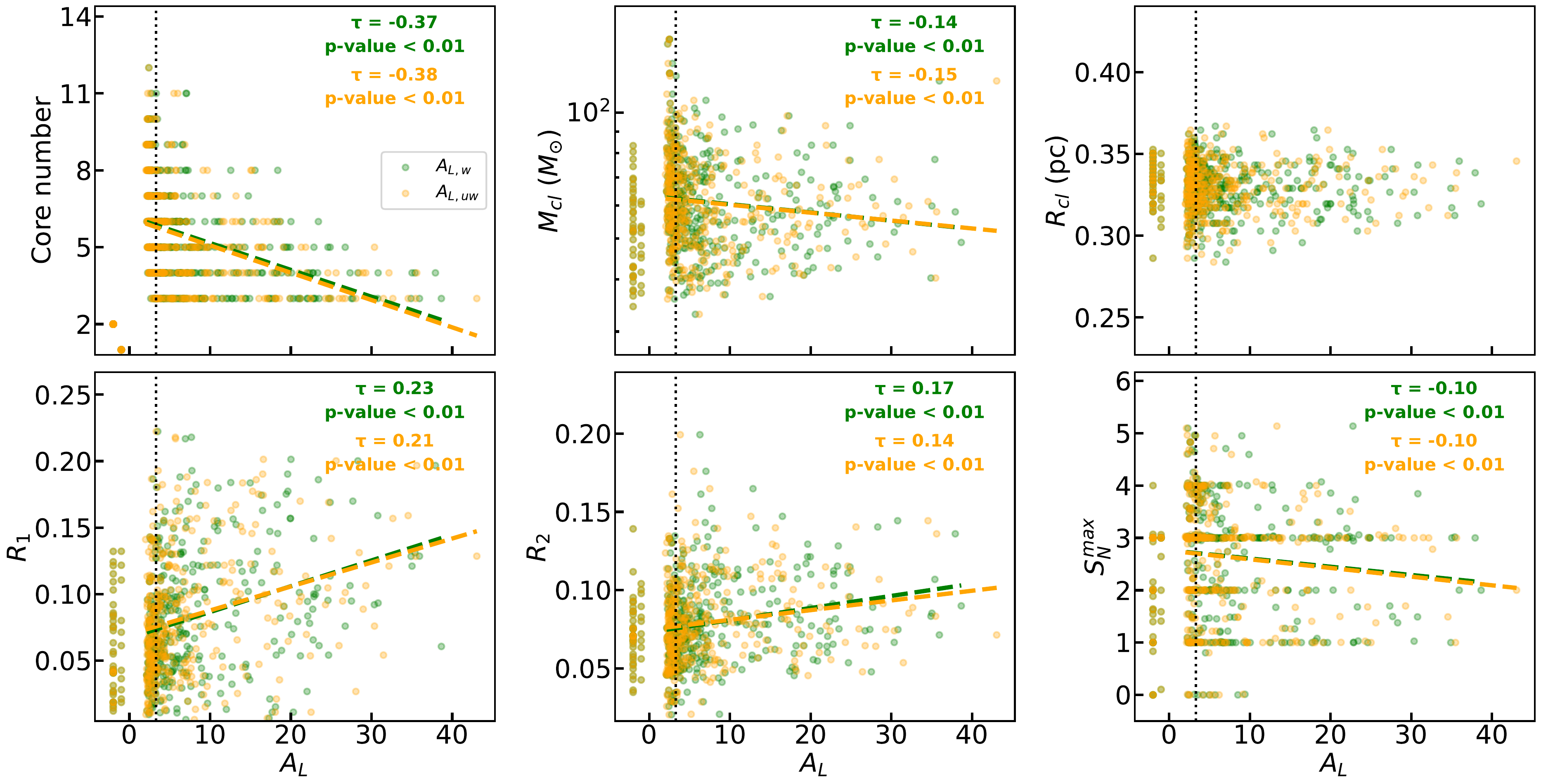}
	\caption{Relationships between \ALuw (orange) and \ALw (green) with identified core number obtained from Section~\ref{sec:GetAL} and clump properties derived in Sections~\ref{sec:MR} to \ref{sec:SNsub}. The Kendall's $\tau$ and corresponding $p$-value are displayed for the correlations with $p$-value $< 0.05$ and a linear function is fitted to visualize the trend. The vertical dashed line indicates \AL = 3.3. Note that clumps with only one or two cores are assigned \AL values of -1 and -2, respectively, and are excluded from the correlation test.}
	\label{fig:MetaCorrAL}
\end{figure*}

We first analyze the correlations between \AL and clump properties from Sections~\ref{sec:MR} to \ref{sec:SNsub}.
Additionally, the core number derived in Section~\ref{sec:GetAL} is also included here.
The relations are presented in Figure~\ref{fig:MetaCorrAL}, with \ALuw represented in orange and \ALw in green. 
For correlations with $p$-values less than 0.05, indicating a significant association, the statistical results and a fitted linear function are shown to visualize the relationship.

Among these clump properties, only the core number, as identified in Section~\ref{sec:GetAL}, exhibits a moderate correlation with both \ALuw and \ALw, with absolute $\tau$ values of $\sim 0.4$. 
This correlation is expected, as a smaller number of cores increases the likelihood of an aligned core configuration, as discussed in \cite{2025ApJ...979...67C}.
The remaining correlations between alignment parameters and clump properties are all significant but weak.

Interestingly, we observe a weak negative correlation with $M_{cl}$. 
This tentative trend suggests that more massive clumps may tend to have lower \AL values, implying a more clustered core configuration. 
Considering the lack of correlation with $R_{cl}$ and the observed trend with $M_{cl}$, these findings likely contribute to the weak correlations with clump density also reported by \cite{2025ApJ...979...67C}.

Regarding the correlations with $R_1$ and $R_2$, both are found to be weak, suggesting that neither the elongation nor central condensation of a clump significantly influences core distribution. 
However, a tentative trend suggests that circular clumps (with lower $R_1$) may exhibit more clustered core configurations. 
A similar and weaker trend is also seen with flatter clumps (with $R_2 \sim 0$).

Finally, the negative correlation with $S_N^{max}$ suggests that high \AL values or aligned cores are more likely to be associated with filamentary structures ($S_N^{max} = 1 \sim 2$) than with hubs ($S_N^{max} \geq 3$). 
While this is expected from filament fragmentation \citep{2016MNRAS.458..319C, 2017MNRAS.468.2489C, 2020MNRAS.497.4390C}, it is important to note that high \AL values can still be found in hubs, indicating that considering only local filamentary structures is insufficient to fully explain core distribution.

In summary, the weak and no correlations found here are consistent with the results from \cite{2025ApJ...979...67C} using ASHES sample. 
Therefore, these findings suggest that the instantaneous core alignment is not strongly determined by factors such as clump mass, elongation, or the environment in which the clump is formed (filament or hub).

\subsection{Correlations with Energy Ratios and Turbulence} \label{sec:CorrelationALEnergy}
In this section, we investigate the correlations between core distribution and energy-related parameters derived in Section~\ref{sec:EnergyTerms}. 
These comparisons enable us to explore how gravity, magnetic fields, and turbulence influence core distribution. 

\begin{figure*}[htb!]
	\includegraphics[width=1.0\textwidth]{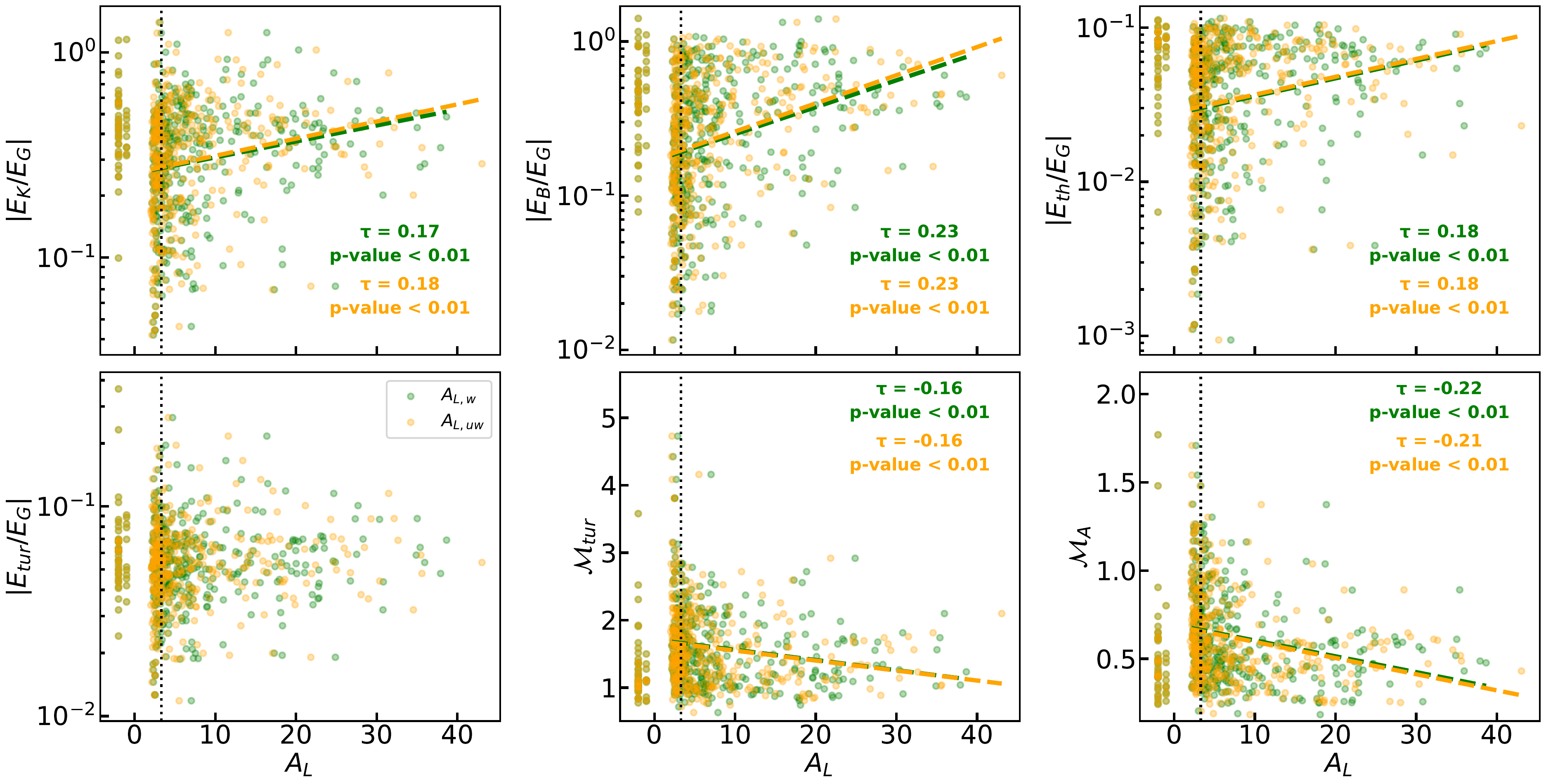}
	\caption{Relationships between \ALuw (orange) and \ALw (green) with clump properties derived in Section \ref{sec:EnergyTerms}. The Kendall's $\tau$ and corresponding $p$-value are displayed for the correlations with $p$-value $< 0.05$ and a linear function is fitted to visualize the trend. The vertical dashed line indicates \AL = 3.3. Note that clumps with only one or two cores are assigned \AL values of -1 and -2, respectively, and are excluded from the correlation test.}
	\label{fig:MetaCorrAL05}
\end{figure*}

The results are shown in Figure~\ref{fig:MetaCorrAL05}, with the same panel configuration as Figure~\ref{fig:MetaCorrAL}. 
Notably, a positive and significant correlation was observed with $|E_B/E_G|$, seemingly aligning with previous observational findings and representing one of the strongest correlations observed in this analysis (excluding the correlation with core number).
This suggests that a more aligned core configuration is more likely to be found in environments with significant magnetic energy ($|E_B/E_G| \gtrsim 0.1$).
However, in this regime, there are still several clumps with \AL $< 3.3$, indicating a clustered configuration is still populated.

Additionally, for clumps with only one or two cores (manually assigned \AL = -1 and -2, and they are excluded from the correlation test), a distribution at higher values of $|E_B/E_G|$ was observed compared to other data points. 
Using the two-sample KS test, the distributions of \AL $< 0$ and \AL $> 0$ were compared. 
For both \ALuw and \ALw, the KS test yielded a statistic of 0.3 and $p$-values of 0.09 and 0.07, respectively, suggesting a possible difference between the distributions.
This shows that high $|E_B/E_G|$ environments can also suppress fragmentation, which is consistent with other studies \citep{2011ApJ...742L...9C, 2013ApJ...779...96T, 2017ApJ...848....2H, 2018A&A...614A..64B}.

Interestingly, only the correlation with $|E_{tur}/E_G|$ was found to be insignificant, even the less fragmented clumps were distributing at the similar range. 
This suggests that small-scale turbulence plays a similar role in the energy balance of clumps regardless of their core distribution (no, aligned or clustered fragmentation).
Therefore, small-scale turbulence has little effect on maintaining core alignment.

Finally, significant but weak correlations are observed between \AL and $\mathcal{M}_{tur}$. 
Notably, highly turbulent clumps are exclusively found to have clustered core distributions. 
In contrast, aligned core configurations are only observed in clumps with trans-sonic turbulence. 
This trend is further supported by the analysis of $\mathcal{M}_{A}$, which considers the influence of the magnetic field. 
The results indicate that super-Alfv\'enic clumps tend to exhibit more clustered core configurations.

\subsection{Direct View on \AL Evolution} \label{sec:EvoCluster}
Thus far, we have only discussed the behavior and correlations of \AL using the entire sample and interpreted its statistical properties. 
To gain a different perspective on its dynamic feature, we now utilize the energy terms derived in Section~\ref{sec:EnergyTerms} and examine how the energetically similar clumps and their \AL evolve over time.

To achieve this, we first selected clumps with $-0.3$ Myr $< t_{age} < -0.2$ Myr from all models. 
We then extracted the corresponding values of $E_G$, $E_K$, $E_B$, and $E_{tur}$ for these clumps and normalized each term to have a mean of 0 and a standard deviation of 1. 
Next, we applied the Density-Based Spatial Clustering of Applications with Noise (DBSCAN) algorithm \citep{DBSCAN1996, DBSCAN2017} from the \texttt{scikit-learn} Python package to these normalized energy terms. 
DBSCAN is a clustering algorithm that groups data points based on their density distribution in parameter space. 
We used \textit{eps} = 0.3 and \textit{min\_samples} = 1 for the DBSCAN parameters.
Finally, DBSCAN identified 11 clumps that met the above criteria. 
The energy ranges for these clumps in real space are: $10^{44.6 - 44.8}$ erg for $|E_G|$, $10^{44.2 - 44.4}$ erg for $E_K$, $10^{43.8 - 44.4}$ erg for $E_B$, and $10^{43.4 - 43.6}$ erg for $E_{tur}$.

\begin{figure}[htb!]
	\includegraphics[width=0.45\textwidth]{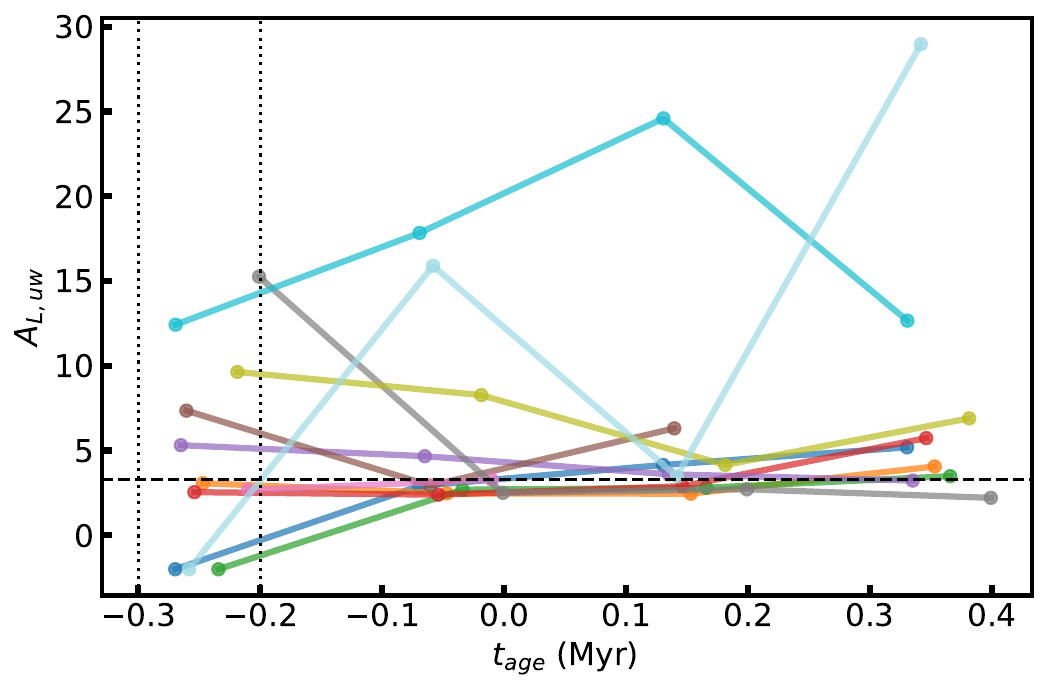}
	\caption{Evolution of \ALuw for the 11 clumps sampled in Section~\ref{sec:EvoCluster}. Each colored line represents the evolutionary path of a single clump. The two vertical dotted lines mark the time interval during which the clumps were sampled, and the dashed horizontal line indicates the threshold of \ALuw = 3.3.}
	\label{fig:EvoClusterALuw}
\end{figure}

Figure~\ref{fig:EvoClusterALuw} presents the evolution of \ALuw for these 11 clumps. 
The two vertical dotted lines mark the time interval during which they were sampled, and the dashed horizontal line indicates the threshold of \ALuw = 3.3. 
Despite sharing similar energetic properties, these clumps exhibit significant variation in their core distribution: five are aligned (\ALuw $>$ 3.3), three are clustered (0 $<$ \ALuw $<$ 3.3), and the remaining three are less fragmented (\ALuw $<$ 0). 
This reinforces the reliability of the weak correlations found in Section~\ref{sec:CorrelationALEnergy} and supports the conclusion that clump properties alone can't fully determine core distribution (Section~\ref{sec:Chaotic}).
Furthermore, different clumps exhibit diverse evolutionary paths: some show minimal variation, while others experience drastic changes in \ALuw. 
A similar trend was observed when using \ALw for these 11 clumps, highlighting that even when clumps are energetically similar at a specific time, their \textquote{current} properties are not sufficient to predict their core distribution.

\subsection{Chaotic and Dynamic Nature} \label{sec:Chaotic}
The lack of strong correlations between alignment parameters and clump properties observed in Sections~\ref{sec:CorrelationALObs} and \ref{sec:CorrelationALEnergy} suggests a \textquote{chaotic} nature in core distribution, consistent with the findings of \cite{2025ApJ...979...67C} using observation data.
Furthermore, the absence of a simple monotonic evolution in alignment parameters in Section~\ref{sec:GetAL}, along with the highly varying core distributions observed for energetically similar clumps in Section~\ref{sec:EvoCluster}, indicates a time-dependent and \textquote{dynamic} nature of core formation as clumps evolve.
These features suggest that instantaneous measurements of core distributions may not always accurately reflect the underlying clump properties.

Turbulence, due to its chaotic nature, may contribute to the lack of correlation.
The bottom right panel of Figure~\ref{fig:EvoEnergy} shows that, unlike other energy components, $E_{tur}$ is co-evolving with gravity before the formation of stars. 
With the supersonic nature of clumps after sink particle formation (Figure~\ref{fig:EvoLine_Mach}), it suggests that turbulence plays a significant role in the system.
Firstly, turbulence can introduce non-linear density perturbations in the gas, which may serve as seeds for new structure formation. 
While the clump is globally collapsing, any density enhancements caused by turbulence would have an even shorter collapse timescale than the background medium, suggesting that ongoing fragmentation is possible within our evolutionary sequences (see \citealp{2019MNRAS.490.3061V} for the discussion of non-homologous collapse). 
This conclusion is consistent with the trend of increasing core numbers shown in Figure~\ref{fig:EvoCDF_CoreNum} and is more evident by observing the second and third row of Figure~\ref{fig:ALContEvo}.
Since alignment parameters, which measure the (weighted) instantaneous core separation, are used to trace the core distribution, the formation of new cores can significantly alter these values, leading to both chaotic and dynamic features. 
This mechanism of turbulence-induced fragmentation within a globally collapsing clump has direct implications for the likelihood of fragmentation and the resulting core masses. 
By constantly generating local density enhancements, turbulence makes fragmentation an efficient and ongoing process rather than a single, monolithic event. 
This could naturally lead to the formation of a population of cores with a spectrum of masses, consistent with the idea of turbulent fragmentation \citep{2004RvMP...76..125M, 2025A&A...695L..25I}. 
While a full analysis of the core mass function and its evolution is beyond the scope of this paper, which focuses on the core spatial distribution, the physical process we identify provides a clear framework for understanding these outcomes.

Secondly, turbulence is stochastic in nature. 
This sensitivity of star-forming regions to turbulence is also discussed in \cite{2022MNRAS.511.2702J}, where different turbulence realizations with statistically identical initial conditions were implemented in star cluster simulations and found to yield $\sim 50\%$ differences in star number and star formation efficiency, highlighting the chaotic nature of star formation.


Moreover, the evolution of clumps and cores also plays a role in core distribution, as shown in Section~\ref{sec:GetAL}. 
While the overall \AL values exhibit dynamic evolution for individual clumps, there is a tentative trend towards decreasing values when considering the entire population.
This suggests that core distributions generally become more clustered as they evolve, a trend that can be explained by two concurrent physical processes. 
First, as the system is gravity-dominated, the formed cores migrate towards the clump's potential minimum or even merge with each other, causing an initially aligned pattern to evolve into a more compact cluster dynamically (e.g., the fourth row of Figure~\ref{fig:ALContEvo}). 
Second, the ongoing turbulence-induced fragmentation we describe can create new cores at later times (e.g., the second and third row of Figure~\ref{fig:ALContEvo}). 
The stochastic nature of this process and the growing turbulence strength mean these new cores are unlikely to follow the initial alignment, further decreasing the measured \AL value. 
The combination of gravitational migration and continuous, disordered fragmentation provides a theoretical basis for the observed trend towards more clustered core configurations in more evolved clumps.
Observations of star-forming clumps at different evolutionary stages have indeed revealed this trend, with more evolved clumps exhibiting shorter core separations and more compact core distributions \citep{2018A&A...617A.100B, 2021A&A...649A.113B, 2023MNRAS.520.2306T, 2024ApJS..270....9X, 2024arXiv240706845I, 2025arXiv251012892W}.

These results also rule out the possibility that the lack of strong correlations in \cite{2025ApJ...979...67C} is due to the missing key clump properties. 
Our findings suggest that properly characterizing turbulence and tracing the evolution of clumps and cores is crucial for understanding core distribution.
While models with higher $\theta_B$ can produce clumps with high \AL values, the combined effects of turbulence and gravitational evolution can obscure the correlation between clump properties and core distribution.


\section{Caveats} \label{sec:Caveats}
The simulations presented in this work did not consider stellar feedback, which is a crucial mechanism affecting gas dynamic in high-mass star formation regions and can potentially influence the later fragmentation process \citep{2007ApJ...656..959K}.
However, even when only considering clumps before the formation of sink particles or with $t_{age} < 0$ (as shown in the Appendix~\ref{app:CorrBeforeSink}), the correlations found in Figure~\ref{fig:MetaCorrAL} and Figure~\ref{fig:MetaCorrAL05} persist and remain weak. 
This suggests that our analysis remains valid for including the clumps after $t_{age} = 0$.

Furthermore, the parameter space explored in our simulations is still relatively narrow, resulting in clumps with similar properties (e.g., masses varying by only a factor of two). 
Additionally, the clumps in our simulations are typically low-mass, with masses $\lesssim 10^2 M_{\odot}$. 
This could be due to the definition of clump mass in Section~\ref{sec:MR}, where we only consider the gas mass within a sphere of radius 0.5 pc. 
These values are not directly comparable to the IRDCs investigated in the ASHES survey, which typically have masses of $\sim 10^{2-4} M_{\odot}$ \citep{2019ApJ...886..102S, 2023ApJ...950..148M}.
This difference is primarily attributed to the chosen initial parameters (i.e., $n_0$ and $v_0$) for the convergent flows. 
Increasing these values to enhance the accretion rate of the sheet-like structure could lead to the formation of higher-mass clumps. 
However, this would significantly require more computational resources due to the increasing number of cells with high refinement levels, limiting our ability to obtain a sufficient number of turbulence realizations for statistical analysis.
And with higher mass, the clumps could collapse faster, resulting in a shorter pre-feedback phase. 
This raises further concerns regarding the exclusion of stellar feedback in our simulations.
Regarding the magnetic field, we only considered a single value for $B_0$ in this study. 
However, it remains unclear whether a stronger magnetic field, ideally capable of further suppressing turbulence and sustaining large-scale structures, could preserve or suppress the aligned core configuration. 
In addition to these points, our study was also limited to a single type of initial turbulence ( Burgers turbulence). A systematic investigation into how different turbulence modes (e.g., solenoidal vs. compressive) or power spectra influence the fragmentation process and its predictability would be a valuable future study, though it is beyond the computational scope of the current work.
To avoid potential biases arising from the limited parameter space, which may lead to similar fragmentation processes, future studies should explore a wider range of initial conditions.

Our analysis of the energy budget in Section~\ref{sec:EnergyTerms} focuses on the internal gravitational, kinetic, and magnetic energies of the clumps. 
It does not explicitly quantify the impact of the external gravitational field from the larger-scale environment in which the clumps are embedded. 
The anisotropic tidal forces within a filamentary network can significantly deform and shear a collapsing clump on a timescale comparable to its free-fall time \citep{2024MNRAS.528.3630G}. 
This tidal influence could directly affect the fragmentation process. 
While a detailed analysis of the tidal field is beyond the scope of this study, we acknowledge that it is an important physical mechanism that warrants dedicated investigation in future work.

Finally, in the simulations, we adopted a time interval of 0.2 Myr for dumping snapshots, which was subsequently used to analyze the evolutionary sequence.
Given the chaotic and dynamic nature of core distribution observed in our analysis, tracing the motion of individual cores can provide valuable insights into their evolution. 
Using the typical values of $\mathcal{M}_{tur}$, which ranges from 1 to 3 (upper panel of Figure~\ref{fig:EvoLine_Mach}), we can estimate that the gas displacement within the clump, which is $\sim 0.05 - 0.1$ pc for a time interval of 0.2 Myr. 
This displacement is comparable to the size of the cores (check Figure~\ref{fig:SYNGallery}).
The issue arises, as the core distribution becomes more clustered and compact, cores may be too close to each other or merge, making it challenging to track their individual motions. 
While this study primarily focused on statistical properties and their correlations with clump-scale properties, future work could delve deeper into tracing the motion of individual cores by choosing a smaller time interval between snapshots.

\section{Conclusion} \label{sec:Conclusion}
In this study, we investigated the correlation between core distribution in star-forming regions and the properties of their host clumps. 
We employed simulations of convergent flows with three magnetic field inclinations ($\theta_B$) and five turbulence realizations.

Convergent flows generated sheet-like structures, which subsequently underwent turbulent instability and formed massive clumps. 
To mimic observational studies, we smoothed the structures and used dendrograms to identify star-forming clumps. 
Various techniques were then applied to estimate clump properties (e.g., mass, size, energy, etc.), and radiative transfer simulations were used to generate synthetic images that mimic ALMA observations.
Cores were identified within these synthetic images, and \textquote{alignment parameters (\AL)} were used to quantify core distribution.

The evolution of these properties with respect to different $\theta_B$ was discussed, and correlations between alignment parameters were also investigated.
By applying a consistent and physically-motivated analysis framework to our large sample of simulated clumps, we can summarize our investigation as follows.
The major findings of this study are:
\begin{enumerate}
	\item In Section~\ref{sec:SheetEvo}, we demonstrate that an inclined magnetic field introduces a perpendicular component relative to the gas flows. As the sheet evolves and accretes mass, this component is subsequently compressed and amplified, influencing the evolution of substructures. Furthermore, this component can also facilitate accretion onto the substructures, contributing to the formation of filamentary structures.
	\item For the clump properties, only mass (Section~\ref{sec:MR}), energy-related parameters and turbulence (Section~\ref{sec:EnergyTerms}) exhibited a clear evolutionary trend. Specifically, clumps generally became more massive, gravity-dominated, and more turbulent over time. 
	\item The core distribution, as quantified by \ALuw and \ALw in Section~\ref{sec:GetAL}, does not exhibit a monotonic trend but rather a \textquote{dynamic} evolution. While examining all clumps in the simulations, we observed a tentative trend towards lower \AL, suggesting that clumps become more clustered at later stages. 
	\item Our analysis revealed weak correlations between core distribution, as quantified by alignment parameters, and clump properties (Sections~\ref{sec:CorrelationALObs} and \ref{sec:CorrelationALEnergy}). This suggests a \textquote{chaotic} nature in core distribution, consistent with the findings of \cite{2025ApJ...979...67C}.
	\item Turbulence is believed to contribute to the \textquote{chaotic and dynamic} behavior by stochastically introducing density perturbations that can lead to ongoing fragmentation and statistically scattering in core distribution. Additionally, as clumps evolve, cores can gravitationally migrate or merge, further influencing core distribution. 
	\item We conclude that instantaneous, large-scale properties of clumps are not sufficient to reliably determine the spatial distribution of their cores. Clumps with similar energetic properties can exhibit diverse core distribution and evolutionary paths (Section~\ref{sec:EvoCluster}).
\end{enumerate}

\begin{acknowledgments}
	We thank the anonymous referee for prompt reviews and constructive suggestions to improve this paper.
    Additionally, we acknowledge Yueh-Ning Lee and Nai-Chieh Lin for their initial support for the simulation setups.
	This work is supported by NSTC grants 112-2112-M-001-066- and NSTC 114-2112-M-001-042 -. 
	We acknowledge Hsi-Yu Schive and the GAMER team for their support of using the GAMER code.
	The computation resources utilized in this work were supported by the TIARA Cluster at the Academia Sinica Institute of Astronomy and Astrophysics (ASIAA).
\end{acknowledgments}

\software{NumPy \citep{harris2020array},
	SciPy \citep{2020SciPy-NMeth},
	Matplotlib \citep{Hunter:2007},
	scikit-learn \citep{scikit-learn}, 
	Astropy \citep{astropy:2013, astropy:2018, astropy:2022}, 
	astrodendro \citep{2019ascl.soft07016R}, 
	pytreegrav \citep{2021JOSS....6.3675G}, 
	R-J plots \citep{2022MNRAS.516.2782C}, 
	DisPerSE \citep{2011MNRAS.414..350S},
	Alignment Parameters \citep{ALPara, 2025ApJ...979...67C}.}

\appendix
\section{Comparisons Between Clumps Showing Different Fragmentation Status} \label{app:CorrALThers}
Figures~\ref{fig:CompareALThres1} and \ref{fig:CompareALThres2} show the comparison of CDFs between clumps classified as aligned (\AL $>$ 3.3) and clustered (\AL $\leq$ 3.3) (Note that clumps with \AL $<$ 0 are excluded). 
The two-sample KS test was used to assess the differences between the distributions. 
For parameters that exhibited only weak correlations in Sections~\ref{sec:CorrelationALObs} and \ref{sec:CorrelationALEnergy}, the difference between the aligned and clustered groups is clearer and more significant (with KS statistics $\sim 0.4$) when using the threshold of \AL = 3.3. 
This is observed for $M_{cl}$, $R_1$, $|E_{th}/E_G|$, $\mathcal{M}_{tur}$, and $\mathcal{M}_{A}$.

Similarly, we can split the sample by \AL = 0 and examine the difference between less fragmented and well fragmented clumps. 
While the detailed results are not shown here, we note that in addition to the parameters that showed significant differences when using \AL = 3.3 (as mentioned above), the correlation for $|E_B/E_G|$ also becomes more pronounced when using this threshold.

These results seemingly suggest that clustered/aligned core configurations and well/less fragmented clumps exhibit distinct large-scale properties. 
However, it is important to note that the observed differences when using the threshold of \AL = 3.3 may be influenced by overall evolutionary trends. 
As discussed in Section~\ref{sec:GetAL}, \AL tends to decrease over time, and some parameters, such as $M_{cl}$ (Section~\ref{sec:MR}), also exhibit evolutionary trends. 
Consequently, the differences observed here could be partially attributed to these evolutionary effects. 
This is supported by the fact that the differences are primarily driven by extreme values, as seen in the tails of the distributions.

\begin{figure*}[htb!]
	\includegraphics[width=1.0\textwidth]{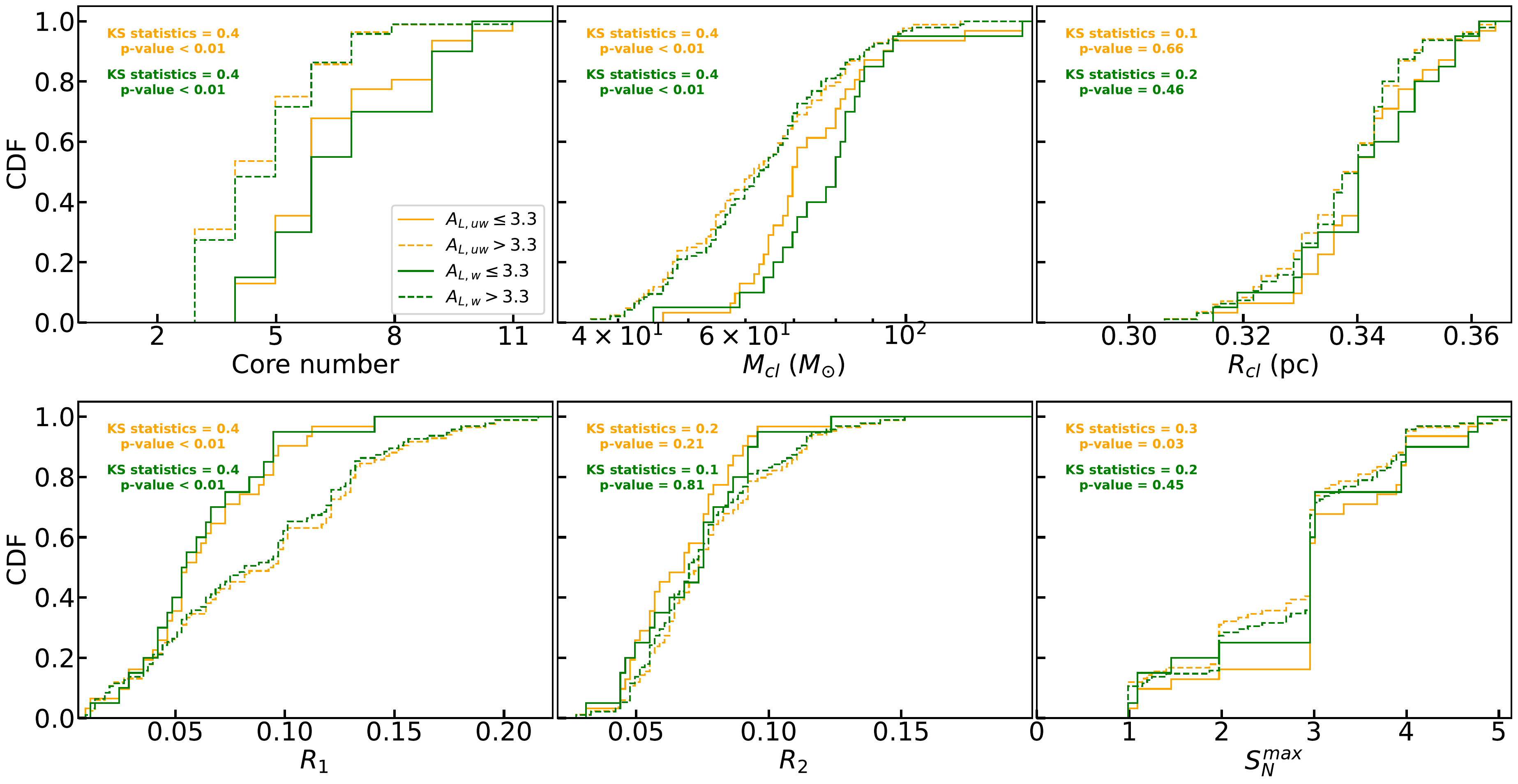}
	\caption{Comparison of CDFs between clumps classified as aligned (\AL $>$ 3.3) and clustered (\AL $\leq$ 3.3) for the parameters shown in Figure~\ref{fig:MetaCorrAL}. Note that clumps with \AL $<$ 0 are excluded. The two-sample KS test was used to assess the difference between the distributions, and the results are shown in each panel.}
	\label{fig:CompareALThres1}
\end{figure*}

\begin{figure*}[htb!]
	\includegraphics[width=1.0\textwidth]{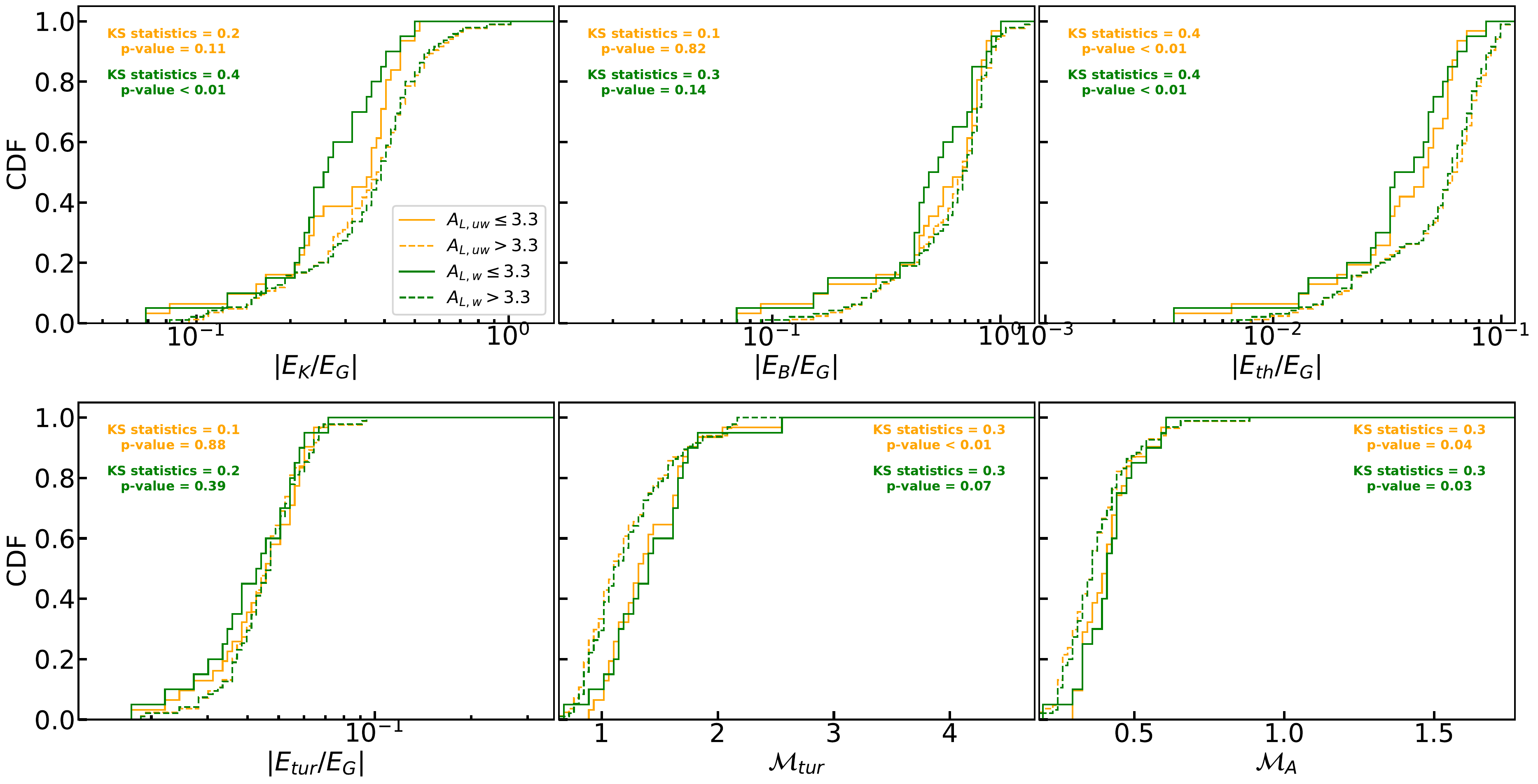}
	\caption{Continuation of Figure~\ref{fig:CompareALThres1}, but for the parameters shown in Figure~\ref{fig:MetaCorrAL05}.}
	\label{fig:CompareALThres2}
\end{figure*}

\section{Correlations for Clumps Before the Formation of Sink Particle} \label{app:CorrBeforeSink}
\begin{figure*}[htb!]
	\includegraphics[width=1.0\textwidth]{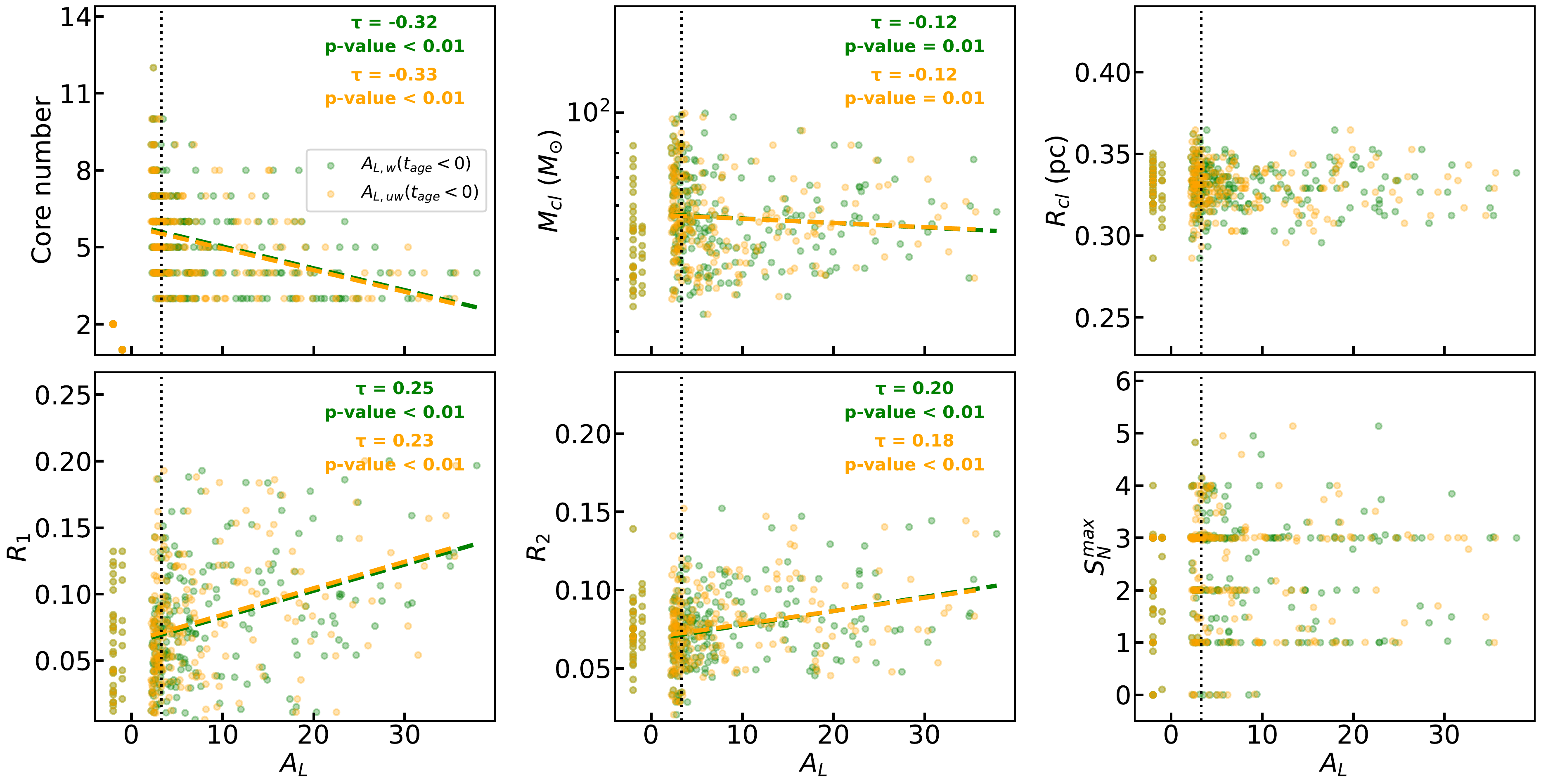}
	\caption{Same as Figure~\ref{fig:MetaCorrAL}, but only consider clumps with $t_{age} < 0$.}
	\label{fig:MetaCorrALBeforeSink}
\end{figure*}

\begin{figure*}[htb!]
	\includegraphics[width=1.0\textwidth]{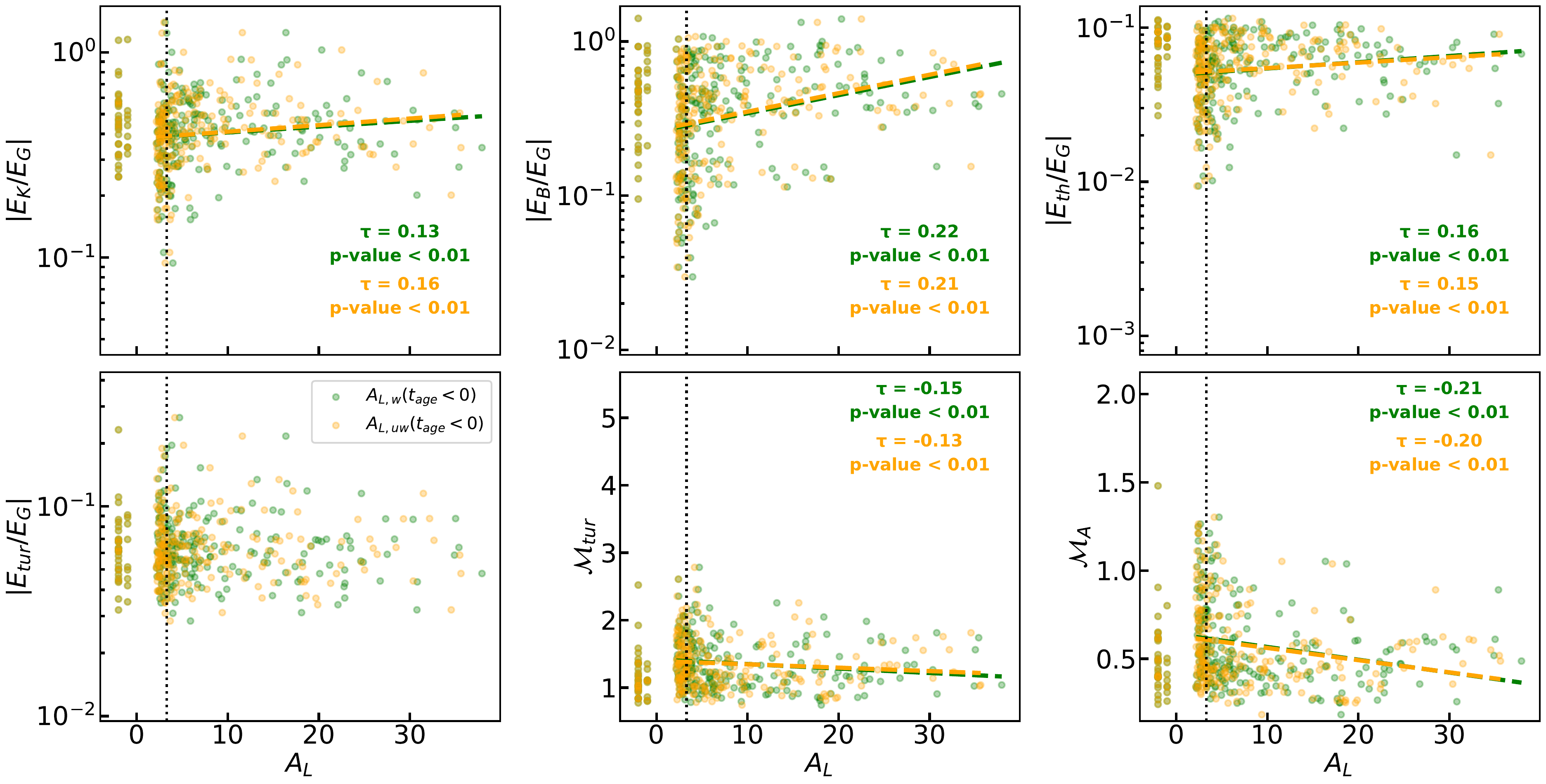}
	\caption{Same as Figure~\ref{fig:MetaCorrAL05}, but only consider clumps with $t_{age} < 0$.}
	\label{fig:MetaCorrAL05BeforeSink}
\end{figure*}

Figure~\ref{fig:MetaCorrALBeforeSink} and \ref{fig:MetaCorrAL05BeforeSink} present the results of similar correlation analyses conducted in Sections~\ref{sec:CorrelationALObs} and \ref{sec:CorrelationALEnergy}, but only focusing on clumps identified before the formation of sink particles (i.e., clumps with $t_{age} < 0$).
In this case, there are 263 clumps for \AL and 227 clumps after excluding \AL $<$ 0. 
Overall, we find that the correlations persist and remain weak, suggesting that the inclusion of more evolved clumps in our simulations, despite the absence of stellar feedback, would not significantly alter the results.

\section{Decomposition of Gravitational Energy Evolution} \label{app:GravDecom}
In Section~\ref{sec:EnergyTerms}, we note that the gravitational energy ($|E_G|$) is always dominant during clump evolution. 
To disentangle the physical processes driving this feature, we perform a differential analysis on clumps between two adjacent snapshots, $t_1$ and $t_2$ (where $t_2 > t_1$), chosen such that no new sink particles form in the interval.

We start with the clump at $t_2$ and recalculate its gravitational energy four separate times. 
In each test, we revert a single component of the clump to its state at $t_1$, keeping all other components at their $t_2$ values. 
The magnitude of the change in $E_G$ relative to its original $t_2$ value reveals the importance of that component. 
The four cases are:
\begin{enumerate}
    \item The total gas mass in the clump is reset to its $t_1$ value, but the $t_2$ gas density profile and all $t_2$ sink properties (masses and locations) are used.
    \item The mass of each sink particle is reset to its $t_1$ value, but the $t_2$ gas properties and $t_2$ sink locations are used.
    \item The gas density profile is reset to its $t_1$ shape, but the $t_2$ total gas mass and all $t_2$ sink properties are used.
    \item The spatial locations of the sink particles are reset to their $t_1$ positions, but the $t_2$ gas properties and $t_2$ sink masses are used.
\end{enumerate}

We must note that these components are dynamically coupled in the simulation. 
This decomposition is a diagnostic tool intended only to estimate the relative importance of each effect.

\begin{figure}[htb!]
	\epsscale{1.1}
	\centering
	\includegraphics[width=0.45\textwidth]{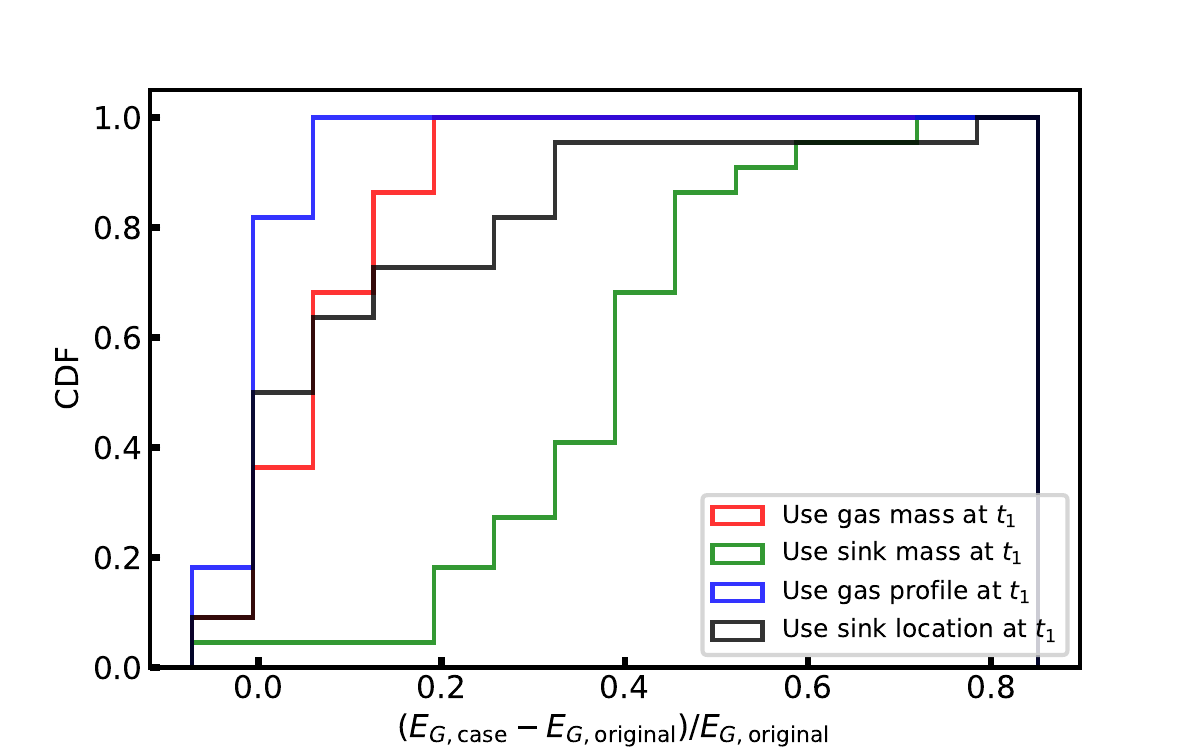}
	\caption{CDFs of the fractional change in gravitational energy ($(E_{G, \rm{case}} - E_{G, \rm{original}})/E_{G, \rm{original}}$, where $E_{G, \rm{original}}$ is the original gravitational energy for the clump at $t_2$) for the four test cases described in Appendix C. The lines correspond to Case 1 (red), Case 2 (green), Case 3 (blue), and Case 4 (black).}
	\label{fig:GravDecom_CDF}
\end{figure}

We applied this analysis to 22 clumps that fulfill the condition (e.g., no new particle forming between snapshots). 
The results, shown in Figure~\ref{fig:GravDecom_CDF}, clearly indicate that Case 2 (green) results in the most significant change compared to the original gravitational energy. 
The contributions from large-scale gas accretion (Case 1, red), gas redistribution (Case 3, blue), and sink clustering (Case 4, black) are all comparatively minor.

Given that our simulations do not include feedback and the accreted gas will eventually go into sink particles, we conclude that the increase in the dominance of gravity is primarily driven by gas accretion on to the clump, rather than by the internal clustering of cores or the large-scale redistribution of gas within the clump.

\bibliography{main}{}

@ARTICLE{2018ApJS..236...50Z,
	author = {{Zhang}, Ui-Han and {Schive}, Hsi-Yu and {Chiueh}, Tzihong},
	title = "{Magnetohydrodynamics with GAMER}",
	journal = {\apjs},
	year = 2018,
	month = jun,
	volume = {236},
	number = {2},
	eid = {50},
	pages = {50},
	doi = {10.3847/1538-4365/aac49e},
	archivePrefix = {arXiv},
	eprint = {1804.03479},
	primaryClass = {physics.comp-ph},
	adsurl = {https://ui.adsabs.harvard.edu/abs/2018ApJS..236...50Z}
}

@ARTICLE{2018MNRAS.481.4815S,
	author = {{Schive}, Hsi-Yu and {ZuHone}, John A. and {Goldbaum}, Nathan J. and {Turk}, Matthew J. and {Gaspari}, Massimo and {Cheng}, Chin-Yu},
	title = "{GAMER-2: a GPU-accelerated adaptive mesh refinement code - accuracy, performance, and scalability}",
	journal = {\mnras},
	year = 2018,
	month = dec,
	volume = {481},
	number = {4},
	pages = {4815-4840},
	doi = {10.1093/mnras/sty2586},
	archivePrefix = {arXiv},
	eprint = {1712.07070},
	primaryClass = {astro-ph.IM},
	adsurl = {https://ui.adsabs.harvard.edu/abs/2018MNRAS.481.4815S}
}

@book{toro2009riemann,
	title={Riemann Solvers and Numerical Methods for Fluid Dynamics: A Practical Introduction},
	author={Toro, E.F.},
	isbn={9783540498346},
	lccn={2009921818},
	year={2009},
	publisher={Springer Berlin Heidelberg}
}

@ARTICLE{MIYOSHI2005315,
	author = {{Miyoshi}, Takahiro and {Kusano}, Kanya},
	title = "{A multi-state HLL approximate Riemann solver for ideal magnetohydrodynamics}",
	journal = {Journal of Computational Physics},
	year = 2005,
	month = sep,
	volume = {208},
	number = {1},
	pages = {315-344},
	doi = {10.1016/j.jcp.2005.02.017},
	adsurl = {https://ui.adsabs.harvard.edu/abs/2005JCoPh.208..315M}
}

@ARTICLE{FFTW3,
	author={Frigo, M. and Johnson, S.G.},
	journal={Proceedings of the IEEE}, 
	title={The Design and Implementation of FFTW3}, 
	year={2005},
	volume={93},
	number={2},
	pages={216-231},
	doi={10.1109/JPROC.2004.840301}}

@book{SOR,
	title={Numerical Recipes 3rd Edition: The Art of Scientific Computing},
	author={Press, W.H.},
	isbn={9780521880688},
	lccn={2007062003},
	series={Numerical Recipes: The Art of Scientific Computing},
	year={2007},
	publisher={Cambridge University Press}
}

@ARTICLE{1988ApJ...332..659E,
	author = {{Evans}, Charles R. and {Hawley}, John F.},
	title = "{Simulation of Magnetohydrodynamic Flows: A Constrained Transport Model}",
	journal = {\apj},
	year = 1988,
	month = sep,
	volume = {332},
	pages = {659},
	doi = {10.1086/166684},
	adsurl = {https://ui.adsabs.harvard.edu/abs/1988ApJ...332..659E}
}

@ARTICLE{2008JCoPh.227.4123G,
	author = {{Gardiner}, Thomas A. and {Stone}, James M.},
	title = "{An unsplit Godunov method for ideal MHD via constrained transport in three dimensions}",
	journal = {Journal of Computational Physics},
	year = 2008,
	month = apr,
	volume = {227},
	number = {8},
	pages = {4123-4141},
	doi = {10.1016/j.jcp.2007.12.017},
	archivePrefix = {arXiv},
	eprint = {0712.2634},
	primaryClass = {astro-ph},
	adsurl = {https://ui.adsabs.harvard.edu/abs/2008JCoPh.227.4123G}
}

@ARTICLE{1997ApJ...489L.179T,
	author = {{Truelove}, J. Kelly and {Klein}, Richard I. and {McKee}, Christopher F. and {Holliman}, John H., II and {Howell}, Louis H. and {Greenough}, Jeffrey A.},
	title = "{The Jeans Condition: A New Constraint on Spatial Resolution in Simulations of Isothermal Self-gravitational Hydrodynamics}",
	journal = {\apjl},
	year = 1997,
	month = nov,
	volume = {489},
	number = {2},
	pages = {L179-L183},
	doi = {10.1086/310975},
	adsurl = {https://ui.adsabs.harvard.edu/abs/1997ApJ...489L.179T}
}

@ARTICLE{2010ApJ...713..269F,
	author = {{Federrath}, Christoph and {Banerjee}, Robi and {Clark}, Paul C. and {Klessen}, Ralf S.},
	title = "{Modeling Collapse and Accretion in Turbulent Gas Clouds: Implementation and Comparison of Sink Particles in AMR and SPH}",
	journal = {\apj},
	year = 2010,
	month = apr,
	volume = {713},
	number = {1},
	pages = {269-290},
	doi = {10.1088/0004-637X/713/1/269},
	archivePrefix = {arXiv},
	eprint = {1001.4456},
	primaryClass = {astro-ph.SR},
	adsurl = {https://ui.adsabs.harvard.edu/abs/2010ApJ...713..269F}
}

@ARTICLE{2017MNRAS.468.2489C,
	author = {{Clarke}, S.~D. and {Whitworth}, A.~P. and {Duarte-Cabral}, A. and {Hubber}, D.~A.},
	title = "{Filamentary fragmentation in a turbulent medium}",
	journal = {\mnras},
	year = 2017,
	month = jun,
	volume = {468},
	number = {2},
	pages = {2489-2505},
	doi = {10.1093/mnras/stx637},
	archivePrefix = {arXiv},
	eprint = {1703.04473},
	primaryClass = {astro-ph.GA},
	adsurl = {https://ui.adsabs.harvard.edu/abs/2017MNRAS.468.2489C}
}

@BOOK{1988csup.book.....H,
	author = {{Hockney}, R.~W. and {Eastwood}, J.~W.},
	title = "{Computer simulation using particles}",
	year = 1988,
	adsurl = {https://ui.adsabs.harvard.edu/abs/1988csup.book.....H}
}

@ARTICLE{1998ApJ...495..346M,
	author = {{Masunaga}, Hirohiko and {Miyama}, Shoken M. and {Inutsuka}, Shu-ichiro},
	title = "{A Radiation Hydrodynamic Model for Protostellar Collapse. I. The First Collapse}",
	journal = {\apj},
	year = 1998,
	month = mar,
	volume = {495},
	number = {1},
	pages = {346-369},
	doi = {10.1086/305281},
	adsurl = {https://ui.adsabs.harvard.edu/abs/1998ApJ...495..346M}
}

@ARTICLE{2000ApJ...531..350M,
	author = {{Masunaga}, Hirohiko and {Inutsuka}, Shu-ichiro},
	title = "{A Radiation Hydrodynamic Model for Protostellar Collapse. II. The Second Collapse and the Birth of a Protostar}",
	journal = {\apj},
	year = 2000,
	month = mar,
	volume = {531},
	number = {1},
	pages = {350-365},
	doi = {10.1086/308439},
	adsurl = {https://ui.adsabs.harvard.edu/abs/2000ApJ...531..350M}
}

@ARTICLE{WOODWARD1984115,
	author = {{Woodward}, P. and {Colella}, P.},
	title = "{The Numerical Stimulation of Two-Dimensional Fluid Flow with Strong Shocks}",
	journal = {Journal of Computational Physics},
	year = 1984,
	month = apr,
	volume = {54},
	number = {1},
	pages = {115-173},
	doi = {10.1016/0021-9991(84)90142-6},
	adsurl = {https://ui.adsabs.harvard.edu/abs/1984JCoPh..54..115W}
}

@INPROCEEDINGS{2023ASPC..534..233P,
	author = {{Pineda}, J.~E. and {Arzoumanian}, D. and {Andre}, P. and {Friesen}, R.~K. and {Zavagno}, A. and {Clarke}, S.~D. and {Inoue}, T. and {Chen}, C. and {Lee}, Y. and {Soler}, J.~D. and {Kuffmeier}, M.},
	title = "{From Bubbles and Filaments to Cores and Disks: Gas Gathering and Growth of Structure Leading to the Formation of Stellar Systems}",
	booktitle = {Protostars and Planets VII},
	year = 2023,
	editor = {{Inutsuka}, S. and {Aikawa}, Y. and {Muto}, T. and {Tomida}, K. and {Tamura}, M.},
	series = {Astronomical Society of the Pacific Conference Series},
	volume = {534},
	month = jul,
	pages = {233},
	doi = {10.48550/arXiv.2205.03935},
	archivePrefix = {arXiv},
	eprint = {2205.03935},
	primaryClass = {astro-ph.GA},
	adsurl = {https://ui.adsabs.harvard.edu/abs/2023ASPC..534..233P}
}

@ARTICLE{2015MNRAS.449..662L,
	author = {{Lomax}, O. and {Whitworth}, A.~P. and {Hubber}, D.~A.},
	title = "{On the effects of solenoidal and compressive turbulence in pre-stellar cores}",
	journal = {\mnras},
	year = 2015,
	month = may,
	volume = {449},
	number = {1},
	pages = {662-669},
	doi = {10.1093/mnras/stv310},
	archivePrefix = {arXiv},
	eprint = {1502.04009},
	primaryClass = {astro-ph.SR},
	adsurl = {https://ui.adsabs.harvard.edu/abs/2015MNRAS.449..662L}
}

@software{2012ascl.soft02015D,
	author = {{Dullemond}, C.~P. and {Juhasz}, A. and {Pohl}, A. and {Sereshti}, F. and {Shetty}, R. and {Peters}, T. and {Commercon}, B. and {Flock}, M.},
	title = "{RADMC-3D: A multi-purpose radiative transfer tool}",
	howpublished = {Astrophysics Source Code Library, record ascl:1202.015},
	year = 2012,
	month = feb,
	eid = {ascl:1202.015},
	adsurl = {https://ui.adsabs.harvard.edu/abs/2012ascl.soft02015D}
}

@ARTICLE{1994A&A...291..943O,
	author = {{Ossenkopf}, V. and {Henning}, Th.},
	title = "{Dust opacities for protostellar cores.}",
	journal = {\aap},
	year = 1994,
	month = nov,
	volume = {291},
	pages = {943-959},
	adsurl = {https://ui.adsabs.harvard.edu/abs/1994A&A...291..943O}
}

@ARTICLE{2009A&A...504..415S,
	author = {{Schuller}, F. and {Menten}, K.~M. and {Contreras}, Y. and {Wyrowski}, F. and {Schilke}, P. and {Bronfman}, L. and {Henning}, T. and {Walmsley}, C.~M. and {Beuther}, H. and {Bontemps}, S. and {Cesaroni}, R. and {Deharveng}, L. and {Garay}, G. and {Herpin}, F. and {Lefloch}, B. and {Linz}, H. and {Mardones}, D. and {Minier}, V. and {Molinari}, S. and {Motte}, F. and {Nyman}, L. -{\r{A}}. and {Reveret}, V. and {Risacher}, C. and {Russeil}, D. and {Schneider}, N. and {Testi}, L. and {Troost}, T. and {Vasyunina}, T. and {Wienen}, M. and {Zavagno}, A. and {Kovacs}, A. and {Kreysa}, E. and {Siringo}, G. and {Wei{\ss}}, A.},
	title = "{ATLASGAL - The APEX telescope large area survey of the galaxy at 870 {\ensuremath{\mu}}m}",
	journal = {\aap},
	year = 2009,
	month = sep,
	volume = {504},
	number = {2},
	pages = {415-427},
	doi = {10.1051/0004-6361/200811568},
	archivePrefix = {arXiv},
	eprint = {0903.1369},
	primaryClass = {astro-ph.GA},
	adsurl = {https://ui.adsabs.harvard.edu/abs/2009A&A...504..415S}
}

@software{2019ascl.soft07016R,
	author = {{Robitaille}, Thomas and {Rice}, Tom and {Beaumont}, Chris and {Ginsburg}, Adam and {MacDonald}, Braden and {Rosolowsky}, Erik},
	title = "{astrodendro: Astronomical data dendrogram creator}",
	howpublished = {Astrophysics Source Code Library, record ascl:1907.016},
	year = 2019,
	month = jul,
	eid = {ascl:1907.016},
	adsurl = {https://ui.adsabs.harvard.edu/abs/2019ascl.soft07016R}
}

@ARTICLE{2019ApJ...886..102S,
	author = {{Sanhueza}, Patricio and {Contreras}, Yanett and {Wu}, Benjamin and {Jackson}, James M. and {Guzm{\'a}n}, Andr{\'e}s E. and {Zhang}, Qizhou and {Li}, Shanghuo and {Lu}, Xing and {Silva}, Andrea and {Izumi}, Natsuko and {Liu}, Tie and {Miura}, Rie E. and {Tatematsu}, Ken'ichi and {Sakai}, Takeshi and {Beuther}, Henrik and {Garay}, Guido and {Ohashi}, Satoshi and {Saito}, Masao and {Nakamura}, Fumitaka and {Saigo}, Kazuya and {Veena}, V.~S. and {Nguyen-Luong}, Quang and {Tafoya}, Daniel},
	title = "{The ALMA Survey of 70 {\ensuremath{\mu}}m Dark High-mass Clumps in Early Stages (ASHES). I. Pilot Survey: Clump Fragmentation}",
	journal = {\apj},
	year = 2019,
	month = dec,
	volume = {886},
	number = {2},
	eid = {102},
	pages = {102},
	doi = {10.3847/1538-4357/ab45e9},
	archivePrefix = {arXiv},
	eprint = {1909.07985},
	primaryClass = {astro-ph.GA},
	adsurl = {https://ui.adsabs.harvard.edu/abs/2019ApJ...886..102S}
}

@ARTICLE{2023ApJ...950..148M,
	author = {{Morii}, Kaho and {Sanhueza}, Patricio and {Nakamura}, Fumitaka and {Zhang}, Qizhou and {Sabatini}, Giovanni and {Beuther}, Henrik and {Lu}, Xing and {Li}, Shanghuo and {Garay}, Guido and {Jackson}, James M. and {Olguin}, Fernando A. and {Tafoya}, Daniel and {Tatematsu}, Ken'ichi and {Izumi}, Natsuko and {Sakai}, Takeshi and {Silva}, Andrea},
	title = "{The ALMA Survey of 70 {\ensuremath{\mu}}m Dark High-mass Clumps in Early Stages (ASHES). IX. Physical Properties and Spatial Distribution of Cores in IRDCs}",
	journal = {\apj},
	year = 2023,
	month = jun,
	volume = {950},
	number = {2},
	eid = {148},
	pages = {148},
	doi = {10.3847/1538-4357/acccea},
	archivePrefix = {arXiv},
	eprint = {2304.01757},
	primaryClass = {astro-ph.GA},
	adsurl = {https://ui.adsabs.harvard.edu/abs/2023ApJ...950..148M}
}

@INPROCEEDINGS{2007ASPC..376..127M,
	author = {{McMullin}, J.~P. and {Waters}, B. and {Schiebel}, D. and {Young}, W. and {Golap}, K.},
	title = "{CASA Architecture and Applications}",
	booktitle = {Astronomical Data Analysis Software and Systems XVI},
	year = 2007,
	editor = {{Shaw}, R.~A. and {Hill}, F. and {Bell}, D.~J.},
	series = {Astronomical Society of the Pacific Conference Series},
	volume = {376},
	month = oct,
	pages = {127},
	adsurl = {https://ui.adsabs.harvard.edu/abs/2007ASPC..376..127M}
}

@ARTICLE{scikit-learn,
	title={Scikit-learn: Machine Learning in {P}ython},
	author={Pedregosa, F. and Varoquaux, G. and Gramfort, A. and Michel, V.
	and Thirion, B. and Grisel, O. and Blondel, M. and Prettenhofer, P.
	and Weiss, R. and Dubourg, V. and Vanderplas, J. and Passos, A. and
	Cournapeau, D. and Brucher, M. and Perrot, M. and Duchesnay, E.},
	journal={Journal of Machine Learning Research},
	volume={12},
	pages={2825--2830},
	year={2011}
}

@ARTICLE{2022MNRAS.516.2782C,
	author = {{Clarke}, S.~D. and {Jaffa}, S.~E. and {Whitworth}, A.~P.},
	title = "{RJ-plots: An improved method to classify structures objectively}",
	journal = {\mnras},
	year = 2022,
	month = oct,
	volume = {516},
	number = {2},
	pages = {2782-2791},
	doi = {10.1093/mnras/stac2318},
	archivePrefix = {arXiv},
	eprint = {2208.07509},
	primaryClass = {astro-ph.GA},
	adsurl = {https://ui.adsabs.harvard.edu/abs/2022MNRAS.516.2782C}
}

@ARTICLE{2011MNRAS.414..350S,
	author = {{Sousbie}, T.},
	title = "{The persistent cosmic web and its filamentary structure - I. Theory and implementation}",
	journal = {\mnras},
	year = 2011,
	month = jun,
	volume = {414},
	number = {1},
	pages = {350-383},
	doi = {10.1111/j.1365-2966.2011.18394.x},
	archivePrefix = {arXiv},
	eprint = {1009.4015},
	primaryClass = {astro-ph.CO},
	adsurl = {https://ui.adsabs.harvard.edu/abs/2011MNRAS.414..350S}
}

@ARTICLE{2020MNRAS.497.4390C,
	author = {{Clarke}, S.~D. and {Williams}, G.~M. and {Walch}, S.},
	title = "{The hierarchical fragmentation of filaments and the role of sub-filaments}",
	journal = {\mnras},
	year = 2020,
	month = oct,
	volume = {497},
	number = {4},
	pages = {4390-4406},
	doi = {10.1093/mnras/staa2298},
	archivePrefix = {arXiv},
	eprint = {2007.15358},
	primaryClass = {astro-ph.GA},
	adsurl = {https://ui.adsabs.harvard.edu/abs/2020MNRAS.497.4390C}
}

@ARTICLE{2019ApJ...878...10T,
	author = {{Tang}, Ya-Wen and {Koch}, Patrick M. and {Peretto}, Nicolas and {Novak}, Giles and {Duarte-Cabral}, Ana and {Chapman}, Nicholas L. and {Hsieh}, Pei-Ying and {Yen}, Hsi-Wei},
	title = "{Gravity, Magnetic Field, and Turbulence: Relative Importance and Impact on Fragmentation in the Infrared Dark Cloud G34.43+00.24}",
	journal = {\apj},
	year = 2019,
	month = jun,
	volume = {878},
	number = {1},
	eid = {10},
	pages = {10},
	doi = {10.3847/1538-4357/ab1484},
	archivePrefix = {arXiv},
	eprint = {1903.12397},
	primaryClass = {astro-ph.GA},
	adsurl = {https://ui.adsabs.harvard.edu/abs/2019ApJ...878...10T}
}

@ARTICLE{2022AJ....164..175C,
	author = {{Chung}, Eun Jung and {Lee}, Chang Won and {Kwon}, Woojin and {Yoo}, Hyunju and {Soam}, Archana and {Cho}, Jungyeon},
	title = "{Evolution of the Hub-filament Structures in IC 5146 in the Context of the Energy Balance of Gravity, Turbulence, and Magnetic Field}",
	journal = {\aj},
	year = 2022,
	month = nov,
	volume = {164},
	number = {5},
	eid = {175},
	pages = {175},
	doi = {10.3847/1538-3881/ac8a43},
	archivePrefix = {arXiv},
	eprint = {2208.07891},
	primaryClass = {astro-ph.GA},
	adsurl = {https://ui.adsabs.harvard.edu/abs/2022AJ....164..175C}
}

@ARTICLE{2023ApJ...951...68C,
	author = {{Chung}, Eun Jung and {Lee}, Chang Won and {Kwon}, Woojin and {Tafalla}, Mario and {Kim}, Shinyoung},
	title = "{Magnetic Fields and Fragmentation of Filaments in the Hub of California-X}",
	journal = {\apj},
	year = 2023,
	month = jul,
	volume = {951},
	number = {1},
	eid = {68},
	pages = {68},
	doi = {10.3847/1538-4357/acd540},
	archivePrefix = {arXiv},
	eprint = {2305.09949},
	primaryClass = {astro-ph.GA},
	adsurl = {https://ui.adsabs.harvard.edu/abs/2023ApJ...951...68C}
}

@ARTICLE{Rosolowsky2008,
	author = {{Rosolowsky}, E.~W. and {Pineda}, J.~E. and {Kauffmann}, J. and {Goodman}, A.~A.},
	title = "{Structural Analysis of Molecular Clouds: Dendrograms}",
	journal = {\apj},
	year = 2008,
	month = jun,
	volume = {679},
	number = {2},
	pages = {1338-1351},
	doi = {10.1086/587685},
	archivePrefix = {arXiv},
	eprint = {0802.2944},
	primaryClass = {astro-ph},
	adsurl = {https://ui.adsabs.harvard.edu/abs/2008ApJ...679.1338R}
}

@ARTICLE{2013ApJ...774L..31I,
	author = {{Inoue}, Tsuyoshi and {Fukui}, Yasuo},
	title = "{Formation of Massive Molecular Cloud Cores by Cloud-Cloud Collision}",
	journal = {\apjl},
	year = 2013,
	month = sep,
	volume = {774},
	number = {2},
	eid = {L31},
	pages = {L31},
	doi = {10.1088/2041-8205/774/2/L31},
	archivePrefix = {arXiv},
	eprint = {1305.4655},
	primaryClass = {astro-ph.GA},
	adsurl = {https://ui.adsabs.harvard.edu/abs/2013ApJ...774L..31I}
}

@ARTICLE{2018PASJ...70S..53I,
	author = {{Inoue}, Tsuyoshi and {Hennebelle}, Patrick and {Fukui}, Yasuo and {Matsumoto}, Tomoaki and {Iwasaki}, Kazunari and {Inutsuka}, Shu-ichiro},
	title = "{The formation of massive molecular filaments and massive stars triggered by a magnetohydrodynamic shock wave}",
	journal = {\pasj},
	year = 2018,
	month = may,
	volume = {70},
	eid = {S53},
	pages = {S53},
	doi = {10.1093/pasj/psx089},
	archivePrefix = {arXiv},
	eprint = {1707.02035},
	primaryClass = {astro-ph.GA},
	adsurl = {https://ui.adsabs.harvard.edu/abs/2018PASJ...70S..53I}
}

@ARTICLE{2022ApJ...934..174I,
	author = {{Iwasaki}, Kazunari and {Tomida}, Kengo},
	title = "{Universal Properties of Dense Clumps in Magnetized Molecular Clouds Formed through Shock Compression of Two-phase Atomic Gases}",
	journal = {\apj},
	year = 2022,
	month = aug,
	volume = {934},
	number = {2},
	eid = {174},
	pages = {174},
	doi = {10.3847/1538-4357/ac75cc},
	archivePrefix = {arXiv},
	eprint = {2206.03627},
	primaryClass = {astro-ph.GA},
	adsurl = {https://ui.adsabs.harvard.edu/abs/2022ApJ...934..174I}
}

@ARTICLE{2021ApJ...916...83A,
	author = {{Abe}, Daisei and {Inoue}, Tsuyoshi and {Inutsuka}, Shu-ichiro and {Matsumoto}, Tomoaki},
	title = "{Classification of Filament Formation Mechanisms in Magnetized Molecular Clouds}",
	journal = {\apj},
	year = 2021,
	month = aug,
	volume = {916},
	number = {2},
	eid = {83},
	pages = {83},
	doi = {10.3847/1538-4357/ac07a1},
	archivePrefix = {arXiv},
	eprint = {2012.02205},
	primaryClass = {astro-ph.GA},
	adsurl = {https://ui.adsabs.harvard.edu/abs/2021ApJ...916...83A}
}

@article{KendallTest,
	author = {Kendall, M. G.},
	title = "{A NEW MEASURE OF RANK CORRELATION}",
	journal = {Biometrika},
	volume = {30},
	number = {1-2},
	pages = {81-93},
	year = {1938},
	month = {06},
	issn = {0006-3444},
	doi = {10.1093/biomet/30.1-2.81},
	url = {https://doi.org/10.1093/biomet/30.1-2.81},
	eprint = {https://academic.oup.com/biomet/article-pdf/30/1-2/81/423380/30-1-2-81.pdf},
}

@ARTICLE{2014ApJ...791..124G,
	author = {{G{\'o}mez}, Gilberto C. and {V{\'a}zquez-Semadeni}, Enrique},
	title = "{Filaments in Simulations of Molecular Cloud Formation}",
	journal = {\apj},
	year = 2014,
	month = aug,
	volume = {791},
	number = {2},
	eid = {124},
	pages = {124},
	doi = {10.1088/0004-637X/791/2/124},
	archivePrefix = {arXiv},
	eprint = {1308.6298},
	primaryClass = {astro-ph.GA},
	adsurl = {https://ui.adsabs.harvard.edu/abs/2014ApJ...791..124G}
}

@ARTICLE{2024MNRAS.528.1460R,
	author = {{Rawat}, Vineet and {Samal}, M.~R. and {Eswaraiah}, Chakali and {Wang}, Jia-Wei and {Elia}, Davide and {Panigrahy}, Sandhyarani and {Zavagno}, A. and {Yadav}, R.~K. and {Walker}, D.~L. and {Jose}, J. and {Ojha}, D.~K. and {Zhang}, C.~P. and {Dutta}, S.},
	title = "{Understanding the relative importance of magnetic field, gravity, and turbulence in star formation at the hub of the giant molecular cloud G148.24+00.41}",
	journal = {\mnras},
	year = 2024,
	month = feb,
	volume = {528},
	number = {2},
	pages = {1460-1475},
	doi = {10.1093/mnras/stae053},
	archivePrefix = {arXiv},
	eprint = {2401.05310},
	primaryClass = {astro-ph.GA},
	adsurl = {https://ui.adsabs.harvard.edu/abs/2024MNRAS.528.1460R}
}

@ARTICLE{2018A&A...613A..11W,
	author = {{Williams}, G.~M. and {Peretto}, N. and {Avison}, A. and {Duarte-Cabral}, A. and {Fuller}, G.~A.},
	title = "{Gravity drives the evolution of infrared dark hubs: JVLA observations of SDC13}",
	journal = {\aap},
	year = 2018,
	month = may,
	volume = {613},
	eid = {A11},
	pages = {A11},
	doi = {10.1051/0004-6361/201731587},
	archivePrefix = {arXiv},
	eprint = {1801.07253},
	primaryClass = {astro-ph.GA},
	adsurl = {https://ui.adsabs.harvard.edu/abs/2018A&A...613A..11W}
}

@ARTICLE{2020A&A...642A..87K,
	author = {{Kumar}, M.~S.~N. and {Palmeirim}, P. and {Arzoumanian}, D. and {Inutsuka}, S.~I.},
	title = "{Unifying low- and high-mass star formation through density-amplified hubs of filaments. The highest mass stars (>100 M$_{{\ensuremath{\odot}}}$) form only in hubs}",
	journal = {\aap},
	year = 2020,
	month = oct,
	volume = {642},
	eid = {A87},
	pages = {A87},
	doi = {10.1051/0004-6361/202038232},
	archivePrefix = {arXiv},
	eprint = {2008.00295},
	primaryClass = {astro-ph.GA},
	adsurl = {https://ui.adsabs.harvard.edu/abs/2020A&A...642A..87K}
}

@ARTICLE{2008ApJ...683..786H,
	author = {{Heitsch}, Fabian and {Hartmann}, Lee W. and {Burkert}, Andreas},
	title = "{Fragmentation of Shocked Flows: Gravity, Turbulence, and Cooling}",
	journal = {\apj},
	year = 2008,
	month = aug,
	volume = {683},
	number = {2},
	pages = {786-795},
	doi = {10.1086/589919},
	archivePrefix = {arXiv},
	eprint = {0805.0801},
	primaryClass = {astro-ph},
	adsurl = {https://ui.adsabs.harvard.edu/abs/2008ApJ...683..786H}
}

@ARTICLE{2017MNRAS.465.3483B,
	author = {{Balfour}, S.~K. and {Whitworth}, A.~P. and {Hubber}, D.~A.},
	title = "{Star formation triggered by non-head-on cloud-cloud collisions, and clouds with pre-collision sub-structure}",
	journal = {\mnras},
	year = 2017,
	month = mar,
	volume = {465},
	number = {3},
	pages = {3483-3494},
	doi = {10.1093/mnras/stw2956},
	adsurl = {https://ui.adsabs.harvard.edu/abs/2017MNRAS.465.3483B}
}

@ARTICLE{2015MNRAS.453.2471B,
	author = {{Balfour}, S.~K. and {Whitworth}, A.~P. and {Hubber}, D.~A. and {Jaffa}, S.~E.},
	title = "{Star formation triggered by cloud-cloud collisions}",
	journal = {\mnras},
	year = 2015,
	month = nov,
	volume = {453},
	number = {3},
	pages = {2471-2479},
	doi = {10.1093/mnras/stv1772},
	archivePrefix = {arXiv},
	eprint = {1509.05287},
	primaryClass = {astro-ph.SR},
	adsurl = {https://ui.adsabs.harvard.edu/abs/2015MNRAS.453.2471B}
}

@ARTICLE{1999ApJ...526..279P,
	author = {{Padoan}, Paolo and {Nordlund}, {\r{A}}ke},
	title = "{A Super-Alfv{\'e}nic Model of Dark Clouds}",
	journal = {\apj},
	year = 1999,
	month = nov,
	volume = {526},
	number = {1},
	pages = {279-294},
	doi = {10.1086/307956},
	archivePrefix = {arXiv},
	eprint = {astro-ph/9901288},
	primaryClass = {astro-ph},
	adsurl = {https://ui.adsabs.harvard.edu/abs/1999ApJ...526..279P}
}

@ARTICLE{2015ApJ...801...77M,
	author = {{Matsumoto}, Tomoaki and {Dobashi}, Kazuhito and {Shimoikura}, Tomomi},
	title = "{Star Formation in Turbulent Molecular Clouds with Colliding Flow}",
	journal = {\apj},
	year = 2015,
	month = mar,
	volume = {801},
	number = {2},
	eid = {77},
	pages = {77},
	doi = {10.1088/0004-637X/801/2/77},
	archivePrefix = {arXiv},
	eprint = {1412.5741},
	primaryClass = {astro-ph.SR},
	adsurl = {https://ui.adsabs.harvard.edu/abs/2015ApJ...801...77M}
}

@ARTICLE{2011ApJ...742L...9C,
	author = {{Commer{\c{c}}on}, Beno{\^\i}t and {Hennebelle}, Patrick and {Henning}, Thomas},
	title = "{Collapse of Massive Magnetized Dense Cores Using Radiation Magnetohydrodynamics: Early Fragmentation Inhibition}",
	journal = {\apjl},
	year = 2011,
	month = nov,
	volume = {742},
	number = {1},
	eid = {L9},
	pages = {L9},
	doi = {10.1088/2041-8205/742/1/L9},
	archivePrefix = {arXiv},
	eprint = {1110.2955},
	primaryClass = {astro-ph.SR},
	adsurl = {https://ui.adsabs.harvard.edu/abs/2011ApJ...742L...9C}
}

@ARTICLE{2024arXiv240808299W,
	author = {{Wells}, M.~R.~A. and {Beuther}, H. and {Molinari}, S. and {Schilke}, P. and {Battersby}, C. and {Ho}, P. and {S{\'a}nchez-Monge}, {\'A}. and {Jones}, B. and {Scheuck}, M.~B. and {Syed}, J. and {Gieser}, C. and {Kuiper}, R. and {Elia}, D. and {Coletta}, A. and {Traficante}, A. and {Wallace}, J. and {Rigby}, A.~J. and {Klessen}, R.~S. and {Zhang}, Q. and {Walch}, S. and {Beltr{\'a}n}, M.~T. and {Tang}, Y. and {Fuller}, G.~A. and {Lis}, D.~C. and {M{\"o}ller}, T. and {van der Tak}, F. and {Klaassen}, P.~D. and {Clarke}, S.~D. and {Moscadelli}, L. and {Mininni}, C. and {Zinnecker}, H. and {Maruccia}, Y. and {Pezzuto}, S. and {Benedettini}, M. and {Soler}, J.~D. and {Brogan}, C.~L. and {Avison}, A. and {Sanhueza}, P. and {Schisano}, E. and {Liu}, T. and {Fontani}, F. and {Rygl}, K.~L.~J. and {Wyrowski}, F. and {Bally}, J. and {Walker}, D.~L. and {Ahmadi}, A. and {Koch}, P. and {Merello}, M. and {Law}, C.~Y. and {Testi}, L.},
	title = "{Dynamical Accretion Flows -- ALMAGAL: Flows along filamentary structures in high-mass star-forming clusters}",
	journal = {arXiv e-prints},
	year = 2024,
	month = aug,
	eid = {arXiv:2408.08299},
	pages = {arXiv:2408.08299},
	doi = {10.48550/arXiv.2408.08299},
	archivePrefix = {arXiv},
	eprint = {2408.08299},
	primaryClass = {astro-ph.GA},
	adsurl = {https://ui.adsabs.harvard.edu/abs/2024arXiv240808299W}
}

@ARTICLE{2023MNRAS.526.2278A,
	author = {{Avison}, A. and {Fuller}, G.~A. and {Frimpong}, N. Asabre and {Etoka}, S. and {Hoare}, M. and {Jones}, B.~M. and {Peretto}, N. and {Traficante}, A. and {van der Tak}, F. and {Pineda}, J.~E. and {Beltr{\'a}n}, M. and {Wyrowski}, F. and {Thompson}, M. and {Lumsden}, S. and {Nagy}, Z. and {Hill}, T. and {Viti}, S. and {Fontani}, F. and {Schilke}, P.},
	title = "{Tracing Evolution in Massive Protostellar Objects - I. Fragmentation and emission properties of massive star-forming clumps in a luminosity-limited ALMA sample}",
	journal = {\mnras},
	year = 2023,
	month = dec,
	volume = {526},
	number = {2},
	pages = {2278-2300},
	doi = {10.1093/mnras/stad2824},
	archivePrefix = {arXiv},
	eprint = {2309.05772},
	primaryClass = {astro-ph.GA},
	adsurl = {https://ui.adsabs.harvard.edu/abs/2023MNRAS.526.2278A}
}

@ARTICLE{2023MNRAS.520.2306T,
	author = {{Traficante}, A. and {Jones}, B.~M. and {Avison}, A. and {Fuller}, G.~A. and {Benedettini}, M. and {Elia}, D. and {Molinari}, S. and {Peretto}, N. and {Pezzuto}, S. and {Pillai}, T. and {Rygl}, K.~L.~J. and {Schisano}, E. and {Smith}, R.~J.},
	title = "{The SQUALO project (Star formation in QUiescent And Luminous Objects) I: clump-fed accretion mechanism in high-mass star-forming objects}",
	journal = {\mnras},
	year = 2023,
	month = apr,
	volume = {520},
	number = {2},
	pages = {2306-2327},
	doi = {10.1093/mnras/stad272},
	archivePrefix = {arXiv},
	eprint = {2301.09917},
	primaryClass = {astro-ph.GA},
	adsurl = {https://ui.adsabs.harvard.edu/abs/2023MNRAS.520.2306T}
}

@ARTICLE{2018A&A...617A.100B,
	author = {{Beuther}, H. and {Mottram}, J.~C. and {Ahmadi}, A. and {Bosco}, F. and {Linz}, H. and {Henning}, Th. and {Klaassen}, P. and {Winters}, J.~M. and {Maud}, L.~T. and {Kuiper}, R. and {Semenov}, D. and {Gieser}, C. and {Peters}, T. and {Urquhart}, J.~S. and {Pudritz}, R. and {Ragan}, S.~E. and {Feng}, S. and {Keto}, E. and {Leurini}, S. and {Cesaroni}, R. and {Beltran}, M. and {Palau}, A. and {S{\'a}nchez-Monge}, {\'A}. and {Galvan-Madrid}, R. and {Zhang}, Q. and {Schilke}, P. and {Wyrowski}, F. and {Johnston}, K.~G. and {Longmore}, S.~N. and {Lumsden}, S. and {Hoare}, M. and {Menten}, K.~M. and {Csengeri}, T.},
	title = "{Fragmentation and disk formation during high-mass star formation. IRAM NOEMA (Northern Extended Millimeter Array) large program CORE}",
	journal = {\aap},
	year = 2018,
	month = sep,
	volume = {617},
	eid = {A100},
	pages = {A100},
	doi = {10.1051/0004-6361/201833021},
	archivePrefix = {arXiv},
	eprint = {1805.01191},
	primaryClass = {astro-ph.GA},
	adsurl = {https://ui.adsabs.harvard.edu/abs/2018A&A...617A.100B}
}

@ARTICLE{2021A&A...649A.113B,
	author = {{Beuther}, H. and {Gieser}, C. and {Suri}, S. and {Linz}, H. and {Klaassen}, P. and {Semenov}, D. and {Winters}, J.~M. and {Henning}, Th. and {Soler}, J.~D. and {Urquhart}, J.~S. and {Syed}, J. and {Feng}, S. and {M{\"o}ller}, T. and {Beltr{\'a}n}, M.~T. and {S{\'a}nchez-Monge}, {\'A}. and {Longmore}, S.~N. and {Peters}, T. and {Ballesteros-Paredes}, J. and {Schilke}, P. and {Moscadelli}, L. and {Palau}, A. and {Cesaroni}, R. and {Lumsden}, S. and {Pudritz}, R. and {Wyrowski}, F. and {Kuiper}, R. and {Ahmadi}, A.},
	title = "{Fragmentation and kinematics in high-mass star formation. CORE-extension targeting two very young high-mass star-forming regions}",
	journal = {\aap},
	year = 2021,
	month = may,
	volume = {649},
	eid = {A113},
	pages = {A113},
	doi = {10.1051/0004-6361/202040106},
	archivePrefix = {arXiv},
	eprint = {2104.02420},
	primaryClass = {astro-ph.GA},
	adsurl = {https://ui.adsabs.harvard.edu/abs/2021A&A...649A.113B}
}

@ARTICLE{2024ApJS..270....9X,
	author = {{Xu}, Fengwei and {Wang}, Ke and {Liu}, Tie and {Tang}, Mengyao and {Evans}, Neal J., II and {Palau}, Aina and {Morii}, Kaho and {He}, Jinhua and {Sanhueza}, Patricio and {Liu}, Hong-Li and {Stutz}, Amelia and {Zhang}, Qizhou and {Chen}, Xi and {Li}, Pak Shing and {G{\'o}mez}, Gilberto C. and {V{\'a}zquez-Semadeni}, Enrique and {Li}, Shanghuo and {Mai}, Xiaofeng and {Lu}, Xing and {Liu}, Meizhu and {Chen}, Li and {Li}, Chuanshou and {Shi}, Hongqiong and {Ren}, Zhiyuan and {Li}, Di and {Garay}, Guido and {Bronfman}, Leonardo and {Dewangan}, Lokesh and {Juvela}, Mika and {Lee}, Chang Won and {Zhang}, S. and {Yue}, Nannan and {Wang}, Chao and {Ge}, Yifei and {Jiao}, Wenyu and {Luo}, Qiuyi and {Zhou}, J. -W. and {Tatematsu}, Ken'ichi and {Chibueze}, James O. and {Su}, Keyun and {Sun}, Shenglan and {Ristorcelli}, I. and {Toth}, L. Viktor},
	title = "{The ALMA Survey of Star Formation and Evolution in Massive Protoclusters with Blue Profiles (ASSEMBLE): Core Growth, Cluster Contraction, and Primordial Mass Segregation}",
	journal = {\apjs},
	year = 2024,
	month = jan,
	volume = {270},
	number = {1},
	eid = {9},
	pages = {9},
	doi = {10.3847/1538-4365/acfee5},
	archivePrefix = {arXiv},
	eprint = {2309.14684},
	primaryClass = {astro-ph.GA},
	adsurl = {https://ui.adsabs.harvard.edu/abs/2024ApJS..270....9X}
}

@ARTICLE{2015A&A...581A.119B,
	author = {{Beuther}, H. and {Henning}, Th. and {Linz}, H. and {Feng}, S. and {Ragan}, S.~E. and {Smith}, R.~J. and {Bihr}, S. and {Sakai}, T. and {Kuiper}, R.},
	title = "{Hierarchical fragmentation and collapse signatures in a high-mass starless region}",
	journal = {\aap},
	year = 2015,
	month = sep,
	volume = {581},
	eid = {A119},
	pages = {A119},
	doi = {10.1051/0004-6361/201526759},
	archivePrefix = {arXiv},
	eprint = {1508.01540},
	primaryClass = {astro-ph.GA},
	adsurl = {https://ui.adsabs.harvard.edu/abs/2015A&A...581A.119B}
}

@ARTICLE{2018ApJ...855...24P,
	author = {{Palau}, Aina and {Zapata}, Luis A. and {Rom{\'a}n-Z{\'u}{\~n}iga}, Carlos G. and {S{\'a}nchez-Monge}, {\'A}lvaro and {Estalella}, Robert and {Busquet}, Gemma and {Girart}, Josep M. and {Fuente}, Asunci{\'o}n and {Commer{\c{c}}on}, Benoit},
	title = "{Thermal Jeans Fragmentation within {\ensuremath{\sim}}1000 au in OMC-1S}",
	journal = {\apj},
	year = 2018,
	month = mar,
	volume = {855},
	number = {1},
	eid = {24},
	pages = {24},
	doi = {10.3847/1538-4357/aaad03},
	archivePrefix = {arXiv},
	eprint = {1706.04623},
	primaryClass = {astro-ph.GA},
	adsurl = {https://ui.adsabs.harvard.edu/abs/2018ApJ...855...24P}
}

@ARTICLE{2019ApJ...871..185L,
	author = {{Liu}, Hauyu Baobab and {Chen}, Huei-Ru Vivien and {Rom{\'a}n-Z{\'u}{\~n}iga}, Carlos G. and {Galv{\'a}n-Madrid}, Roberto and {Ginsburg}, Adam and {Ho}, Paul T.~P. and {Minh}, Young Chol and {Jim{\'e}nez-Serra}, Izaskun and {Testi}, Leonardo and {Zhang}, Qizhou},
	title = "{Investigating Fragmentation of Gas Structures in OB Cluster-forming Molecular Clump G33.92+0.11 with 1000 au Resolution Observations of ALMA}",
	journal = {\apj},
	year = 2019,
	month = feb,
	volume = {871},
	number = {2},
	eid = {185},
	pages = {185},
	doi = {10.3847/1538-4357/aaf6b4},
	archivePrefix = {arXiv},
	eprint = {1808.07702},
	primaryClass = {astro-ph.GA},
	adsurl = {https://ui.adsabs.harvard.edu/abs/2019ApJ...871..185L}
}

@ARTICLE{2020ApJ...894L..14L,
	author = {{Lu}, Xing and {Cheng}, Yu and {Ginsburg}, Adam and {Longmore}, Steven N. and {Kruijssen}, J.~M. Diederik and {Battersby}, Cara and {Zhang}, Qizhou and {Walker}, Daniel L.},
	title = "{ALMA Observations of Massive Clouds in the Central Molecular Zone: Jeans Fragmentation and Cluster Formation}",
	journal = {\apjl},
	year = 2020,
	month = may,
	volume = {894},
	number = {2},
	eid = {L14},
	pages = {L14},
	doi = {10.3847/2041-8213/ab8b65},
	archivePrefix = {arXiv},
	eprint = {2004.09532},
	primaryClass = {astro-ph.GA},
	adsurl = {https://ui.adsabs.harvard.edu/abs/2020ApJ...894L..14L}
}

@ARTICLE{2024ApJ...966..171M,
	author = {{Morii}, Kaho and {Sanhueza}, Patricio and {Zhang}, Qizhou and {Nakamura}, Fumitaka and {Li}, Shanghuo and {Sabatini}, Giovanni and {Olguin}, Fernando A. and {Beuther}, Henrik and {Tafoya}, Daniel and {Izumi}, Natsuko and {Tatematsu}, Ken'ichi and {Sakai}, Takeshi},
	title = "{The ALMA Survey of 70 {\ensuremath{\mu}}m Dark High-mass Clumps in Early Stages (ASHES). XI. Statistical Study of Early Fragmentation}",
	journal = {\apj},
	year = 2024,
	month = may,
	volume = {966},
	number = {2},
	eid = {171},
	pages = {171},
	doi = {10.3847/1538-4357/ad32d0},
	archivePrefix = {arXiv},
	eprint = {2403.07058},
	primaryClass = {astro-ph.GA},
	adsurl = {https://ui.adsabs.harvard.edu/abs/2024ApJ...966..171M}
}

@ARTICLE{2024arXiv240706845I,
       author = {{Ishihara}, Kousuke and {Sanhueza}, Patricio and {Nakamura}, Fumitaka and {Saito}, Masao and {Chen}, Huei-Ru Vivien and {Li}, Shanghuo and {Olguin}, Fernando and {Taniguchi}, Kotomi and {Morii}, Kaho and {Lu}, Xing and {Luo}, Qiu-yi and {Sakai}, Takeshi and {Zhang}, Qizhou},
		title = "{Digging into the Interior of Hot Cores with ALMA (DIHCA). IV. Fragmentation in High-mass Star-forming Clumps}",
		journal = {\apj},
		year = 2024,
		month = oct,
		volume = {974},
		number = {1},
		eid = {95},
		pages = {95},
		doi = {10.3847/1538-4357/ad630f},
		archivePrefix = {arXiv},
		eprint = {2407.06845},
		primaryClass = {astro-ph.GA},
		adsurl = {https://ui.adsabs.harvard.edu/abs/2024ApJ...974...95I}
}

@ARTICLE{2016MNRAS.458..319C,
	author = {{Clarke}, S.~D. and {Whitworth}, A.~P. and {Hubber}, D.~A.},
	title = "{Perturbation growth in accreting filaments}",
	journal = {\mnras},
	year = 2016,
	month = may,
	volume = {458},
	number = {1},
	pages = {319-324},
	doi = {10.1093/mnras/stw407},
	archivePrefix = {arXiv},
	eprint = {1602.07651},
	primaryClass = {astro-ph.GA},
	adsurl = {https://ui.adsabs.harvard.edu/abs/2016MNRAS.458..319C}
}

@ARTICLE{2016MNRAS.463.4301H,
	author = {{Heigl}, S. and {Burkert}, A. and {Hacar}, A.},
	title = "{Non-linear dense core formation in the dark cloud L1517}",
	journal = {\mnras},
	year = 2016,
	month = dec,
	volume = {463},
	number = {4},
	pages = {4301-4310},
	doi = {10.1093/mnras/stw2271},
	archivePrefix = {arXiv},
	eprint = {1601.02018},
	primaryClass = {astro-ph.SR},
	adsurl = {https://ui.adsabs.harvard.edu/abs/2016MNRAS.463.4301H}
}

@ARTICLE{2018MNRAS.481L...1H,
	author = {{Heigl}, S. and {Gritschneder}, M. and {Burkert}, A.},
	title = "{Morphology of prestellar cores in pressure-confined filaments}",
	journal = {\mnras},
	year = 2018,
	month = nov,
	volume = {481},
	number = {1},
	pages = {L1-L5},
	doi = {10.1093/mnrasl/sly146},
	archivePrefix = {arXiv},
	eprint = {1808.01284},
	primaryClass = {astro-ph.GA},
	adsurl = {https://ui.adsabs.harvard.edu/abs/2018MNRAS.481L...1H}
}

@ARTICLE{2015MNRAS.452.2410S,
	author = {{Seifried}, D. and {Walch}, S.},
	title = "{The impact of turbulence and magnetic field orientation on star-forming filaments}",
	journal = {\mnras},
	year = 2015,
	month = sep,
	volume = {452},
	number = {3},
	pages = {2410-2422},
	doi = {10.1093/mnras/stv1458},
	archivePrefix = {arXiv},
	eprint = {1503.01659},
	primaryClass = {astro-ph.GA},
	adsurl = {https://ui.adsabs.harvard.edu/abs/2015MNRAS.452.2410S}
}

@ARTICLE{2017ApJ...848....2H,
	author = {{Hanawa}, Tomoyuki and {Kudoh}, Takahiro and {Tomisaka}, Kohji},
	title = "{Fragmentation of a Filamentary Cloud Permeated by a Perpendicular Magnetic Field}",
	journal = {\apj},
	year = 2017,
	month = oct,
	volume = {848},
	number = {1},
	eid = {2},
	pages = {2},
	doi = {10.3847/1538-4357/aa8b6d},
	archivePrefix = {arXiv},
	eprint = {1709.05149},
	primaryClass = {astro-ph.GA},
	adsurl = {https://ui.adsabs.harvard.edu/abs/2017ApJ...848....2H}
}

@ARTICLE{2019ApJ...881...97H,
	author = {{Hanawa}, Tomoyuki and {Kudoh}, Takahiro and {Tomisaka}, Kohji},
	title = "{Fragmentation of a Filamentary Cloud Permeated by a Perpendicular Magnetic Field. II. Dependence on the Initial Density Profile}",
	journal = {\apj},
	year = 2019,
	month = aug,
	volume = {881},
	number = {2},
	eid = {97},
	pages = {97},
	doi = {10.3847/1538-4357/ab2d26},
	archivePrefix = {arXiv},
	eprint = {1907.03384},
	primaryClass = {astro-ph.GA},
	adsurl = {https://ui.adsabs.harvard.edu/abs/2019ApJ...881...97H}
}

@ARTICLE{2009ApJ...700.1609M,
	author = {{Myers}, Philip C.},
	title = "{Filamentary Structure of Star-forming Complexes}",
	journal = {\apj},
	year = 2009,
	month = aug,
	volume = {700},
	number = {2},
	pages = {1609-1625},
	doi = {10.1088/0004-637X/700/2/1609},
	archivePrefix = {arXiv},
	eprint = {0906.2005},
	primaryClass = {astro-ph.GA},
	adsurl = {https://ui.adsabs.harvard.edu/abs/2009ApJ...700.1609M}
}

@ARTICLE{2024MNRAS.527.4244S,
	author = {{Seshadri}, Arun and {Vig}, S. and {Ghosh}, S.~K. and {Ojha}, D.~K.},
	title = "{Massive star formation in the hub-filament system of RCW 117}",
	journal = {\mnras},
	year = 2024,
	month = jan,
	volume = {527},
	number = {2},
	pages = {4244-4259},
	doi = {10.1093/mnras/stad3385},
	archivePrefix = {arXiv},
	eprint = {2311.00477},
	primaryClass = {astro-ph.GA},
	adsurl = {https://ui.adsabs.harvard.edu/abs/2024MNRAS.527.4244S}
}

@ARTICLE{1966MNRAS.133..265M,
	author = {{Mestel}, L.},
	title = "{The magnetic field of a contracting gas cloud. I,Strict flux-freezing}",
	journal = {\mnras},
	year = 1966,
	month = jan,
	volume = {133},
	pages = {265},
	doi = {10.1093/mnras/133.2.265},
	adsurl = {https://ui.adsabs.harvard.edu/abs/1966MNRAS.133..265M}
}

@ARTICLE{2010ApJ...725..466C,
	author = {{Crutcher}, Richard M. and {Wandelt}, Benjamin and {Heiles}, Carl and {Falgarone}, Edith and {Troland}, Thomas H.},
	title = "{Magnetic Fields in Interstellar Clouds from Zeeman Observations: Inference of Total Field Strengths by Bayesian Analysis}",
	journal = {\apj},
	year = 2010,
	month = dec,
	volume = {725},
	number = {1},
	pages = {466-479},
	doi = {10.1088/0004-637X/725/1/466},
	adsurl = {https://ui.adsabs.harvard.edu/abs/2010ApJ...725..466C}
}

@ARTICLE{1958ArM.....3..469H,
	author = {{Hodges}, J.~L.},
	title = "{The significance probability of the smirnov two-sample test}",
	journal = {Arkiv for Matematik},
	year = 1958,
	month = jan,
	volume = {3},
	number = {5},
	pages = {469-486},
	doi = {10.1007/BF02589501},
	adsurl = {https://ui.adsabs.harvard.edu/abs/1958ArM.....3..469H}
}

@ARTICLE{2013ApJ...779...96T,
	author = {{Tan}, Jonathan C. and {Kong}, Shuo and {Butler}, Michael J. and {Caselli}, Paola and {Fontani}, Francesco},
	title = "{The Dynamics of Massive Starless Cores with ALMA}",
	journal = {\apj},
	year = 2013,
	month = dec,
	volume = {779},
	number = {2},
	eid = {96},
	pages = {96},
	doi = {10.1088/0004-637X/779/2/96},
	archivePrefix = {arXiv},
	eprint = {1303.4343},
	primaryClass = {astro-ph.GA},
	adsurl = {https://ui.adsabs.harvard.edu/abs/2013ApJ...779...96T}
}

@ARTICLE{2018A&A...614A..64B,
	author = {{Beuther}, H. and {Soler}, J.~D. and {Vlemmings}, W. and {Linz}, H. and {Henning}, Th. and {Kuiper}, R. and {Rao}, R. and {Smith}, R. and {Sakai}, T. and {Johnston}, K. and {Walsh}, A. and {Feng}, S.},
	title = "{Magnetic fields at the onset of high-mass star formation}",
	journal = {\aap},
	year = 2018,
	month = jun,
	volume = {614},
	eid = {A64},
	pages = {A64},
	doi = {10.1051/0004-6361/201732378},
	archivePrefix = {arXiv},
	eprint = {1802.00005},
	primaryClass = {astro-ph.GA},
	adsurl = {https://ui.adsabs.harvard.edu/abs/2018A&A...614A..64B}
}

@ARTICLE{2022MNRAS.511.2702J,
	author = {{Jaffa}, S.~E. and {Dale}, J. and {Krause}, M. and {Clarke}, S.~D.},
	title = "{Chaotic star formation: error bars for the star formation efficiency and column density PDF}",
	journal = {\mnras},
	year = 2022,
	month = apr,
	volume = {511},
	number = {2},
	pages = {2702-2707},
	doi = {10.1093/mnras/stac131},
	adsurl = {https://ui.adsabs.harvard.edu/abs/2022MNRAS.511.2702J}
}

@ARTICLE{2007ApJ...656..959K,
	author = {{Krumholz}, Mark R. and {Klein}, Richard I. and {McKee}, Christopher F.},
	title = "{Radiation-Hydrodynamic Simulations of Collapse and Fragmentation in Massive Protostellar Cores}",
	journal = {\apj},
	year = 2007,
	month = feb,
	volume = {656},
	number = {2},
	pages = {959-979},
	doi = {10.1086/510664},
	archivePrefix = {arXiv},
	eprint = {astro-ph/0609798},
	primaryClass = {astro-ph},
	adsurl = {https://ui.adsabs.harvard.edu/abs/2007ApJ...656..959K}
}

@INPROCEEDINGS{2014prpl.conf...27A,
	author = {{Andr{\'e}}, P. and {Di Francesco}, J. and {Ward-Thompson}, D. and {Inutsuka}, S. -I. and {Pudritz}, R.~E. and {Pineda}, J.~E.},
	title = "{From Filamentary Networks to Dense Cores in Molecular Clouds: Toward a New Paradigm for Star Formation}",
	booktitle = {Protostars and Planets VI},
	year = 2014,
	editor = {{Beuther}, Henrik and {Klessen}, Ralf S. and {Dullemond}, Cornelis P. and {Henning}, Thomas},
	month = jan,
	pages = {27-51},
	doi = {10.2458/azu_uapress_9780816531240-ch002},
	archivePrefix = {arXiv},
	eprint = {1312.6232},
	primaryClass = {astro-ph.GA},
	adsurl = {https://ui.adsabs.harvard.edu/abs/2014prpl.conf...27A}
}

@misc{ALPara,
	author       = {{Chen}, Wei-An},
	title        = {Alignment Parameters},
	month        = oct,
	year         = 2024,
	publisher    = {Zenodo},
	version      = {v1.1},
	doi          = {10.5281/zenodo.14013689},
	url          = {https://doi.org/10.5281/zenodo.14013689}
}

@ARTICLE{2019MNRAS.490.3061V,
	author = {{V{\'a}zquez-Semadeni}, Enrique and {Palau}, Aina and {Ballesteros-Paredes}, Javier and {G{\'o}mez}, Gilberto C. and {Zamora-Avil{\'e}s}, Manuel},
	title = "{Global hierarchical collapse in molecular clouds. Towards a comprehensive scenario}",
	journal = {\mnras},
	year = 2019,
	month = dec,
	volume = {490},
	number = {3},
	pages = {3061-3097},
	doi = {10.1093/mnras/stz2736},
	archivePrefix = {arXiv},
	eprint = {1903.11247},
	primaryClass = {astro-ph.GA},
	adsurl = {https://ui.adsabs.harvard.edu/abs/2019MNRAS.490.3061V}
}

@ARTICLE{2021ApJ...912..159P,
	author = {{Palau}, Aina and {Zhang}, Qizhou and {Girart}, Josep M. and {Liu}, Junhao and {Rao}, Ramprasad and {Koch}, Patrick M. and {Estalella}, Robert and {Chen}, Huei-Ru Vivien and {Liu}, Hauyu Baobab and {Qiu}, Keping and {Li}, Zhi-Yun and {Zapata}, Luis A. and {Bontemps}, Sylvain and {Ho}, Paul T.~P. and {Beuther}, Henrik and {Ching}, Tao-Chung and {Shinnaga}, Hiroko and {Ahmadi}, Aida},
	title = "{Does the Magnetic Field Suppress Fragmentation in Massive Dense Cores?}",
	journal = {\apj},
	year = 2021,
	month = may,
	volume = {912},
	number = {2},
	eid = {159},
	pages = {159},
	doi = {10.3847/1538-4357/abee1e},
	archivePrefix = {arXiv},
	eprint = {2010.12099},
	primaryClass = {astro-ph.GA},
	adsurl = {https://ui.adsabs.harvard.edu/abs/2021ApJ...912..159P}
}

@ARTICLE{2024A&A...682A..81B,
	author = {{Beuther}, H. and {Gieser}, C. and {Soler}, J.~D. and {Zhang}, Q. and {Rao}, R. and {Semenov}, D. and {Henning}, Th. and {Pudritz}, R. and {Peters}, T. and {Klaassen}, P. and {Beltr{\'a}n}, M.~T. and {Palau}, A. and {M{\"o}ller}, T. and {Johnston}, K.~G. and {Zinnecker}, H. and {Urquhart}, J. and {Kuiper}, R. and {Ahmadi}, A. and {S{\'a}nchez-Monge}, {\'A}. and {Feng}, S. and {Leurini}, S. and {Ragan}, S.~E.},
	title = "{Density distributions, magnetic field structures, and fragmentation in high-mass star formation}",
	journal = {\aap},
	year = 2024,
	month = feb,
	volume = {682},
	eid = {A81},
	pages = {A81},
	doi = {10.1051/0004-6361/202348117},
	archivePrefix = {arXiv},
	eprint = {2311.11874},
	primaryClass = {astro-ph.GA},
	adsurl = {https://ui.adsabs.harvard.edu/abs/2024A&A...682A..81B}
}

@ARTICLE{2019FrASS...6....5H,
	author = {{Hennebelle}, Patrick and {Inutsuka}, Shu-ichiro},
	title = "{The role of magnetic field in molecular cloud formation and evolution}",
	journal = {Frontiers in Astronomy and Space Sciences},
	year = 2019,
	month = mar,
	volume = {6},
	eid = {5},
	pages = {5},
	doi = {10.3389/fspas.2019.00005},
	archivePrefix = {arXiv},
	eprint = {1902.00798},
	primaryClass = {astro-ph.GA},
	adsurl = {https://ui.adsabs.harvard.edu/abs/2019FrASS...6....5H}
}

@ARTICLE{2008MNRAS.385.1820P,
	author = {{Price}, Daniel J. and {Bate}, Matthew R.},
	title = "{The effect of magnetic fields on star cluster formation}",
	journal = {\mnras},
	year = 2008,
	month = apr,
	volume = {385},
	number = {4},
	pages = {1820-1834},
	doi = {10.1111/j.1365-2966.2008.12976.x},
	archivePrefix = {arXiv},
	eprint = {0801.3293},
	primaryClass = {astro-ph},
	adsurl = {https://ui.adsabs.harvard.edu/abs/2008MNRAS.385.1820P}
}

@ARTICLE{1976ApJ...210..326M,
	author = {{Mouschovias}, T. Ch. and {Spitzer}, L., Jr.},
	title = "{Note on the collapse of magnetic interstellar clouds.}",
	journal = {\apj},
	year = 1976,
	month = dec,
	volume = {210},
	pages = {326},
	doi = {10.1086/154835},
	adsurl = {https://ui.adsabs.harvard.edu/abs/1976ApJ...210..326M}
}

@ARTICLE{2025ApJ...979...67C,
	author = {{Chen}, Wei-An and {Tang}, Ya-Wen and {Clarke}, S.~D. and {Sanhueza}, Patricio},
	title = "{Alignment Parameters: Quantifying Dense Core Alignment in Star-forming Regions}",
	journal = {\apj},
	year = 2025,
	month = jan,
	volume = {979},
	number = {1},
	eid = {67},
	pages = {67},
	doi = {10.3847/1538-4357/ad9a5b},
	archivePrefix = {arXiv},
	eprint = {2412.02243},
	primaryClass = {astro-ph.GA},
	adsurl = {https://ui.adsabs.harvard.edu/abs/2025ApJ...979...67C}
}

@ARTICLE{2019ApJ...873....6I,
	author = {{Iwasaki}, Kazunari and {Tomida}, Kengo and {Inoue}, Tsuyoshi and {Inutsuka}, Shu-ichiro},
	title = "{The Early Stage of Molecular Cloud Formation by Compression of Two-phase Atomic Gases}",
	journal = {\apj},
	year = 2019,
	month = mar,
	volume = {873},
	number = {1},
	eid = {6},
	pages = {6},
	doi = {10.3847/1538-4357/ab02ff},
	archivePrefix = {arXiv},
	eprint = {1806.03824},
	primaryClass = {astro-ph.GA},
	adsurl = {https://ui.adsabs.harvard.edu/abs/2019ApJ...873....6I}
}

@inproceedings{DBSCAN1996,
	author = {Ester, Martin and Kriegel, Hans-Peter and Sander, J\"{o}rg and Xu, Xiaowei},
	title = {A density-based algorithm for discovering clusters in large spatial databases with noise},
	year = {1996},
	publisher = {AAAI Press},
	booktitle = {Proceedings of the Second International Conference on Knowledge Discovery and Data Mining},
	pages = {226–231},
	numpages = {6},
	location = {Portland, Oregon},
	series = {KDD'96}
}

@article{DBSCAN2017,
	author = {Schubert, Erich and Sander, J\"{o}rg and Ester, Martin and Kriegel, Hans Peter and Xu, Xiaowei},
	title = {DBSCAN Revisited, Revisited: Why and How You Should (Still) Use DBSCAN},
	year = {2017},
	issue_date = {September 2017},
	publisher = {Association for Computing Machinery},
	address = {New York, NY, USA},
	volume = {42},
	number = {3},
	issn = {0362-5915},
	url = {https://doi.org/10.1145/3068335},
	doi = {10.1145/3068335},
	journal = {ACM Trans. Database Syst.},
	month = jul,
	articleno = {19},
	numpages = {21}
}

@Article{         harris2020array,
	title         = {Array programming with {NumPy}},
	author        = {Charles R. Harris and K. Jarrod Millman and St{\'{e}}fan J.
	van der Walt and Ralf Gommers and Pauli Virtanen and David
	Cournapeau and Eric Wieser and Julian Taylor and Sebastian
	Berg and Nathaniel J. Smith and Robert Kern and Matti Picus
	and Stephan Hoyer and Marten H. van Kerkwijk and Matthew
	Brett and Allan Haldane and Jaime Fern{\'{a}}ndez del
	R{\'{i}}o and Mark Wiebe and Pearu Peterson and Pierre
	G{\'{e}}rard-Marchant and Kevin Sheppard and Tyler Reddy and
	Warren Weckesser and Hameer Abbasi and Christoph Gohlke and
	Travis E. Oliphant},
	year          = {2020},
	month         = sep,
	journal       = {Nature},
	volume        = {585},
	number        = {7825},
	pages         = {357--362},
	doi           = {10.1038/s41586-020-2649-2},
	publisher     = {Springer Science and Business Media {LLC}},
	url           = {https://doi.org/10.1038/s41586-020-2649-2}
}

@ARTICLE{2020SciPy-NMeth,
	author  = {Virtanen, Pauli and Gommers, Ralf and Oliphant, Travis E. and
	Haberland, Matt and Reddy, Tyler and Cournapeau, David and
	Burovski, Evgeni and Peterson, Pearu and Weckesser, Warren and
	Bright, Jonathan and {van der Walt}, St{\'e}fan J. and
	Brett, Matthew and Wilson, Joshua and Millman, K. Jarrod and
	Mayorov, Nikolay and Nelson, Andrew R. J. and Jones, Eric and
	Kern, Robert and Larson, Eric and Carey, C J and
	Polat, {\.I}lhan and Feng, Yu and Moore, Eric W. and
	{VanderPlas}, Jake and Laxalde, Denis and Perktold, Josef and
	Cimrman, Robert and Henriksen, Ian and Quintero, E. A. and
	Harris, Charles R. and Archibald, Anne M. and
	Ribeiro, Ant{\^o}nio H. and Pedregosa, Fabian and
	{van Mulbregt}, Paul and {SciPy 1.0 Contributors}},
	title   = {{{SciPy} 1.0: Fundamental Algorithms for Scientific
	Computing in Python}},
	journal = {Nature Methods},
	year    = {2020},
	volume  = {17},
	pages   = {261--272},
	adsurl  = {https://rdcu.be/b08Wh},
	doi     = {10.1038/s41592-019-0686-2},
}

@Article{Hunter:2007,
	Author    = {Hunter, J. D.},
	Title     = {Matplotlib: A 2D graphics environment},
	Journal   = {Computing in Science \& Engineering},
	Volume    = {9},
	Number    = {3},
	Pages     = {90--95},
	publisher = {IEEE COMPUTER SOC},
	doi       = {10.1109/MCSE.2007.55},
	year      = 2007
}

@article{astropy:2013,
	Adsurl = {http://adsabs.harvard.edu/abs/2013A%26A...558A..33A},
	Archiveprefix = {arXiv},
	Author = {{Astropy Collaboration} and {Robitaille}, T.~P. and {Tollerud}, E.~J. and {Greenfield}, P. and {Droettboom}, M. and {Bray}, E. and {Aldcroft}, T. and {Davis}, M. and {Ginsburg}, A. and {Price-Whelan}, A.~M. and {Kerzendorf}, W.~E. and {Conley}, A. and {Crighton}, N. and {Barbary}, K. and {Muna}, D. and {Ferguson}, H. and {Grollier}, F. and {Parikh}, M.~M. and {Nair}, P.~H. and {Unther}, H.~M. and {Deil}, C. and {Woillez}, J. and {Conseil}, S. and {Kramer}, R. and {Turner}, J.~E.~H. and {Singer}, L. and {Fox}, R. and {Weaver}, B.~A. and {Zabalza}, V. and {Edwards}, Z.~I. and {Azalee Bostroem}, K. and {Burke}, D.~J. and {Casey}, A.~R. and {Crawford}, S.~M. and {Dencheva}, N. and {Ely}, J. and {Jenness}, T. and {Labrie}, K. and {Lim}, P.~L. and {Pierfederici}, F. and {Pontzen}, A. and {Ptak}, A. and {Refsdal}, B. and {Servillat}, M. and {Streicher}, O.},
	Doi = {10.1051/0004-6361/201322068},
	Eid = {A33},
	Eprint = {1307.6212},
	Journal = {\aap},
	Month = oct,
	Pages = {A33},
	Primaryclass = {astro-ph.IM},
	Title = {{Astropy: A community Python package for astronomy}},
	Volume = 558,
	Year = 2013}

@ARTICLE{astropy:2018,
	author = {{Astropy Collaboration} and {Price-Whelan}, A.~M. and
	{Sip{\H{o}}cz}, B.~M. and {G{\"u}nther}, H.~M. and {Lim}, P.~L. and
	{Crawford}, S.~M. and {Conseil}, S. and {Shupe}, D.~L. and
	{Craig}, M.~W. and {Dencheva}, N. and {Ginsburg}, A. and {Vand
	erPlas}, J.~T. and {Bradley}, L.~D. and {P{\'e}rez-Su{\'a}rez}, D. and
	{de Val-Borro}, M. and {Aldcroft}, T.~L. and {Cruz}, K.~L. and
	{Robitaille}, T.~P. and {Tollerud}, E.~J. and {Ardelean}, C. and
	{Babej}, T. and {Bach}, Y.~P. and {Bachetti}, M. and {Bakanov}, A.~V. and
	{Bamford}, S.~P. and {Barentsen}, G. and {Barmby}, P. and
	{Baumbach}, A. and {Berry}, K.~L. and {Biscani}, F. and {Boquien}, M. and
	{Bostroem}, K.~A. and {Bouma}, L.~G. and {Brammer}, G.~B. and
	{Bray}, E.~M. and {Breytenbach}, H. and {Buddelmeijer}, H. and
	{Burke}, D.~J. and {Calderone}, G. and {Cano Rodr{\'\i}guez}, J.~L. and
	{Cara}, M. and {Cardoso}, J.~V.~M. and {Cheedella}, S. and {Copin}, Y. and
	{Corrales}, L. and {Crichton}, D. and {D'Avella}, D. and {Deil}, C. and
	{Depagne}, {\'E}. and {Dietrich}, J.~P. and {Donath}, A. and
	{Droettboom}, M. and {Earl}, N. and {Erben}, T. and {Fabbro}, S. and
	{Ferreira}, L.~A. and {Finethy}, T. and {Fox}, R.~T. and
	{Garrison}, L.~H. and {Gibbons}, S.~L.~J. and {Goldstein}, D.~A. and
	{Gommers}, R. and {Greco}, J.~P. and {Greenfield}, P. and
	{Groener}, A.~M. and {Grollier}, F. and {Hagen}, A. and {Hirst}, P. and
	{Homeier}, D. and {Horton}, A.~J. and {Hosseinzadeh}, G. and {Hu}, L. and
	{Hunkeler}, J.~S. and {Ivezi{\'c}}, {\v{Z}}. and {Jain}, A. and
	{Jenness}, T. and {Kanarek}, G. and {Kendrew}, S. and {Kern}, N.~S. and
	{Kerzendorf}, W.~E. and {Khvalko}, A. and {King}, J. and {Kirkby}, D. and
	{Kulkarni}, A.~M. and {Kumar}, A. and {Lee}, A. and {Lenz}, D. and
	{Littlefair}, S.~P. and {Ma}, Z. and {Macleod}, D.~M. and
	{Mastropietro}, M. and {McCully}, C. and {Montagnac}, S. and
	{Morris}, B.~M. and {Mueller}, M. and {Mumford}, S.~J. and {Muna}, D. and
	{Murphy}, N.~A. and {Nelson}, S. and {Nguyen}, G.~H. and
	{Ninan}, J.~P. and {N{\"o}the}, M. and {Ogaz}, S. and {Oh}, S. and
	{Parejko}, J.~K. and {Parley}, N. and {Pascual}, S. and {Patil}, R. and
	{Patil}, A.~A. and {Plunkett}, A.~L. and {Prochaska}, J.~X. and
	{Rastogi}, T. and {Reddy Janga}, V. and {Sabater}, J. and
	{Sakurikar}, P. and {Seifert}, M. and {Sherbert}, L.~E. and
	{Sherwood-Taylor}, H. and {Shih}, A.~Y. and {Sick}, J. and
	{Silbiger}, M.~T. and {Singanamalla}, S. and {Singer}, L.~P. and
	{Sladen}, P.~H. and {Sooley}, K.~A. and {Sornarajah}, S. and
	{Streicher}, O. and {Teuben}, P. and {Thomas}, S.~W. and
	{Tremblay}, G.~R. and {Turner}, J.~E.~H. and {Terr{\'o}n}, V. and
	{van Kerkwijk}, M.~H. and {de la Vega}, A. and {Watkins}, L.~L. and
	{Weaver}, B.~A. and {Whitmore}, J.~B. and {Woillez}, J. and
	{Zabalza}, V. and {Astropy Contributors}},
	title = "{The Astropy Project: Building an Open-science Project and Status of the v2.0 Core Package}",
	journal = {\aj},
	year = 2018,
	month = sep,
	volume = {156},
	number = {3},
	eid = {123},
	pages = {123},
	doi = {10.3847/1538-3881/aabc4f},
	archivePrefix = {arXiv},
	eprint = {1801.02634},
	primaryClass = {astro-ph.IM},
	adsurl = {https://ui.adsabs.harvard.edu/abs/2018AJ....156..123A}
}

@ARTICLE{astropy:2022,
	author = {{Astropy Collaboration} and {Price-Whelan}, Adrian M. and {Lim}, Pey Lian and {Earl}, Nicholas and {Starkman}, Nathaniel and {Bradley}, Larry and {Shupe}, David L. and {Patil}, Aarya A. and {Corrales}, Lia and {Brasseur}, C.~E. and {N{"o}the}, Maximilian and {Donath}, Axel and {Tollerud}, Erik and {Morris}, Brett M. and {Ginsburg}, Adam and {Vaher}, Eero and {Weaver}, Benjamin A. and {Tocknell}, James and {Jamieson}, William and {van Kerkwijk}, Marten H. and {Robitaille}, Thomas P. and {Merry}, Bruce and {Bachetti}, Matteo and {G{"u}nther}, H. Moritz and {Aldcroft}, Thomas L. and {Alvarado-Montes}, Jaime A. and {Archibald}, Anne M. and {B{'o}di}, Attila and {Bapat}, Shreyas and {Barentsen}, Geert and {Baz{'a}n}, Juanjo and {Biswas}, Manish and {Boquien}, M{'e}d{'e}ric and {Burke}, D.~J. and {Cara}, Daria and {Cara}, Mihai and {Conroy}, Kyle E. and {Conseil}, Simon and {Craig}, Matthew W. and {Cross}, Robert M. and {Cruz}, Kelle L. and {D'Eugenio}, Francesco and {Dencheva}, Nadia and {Devillepoix}, Hadrien A.~R. and {Dietrich}, J{"o}rg P. and {Eigenbrot}, Arthur Davis and {Erben}, Thomas and {Ferreira}, Leonardo and {Foreman-Mackey}, Daniel and {Fox}, Ryan and {Freij}, Nabil and {Garg}, Suyog and {Geda}, Robel and {Glattly}, Lauren and {Gondhalekar}, Yash and {Gordon}, Karl D. and {Grant}, David and {Greenfield}, Perry and {Groener}, Austen M. and {Guest}, Steve and {Gurovich}, Sebastian and {Handberg}, Rasmus and {Hart}, Akeem and {Hatfield-Dodds}, Zac and {Homeier}, Derek and {Hosseinzadeh}, Griffin and {Jenness}, Tim and {Jones}, Craig K. and {Joseph}, Prajwel and {Kalmbach}, J. Bryce and {Karamehmetoglu}, Emir and {Ka{l}uszy{'n}ski}, Miko{l}aj and {Kelley}, Michael S.~P. and {Kern}, Nicholas and {Kerzendorf}, Wolfgang E. and {Koch}, Eric W. and {Kulumani}, Shankar and {Lee}, Antony and {Ly}, Chun and {Ma}, Zhiyuan and {MacBride}, Conor and {Maljaars}, Jakob M. and {Muna}, Demitri and {Murphy}, N.~A. and {Norman}, Henrik and {O'Steen}, Richard and {Oman}, Kyle A. and {Pacifici}, Camilla and {Pascual}, Sergio and {Pascual-Granado}, J. and {Patil}, Rohit R. and {Perren}, Gabriel I. and {Pickering}, Timothy E. and {Rastogi}, Tanuj and {Roulston}, Benjamin R. and {Ryan}, Daniel F. and {Rykoff}, Eli S. and {Sabater}, Jose and {Sakurikar}, Parikshit and {Salgado}, Jes{'u}s and {Sanghi}, Aniket and {Saunders}, Nicholas and {Savchenko}, Volodymyr and {Schwardt}, Ludwig and {Seifert-Eckert}, Michael and {Shih}, Albert Y. and {Jain}, Anany Shrey and {Shukla}, Gyanendra and {Sick}, Jonathan and {Simpson}, Chris and {Singanamalla}, Sudheesh and {Singer}, Leo P. and {Singhal}, Jaladh and {Sinha}, Manodeep and {Sip{H{o}}cz}, Brigitta M. and {Spitler}, Lee R. and {Stansby}, David and {Streicher}, Ole and {{{S}}umak}, Jani and {Swinbank}, John D. and {Taranu}, Dan S. and {Tewary}, Nikita and {Tremblay}, Grant R. and {Val-Borro}, Miguel de and {Van Kooten}, Samuel J. and {Vasovi{'c}}, Zlatan and {Verma}, Shresth and {de Miranda Cardoso}, Jos{'e} Vin{'i}cius and {Williams}, Peter K.~G. and {Wilson}, Tom J. and {Winkel}, Benjamin and {Wood-Vasey}, W.~M. and {Xue}, Rui and {Yoachim}, Peter and {Zhang}, Chen and {Zonca}, Andrea and {Astropy Project Contributors}},
	title = "{The Astropy Project: Sustaining and Growing a Community-oriented Open-source Project and the Latest Major Release (v5.0) of the Core Package}",
	journal = {\apj},
	year = 2022,
	month = aug,
	volume = {935},
	number = {2},
	eid = {167},
	pages = {167},
	doi = {10.3847/1538-4357/ac7c74},
	archivePrefix = {arXiv},
	eprint = {2206.14220},
	primaryClass = {astro-ph.IM},
	adsurl = {https://ui.adsabs.harvard.edu/abs/2022ApJ...935..167A}
}

@ARTICLE{2021JOSS....6.3675G,
	author = {{Grudi{\'c}}, Michael and {Gurvich}, Alexander},
	title = "{pytreegrav: A fast Python gravity solver}",
	journal = {The Journal of Open Source Software},
	year = 2021,
	month = dec,
	volume = {6},
	number = {68},
	eid = {3675},
	pages = {3675},
	doi = {10.21105/joss.03675},
	adsurl = {https://ui.adsabs.harvard.edu/abs/2021JOSS....6.3675G}
}

@ARTICLE{2025A&A...695L..25I,
	author = {{Ishihara}, Kousuke and {Nakamura}, Fumitaka and {Sanhueza}, Patricio and {Saito}, Masao},
	title = "{Turbulent fragmentation as the primary driver of core formation in Polaris Flare and Lupus I}",
	journal = {\aap},
	year = 2025,
	month = mar,
	volume = {695},
	eid = {L25},
	pages = {L25},
	doi = {10.1051/0004-6361/202452427},
	archivePrefix = {arXiv},
	eprint = {2503.06613},
	primaryClass = {astro-ph.GA},
	adsurl = {https://ui.adsabs.harvard.edu/abs/2025A&A...695L..25I}
}

@ARTICLE{2016A&A...586A.138P,
       author = {{Planck Collaboration} and {Ade}, P.~A.~R. and {Aghanim}, N. and {Alves}, M.~I.~R. and {Arnaud}, M. and {Arzoumanian}, D. and {Ashdown}, M. and {Aumont}, J. and {Baccigalupi}, C. and {Banday}, A.~J. and {Barreiro}, R.~B. and {Bartolo}, N. and {Battaner}, E. and {Benabed}, K. and {Beno{\^\i}t}, A. and {Benoit-L{\'e}vy}, A. and {Bernard}, J. -P. and {Bersanelli}, M. and {Bielewicz}, P. and {Bock}, J.~J. and {Bonavera}, L. and {Bond}, J.~R. and {Borrill}, J. and {Bouchet}, F.~R. and {Boulanger}, F. and {Bracco}, A. and {Burigana}, C. and {Calabrese}, E. and {Cardoso}, J. -F. and {Catalano}, A. and {Chiang}, H.~C. and {Christensen}, P.~R. and {Colombo}, L.~P.~L. and {Combet}, C. and {Couchot}, F. and {Crill}, B.~P. and {Curto}, A. and {Cuttaia}, F. and {Danese}, L. and {Davies}, R.~D. and {Davis}, R.~J. and {de Bernardis}, P. and {de Rosa}, A. and {de Zotti}, G. and {Delabrouille}, J. and {Dickinson}, C. and {Diego}, J.~M. and {Dole}, H. and {Donzelli}, S. and {Dor{\'e}}, O. and {Douspis}, M. and {Ducout}, A. and {Dupac}, X. and {Efstathiou}, G. and {Elsner}, F. and {En{\ss}lin}, T.~A. and {Eriksen}, H.~K. and {Falceta-Gon{\c{c}}alves}, D. and {Falgarone}, E. and {Ferri{\`e}re}, K. and {Finelli}, F. and {Forni}, O. and {Frailis}, M. and {Fraisse}, A.~A. and {Franceschi}, E. and {Frejsel}, A. and {Galeotta}, S. and {Galli}, S. and {Ganga}, K. and {Ghosh}, T. and {Giard}, M. and {Gjerl{\o}w}, E. and {Gonz{\'a}lez-Nuevo}, J. and {G{\'o}rski}, K.~M. and {Gregorio}, A. and {Gruppuso}, A. and {Gudmundsson}, J.~E. and {Guillet}, V. and {Harrison}, D.~L. and {Helou}, G. and {Hennebelle}, P. and {Henrot-Versill{\'e}}, S. and {Hern{\'a}ndez-Monteagudo}, C. and {Herranz}, D. and {Hildebrandt}, S.~R. and {Hivon}, E. and {Holmes}, W.~A. and {Hornstrup}, A. and {Huffenberger}, K.~M. and {Hurier}, G. and {Jaffe}, A.~H. and {Jaffe}, T.~R. and {Jones}, W.~C. and {Juvela}, M. and {Keih{\"a}nen}, E. and {Keskitalo}, R. and {Kisner}, T.~S. and {Knoche}, J. and {Kunz}, M. and {Kurki-Suonio}, H. and {Lagache}, G. and {Lamarre}, J. -M. and {Lasenby}, A. and {Lattanzi}, M. and {Lawrence}, C.~R. and {Leonardi}, R. and {Levrier}, F. and {Liguori}, M. and {Lilje}, P.~B. and {Linden-V{\o}rnle}, M. and {L{\'o}pez-Caniego}, M. and {Lubin}, P.~M. and {Mac{\'\i}as-P{\'e}rez}, J.~F. and {Maino}, D. and {Mandolesi}, N. and {Mangilli}, A. and {Maris}, M. and {Martin}, P.~G. and {Mart{\'\i}nez-Gonz{\'a}lez}, E. and {Masi}, S. and {Matarrese}, S. and {Melchiorri}, A. and {Mendes}, L. and {Mennella}, A. and {Migliaccio}, M. and {Miville-Desch{\^e}nes}, M. -A. and {Moneti}, A. and {Montier}, L. and {Morgante}, G. and {Mortlock}, D. and {Munshi}, D. and {Murphy}, J.~A. and {Naselsky}, P. and {Nati}, F. and {Netterfield}, C.~B. and {Noviello}, F. and {Novikov}, D. and {Novikov}, I. and {Oppermann}, N. and {Oxborrow}, C.~A. and {Pagano}, L. and {Pajot}, F. and {Paladini}, R. and {Paoletti}, D. and {Pasian}, F. and {Perotto}, L. and {Pettorino}, V. and {Piacentini}, F. and {Piat}, M. and {Pierpaoli}, E. and {Pietrobon}, D. and {Plaszczynski}, S. and {Pointecouteau}, E. and {Polenta}, G. and {Ponthieu}, N. and {Pratt}, G.~W. and {Prunet}, S. and {Puget}, J. -L. and {Rachen}, J.~P. and {Reinecke}, M. and {Remazeilles}, M. and {Renault}, C. and {Renzi}, A. and {Ristorcelli}, I. and {Rocha}, G. and {Rossetti}, M. and {Roudier}, G. and {Rubi{\~n}o-Mart{\'\i}n}, J.~A. and {Rusholme}, B. and {Sandri}, M. and {Santos}, D. and {Savelainen}, M. and {Savini}, G. and {Scott}, D. and {Soler}, J.~D. and {Stolyarov}, V. and {Sudiwala}, R. and {Sutton}, D. and {Suur-Uski}, A. -S. and {Sygnet}, J. -F. and {Tauber}, J.~A. and {Terenzi}, L. and {Toffolatti}, L. and {Tomasi}, M. and {Tristram}, M. and {Tucci}, M. and {Umana}, G. and {Valenziano}, L. and {Valiviita}, J. and {Van Tent}, B. and {Vielva}, P. and {Villa}, F. and {Wade}, L.~A. and {Wandelt}, B.~D. and {Wehus}, I.~K. and {Ysard}, N. and {Yvon}, D. and {Zonca}, A.},
        title = "{Planck intermediate results. XXXV. Probing the role of the magnetic field in the formation of structure in molecular clouds}",
      journal = {\aap},
         year = 2016,
        month = feb,
       volume = {586},
          eid = {A138},
        pages = {A138},
          doi = {10.1051/0004-6361/201525896},
archivePrefix = {arXiv},
       eprint = {1502.04123},
 primaryClass = {astro-ph.GA},
       adsurl = {https://ui.adsabs.harvard.edu/abs/2016A&A...586A.138P}
}

@ARTICLE{2024MNRAS.528.3630G,
       author = {{Ganguly}, Shashwata and {Walch}, S. and {Clarke}, S.~D. and {Seifried}, D.},
        title = "{SILCC-Zoom: the dynamic balance in molecular cloud substructures}",
      journal = {\mnras},
         year = 2024,
        month = feb,
       volume = {528},
       number = {2},
        pages = {3630-3657},
          doi = {10.1093/mnras/stae032},
archivePrefix = {arXiv},
       eprint = {2204.02511},
 primaryClass = {astro-ph.GA},
       adsurl = {https://ui.adsabs.harvard.edu/abs/2024MNRAS.528.3630G}
}

@ARTICLE{2025arXiv251012892W,
       author = {{Wallace}, Jennifer and {Kolz}, Taevis and {Battersby}, Cara and {Kuznetsova}, Aleksandra and {S{\'a}nchez-Monge}, {\'A}lvaro and {Schisano}, Eugenio and {Coletta}, Alessandro and {Zhang}, Qizhou and {Molinari}, Sergio and {Schilke}, Peter and {Ho}, Paul T.~P. and {Kuiper}, Rolf and {Zhang}, Tianwei and {M{\"o}ller}, Thomas and {Klessen}, Ralf S. and {Beltr{\'a}n}, Maria T. and {van der Tak}, Floris and {Pezzuto}, Stefania and {Beuther}, Henrik and {Traficante}, Alessio and {Elia}, Davide and {Bronfman}, Leonardo and {Klaassen}, Pamela and {Lis}, Dariusz C. and {Moscadelli}, Luca and {Rygl}, Kazi and {Benedettini}, Milena and {Law}, Chi Yan and {Allande}, Jofre and {Nucara}, Alice and {Koch}, Patrick M. and {Kim}, Won-ju and {Sanhueza}, Patricio and {Fuller}, Gary and {Stroud}, Georgie and {Jones}, Beth and {Brogan}, Crystal and {Hunter}, Todd and {Ahmadi}, Aida and {Avison}, Adam and {Johnston}, Katharine and {Liu}, Sheng-Yuan and {Mininni}, Chiara and {Su}, Yu-Nung},
        title = "{ALMAGAL VIII. Cataloging Hierarchical Mass Structure from Cores to Clumps across the Galactic Disk}",
      journal = {arXiv e-prints},
         year = 2025,
        month = oct,
          eid = {arXiv:2510.12892},
        pages = {arXiv:2510.12892},
archivePrefix = {arXiv},
       eprint = {2510.12892},
 primaryClass = {astro-ph.GA},
       adsurl = {https://ui.adsabs.harvard.edu/abs/2025arXiv251012892W}
}

@ARTICLE{2004RvMP...76..125M,
       author = {{Mac Low}, Mordecai-Mark and {Klessen}, Ralf S.},
        title = "{Control of star formation by supersonic turbulence}",
      journal = {Reviews of Modern Physics},
         year = 2004,
        month = jan,
       volume = {76},
       number = {1},
        pages = {125-194},
          doi = {10.1103/RevModPhys.76.125},
archivePrefix = {arXiv},
       eprint = {astro-ph/0301093},
 primaryClass = {astro-ph},
       adsurl = {https://ui.adsabs.harvard.edu/abs/2004RvMP...76..125M}
}
\bibliographystyle{aasjournal}

\end{document}